\documentstyle[preprint,floats,prd,aps,rotating]{revtex}
\def\PRref#1&#2&#3(#4){\unskip\ #1~\bf #2\rm, #3 (#4)}

\def\NIMA{Nucl. Inst. and Meth. A}
\def\NPB{Nucl. Phys. B}
\def\PLB{Phys. Lett. B}
\def\PRL{Phys. Rev. Lett.}
\def\PRD{Phys. Rev. D}
\def\ZPC{Z. Phys. C}
\def\etal{{\it et al.}}
\input psfig
\begin{document}
\tighten
\preprint{\vbox{\hbox{UH-515-848-96 \hfill}
                \hbox{OHSTPY-HEP-E-96-006 \hfill}
                \hbox{\today        \hfill}}}

%
%
\title{Nonleptonic Decays and Lifetimes of \\ b-quark and c-quark Hadrons}
\author{ Thomas~E.~Browder \\ \scriptsize                       
 University of Hawaii at Manoa, Honolulu, Hawaii 96822, U.S.A.
\and \\ \normalsize Klaus~Honscheid \\ \scriptsize
 Ohio State University, Columbus, Ohio 43210, U.S.A.
\and \\ \normalsize Daniele~Pedrini \\ \scriptsize
 Istituto Nazionale di Fisica Nucleare - sezione di Milano, I-20133 Milan, 
 Italy }
\maketitle
\par\vskip 1.cm

\begin{center}
{\bf Abstract}
\end{center}
We review recent experimental results on lifetimes
 and hadronic decays of hadrons that contain $c$ and $b$ quarks.
The theoretical implications of these results are also considered.
An understanding of hadronic decays of heavy quarks is required
to interpret the CP violating asymmetries in $B$ decays that will be observed
in experiments planned for the near future.

\vfill
\centerline{To appear in {\it Annual Review of Nuclear and Particle 
Science, Vol. 46}.}
{\bf 
A copy of this review can be obtained via anonymous ftp or from our WWW
site.\\}
\begin{center}
\begin{tabular}{lll}
Anonymous ftp & Node & ftp.physics.ohio-state.edu \\
 & Directory: &  pub/hepex/kh \\
WWW & URL &
www-physics.mps.ohio-state.edu/$\sim$phys111/b-physics/bphysics.html \\
\end{tabular}
\end{center}
\newpage
\tableofcontents

\newpage
\section{\bf Introduction}

Heavy-flavor physics began in
1974 with the discovery of the $J/\psi$ meson\cite{J/psi}, 
a narrow resonance at a mass of 3.1 GeV. 
The $J/\psi$ was quickly identified as a bound state of  a
charm and anti-charm quarks, a previously 
unobserved quark flavor with a mass around 1.5 GeV.

Charm was not only the first heavy flavor quark, it was
also the first quark whose existence was predicted before its
discovery. In 1970, Glashow, Illiopoulos, and Maiani 
introduced the GIM mechanism and postulated a new type of quark 
in order to explain the absence of flavor-changing 
neutral currents in kaon decay\cite{GIM}.

In 1977,
the second heavy flavor, the bottom (or b)
quark with a mass of $m_{b}\sim 5 \,$
GeV/$c^2$ and a charge of -1/3, was observed at 
Fermilab in the bound states of the $\Upsilon$ family\cite{Upsilon}.

The recent observation of the top quark by the CDF and D0 
collaborations \cite{Top} completes the three quark families of the
Standard Model:
\begin{center}
\begin{math}
\left( \begin{array}{c} u \\ d \end{array} \right)
\left( \begin{array}{c} c \\ s \end{array} \right)
\left( \begin{array}{c} t \\ b \end{array} \right)
\end{math}
\end{center}
The six quarks are divided naturally into heavy and light 
flavors. The $c$, $b$, and $t$ quarks are called heavy 
because their masses are 
larger than the QCD scale, $\Lambda$, while the 
masses of the remaining quarks are lighter.

Weak decays of heavy quarks test the 
Standard Model and can be used
to determine its parameters, including the weak mixing
angles of the Cabibbo-Kobayashi-Maskawa (CKM) matrix\CITE{CKM}. 
In addition, the study of heavy quark decay provides important insight 
into the least well understood sector of the strong interaction: the 
non-perturbative regime which describes the formation of hadrons from 
quarks.

 In the Standard Model the charm (bottom) quark decays through the weak 
charged current into a light quark with a charge of 
$-1/3$ ($+2/3$), {\it i.e.,} an 
s (c) or d (u) quark. The coupling is proportional to the element
$V_{Qq}$ of the 
CKM mixing matrix, where Q denotes a  heavy quark,
either $c$ or $b$.
  In charm decays the CKM
matrix can be approximated by a $2 \times 2$ rotation
matrix with one real angle, the Cabibbo angle
$\theta_c \sim 14^0$. In this approximation,
the $c \rightarrow W s$ transition, proportional
to $\cos\theta_c$, is favoured with respect to the $c \rightarrow W d$ 
transition proportional to $\sin\theta_c$. These two types of transitions are
called Cabibbo-favoured and Cabibbo-suppressed, respectively.

 The lowest order decay diagrams for charm (bottom) mesons 
are shown in Fig. 1. The spectator diagram (Figs. 1 (a) and (b)), 
in which the light antiquark 
does not take part in the weak interaction, is thought to be
dominant. As in muon decay, the decay rate for this diagram is proportional to 
$m_{Q}^5$.
In the external spectator diagram (Fig. 1(a)) color is automatically 
conserved, while the internal spectator amplitude
(Fig. 1(b)) is color suppressed since the color of the
quarks from the virtual W must match the color 
of the quarks from the parent meson. In the na\"\i ve quark model the color
matching factor $\xi$ has a value of $1/N_c = 1/3$, so that the decay
rate should be reduced by a factor $1/18$ ($=(1/3)^2\times (1/\sqrt{2})^2$
for the $\pi^0$ wave function) for a decay such as $\bar{B}^0\to D^0\pi^0$.

 The exchange and annihilation diagrams (Figs. 1(c), (d)) are helicity 
suppressed. This suppression can be somewhat
mitigated by the emission of soft gluons. There is also a further reduction 
in the amplitude which is proportional to the 
magnitude of the wavefunction at the origin. 

In addition, there are small contributions from the
penguin diagram and the box diagrams, which are
responsible for $B^{0}-\overline{B^{0}}$ mixing. These
are shown in Figs. 1(e) and (f), 
respectively. Due to the GIM mechanism
these diagrams are highly suppressed in charm decay.

Decays of heavy baryons containing
charm or bottom quarks are more complex. The annihilation 
amplitude is absent, but the exchange diagram is no longer 
helicity suppressed. The dominant
 hadronic decay mechanisms for charm (bottom) 
baryons are shown in Fig. 2.
The external spectator decay mechanism is shown in Fig. 2(a) while
the diagrams for the
internal spectator contributions are shown in Figs. 2(b) and (c).
Fig. 2(d) shows the W-exchange mechanism. 
The contribution of diagrams other than the external spectator diagram
is expected to
be significant for decays of baryons with heavy quarks.

Decay modes can be subdivided into three categories according to
 the final state particles produced. These are
leptonic, semileptonic and hadronic decays. The first can only proceed
by the annihilation diagram, while semileptonic
decays occur by the spectator diagram.
Hadronic decays may proceed via all the decay mechanisms. 
In contrast to semileptonic and purely leptonic transitions, 
hadronic decays involve an intricate interplay
of  quark rearrangement due to 
soft and hard gluon exchanges. In addition, the hadrons in the final state
can rescatter into one another. 
For example, a $D^0$ can 
decay directly into $\overline{K^0}\pi^0$ or rescatter
via the intermediate state
$K^-\pi^+$, since $K^-\pi^+\rightarrow \overline K^0\pi^0$ 
is an allowed strong interaction. These processes are referred to generically
as final state interactions, (FSI).

Although readily accommodated in the Standard
Model by a complex phase in the CKM matrix,
CP violation remains one of the least well
understood phenomena in physics. 
So far it has only been observed in the decays of kaons. 
While the results from the kaon sector are consistent with 
the Standard Model, the complications introduced by
strong interaction effects make it nearly impossible to 
ascertain whether the complex CKM phase is the sole source
for the observed asymmetries. If the Standard Model is correct,
large CP asymmetries are expected in hadronic $B$ decays
to CP eigenstates. Efforts are now underway at
every major high energy physics laboratory
to observe these CP violating effects in the $B$ sector.

Data samples at least one order of magnitude larger
than those available at present are required to observe
CP asymmetries in the $B$ meson system and to provide
fundamental consistency checks of the Standard Model.
This is the justification for the 
construction of high luminosity $e^+ e^-$
storage rings in the US at SLAC(PEP~II/BABAR) and Cornell
(CESR PHASE~III/CLEO~III) and in
Japan(KEK-B/BELLE), 
as well as for the dedicated fixed target experiment
at the HERA ring at DESY. 
Hadron collider experiments 
dedicated to the study of CP violation have also been proposed 
at Fermilab and at CERN. 
In addition, these new machines will produce large samples
of charm mesons and baryons which can also be studied in detail.

In order to extract information about the weak phase from the
asymmetries that will be observed by these 
experiments in the near future, 
an understanding of the interplay between the weak
and strong interaction  responsible for hadronic decays
and of the lifetimes of particles containing heavy quarks is needed.
In this review we  describe recent experimental results
on lifetimes and decays of mesons and baryons containing heavy quarks
and we report on the progress in interpreting these results. 

Semileptonic and leptonic decays of charm and bottom hadrons have been 
reviewed elsewhere\cite{Review_2,Witherell}.
More detailed reviews 
of B decays are also available\cite{Review_1}.
\section{Experimental Study of Charm and Bottom Decay}

For many years after the discovery of the charm quark in fixed-target
and $e^+e^-$ collisions, $e^+e^-$ colliders provided most of the results in 
the study of charmed hadrons. 
In the mid-eighties, however, the 
introduction of silicon vertex detectors made fixed-target experiments 
competitive once again\cite{Kaplan}. Now Fermilab fixed-target 
experiments dominate several areas of charm physics including lifetime 
measurements and rare decay searches.

Table~\ref{charmepem} gives
the sizes of charm data samples from $e^+e^-$ 
colliding beam experiments\cite{Review_2}. The major advantage offered 
by $e^+e^-$ annihilation is that the fraction of hadronic events 
containing heavy quarks is relatively large and hence backgrounds
are small.  
In fixed target experiments the production cross section is larger
 but the fraction of hadronic 
events that contain charm particles is much smaller. The
charm hadroproduction cross section is on the 
order of $20~\mu b$ (for an incident
 proton momentum of $\sim 400$ GeV/c), but charm 
events represent only about $10^{-3}$ of the total cross 
section\cite{Review_2}. Photoproduction has a smaller charm cross section
but  a larger fraction of charm produced.
 Table~\ref{charmfix} gives the number of 
reconstructed charm decays for several 
fixed-target experiments. The current data samples contain ${\cal{O}}$($10^5$) 
reconstructed charm decays. Samples with  ${\cal{O}}$($10^6$) reconstructed 
events are expected 
during the next few years from  Fermilab experiments E781 (SELEX) 
and E831 (FOCUS), as well as in $e^+e^-$ annihilation from CLEO III at CESR.

 Most of the current knowledge of the decays of $B$ mesons
is based on analyses of data collected by experiments at CESR and
DORIS. These experiments record data at
the $\Upsilon (4S)$ resonance, which is the lowest lying $b \bar{b}$ resonance
above the threshold for $B \bar{B}$ pair production.
The observed 
events originate from the decay of either a $B$ or a $\bar{B}$ meson
as there is not sufficient energy
to produce additional particles. The
$B$ mesons are also produced nearly at rest. The average momentum is about
$330$~MeV so the average decay length is approximately 30 $\mu m$.

In recent years, advances in detector technology, in particular the
introduction of high resolution 
silicon vertex detectors have allowed experiments
at high energy colliders (i.e. LEP, SLC and the TEVATRON) to observe 
decay vertices of $b$ quarks. 
This has led to precise lifetime measurements, as well as to the direct 
observation of time dependent $B-\bar{B}$ mixing and to the 
discovery of new $b$-flavored hadrons.

The first fully reconstructed $B$ mesons were reported in 1983 by the
CLEO~I collaboration. 
Since then the CLEO~1.5 experiment 
has collected a sample with an integrated
luminosity of $212 ~\rm{pb}^{-1}$,
the ARGUS experiment has collected $246 ~\rm{pb}^{-1}$, 
and to date the CLEO~II experiment
has collected about $4 ~\rm{fb}^{-1}$, of which  up to
$3 ~\rm{fb}^{-1}$ have been used 
to obtain the results described in this review.

For quantitative studies of $B$ decays the initial composition of the data
sample must be known. The ratio of 
the production of neutral and charged $B$ mesons from the
$\Upsilon (4S)$ is, therefore, an important
parameter for these experiments. The ratio is denoted ${f_+}/{f_0}$
and is measured by CLEO \cite{dstlnu} to be,
$$
\frac{f_+}{f_0}\; = \; \frac{{\cal B}(\Upsilon(4S) \to B^+B^-)}
{{\cal B}(\Upsilon(4S) \to B^0\bar{B}^0)} 
\; = \; 1.13 \pm 0.14 \pm 0.13 \pm 0.06.
$$
The third error is due to the uncertainty in the ratio of $B^0$ and
$B^+$ lifetimes.
This result is consistent with equal production of 
$B^+ B^-$ and $B^0 \bar{B^0}$ pairs and unless explicitly stated otherwise 
we will assume that $f_+/f_0\, = \, 1$. The assumption 
of equal production of charged and neutral $B$ mesons is further supported
by the near equality of the observed $B^-$ and $\bar{B}^0$ masses.
Older experimental results 
which assumed other values of $f_+$ and $f_0$ have been rescaled.

Two variables are used to isolate  exclusive
hadronic $B$ decay modes at CLEO and ARGUS.
To determine the signal yield and
display the data the beam constrained mass is formed
\begin{equation} M_B^2=E_{beam}^2 - \left(\sum_i{\vec{p_i}}\right)^2,
\label{EBmass}
\end{equation}
where $\vec{p_i}$ is the reconstructed
momentum of the $i$-th daughter of the $B$ candidate.
An example is shown in Fig. \ref{FBM}.
The resolution in this variable is determined by the beam energy spread,
and is about 2.7~MeV for CLEO~II, and about 4.0~MeV for ARGUS.
These resolutions are a factor of ten better
than the resolution in invariant mass obtained without the beam energy
constraint.
The measured sum of charged and neutral energies, $E_{meas}$,
of correctly reconstructed $B$ mesons produced at the
$\Upsilon (4S)$, must also equal the
beam energy, $E_{beam}$, to within the experimental resolution.
Depending on the $B$ decay mode,
$\sigma_{\Delta E}$, the resolution on the energy difference
$\Delta E\; = \; E_{beam} - E_{meas}$
varies between 14 and 46~MeV.
Note that this resolution is usually sufficient to distinguish
the correct $B$ decay mode from a mode with one additional or one fewer
pion.

\subsection {High Energy Collider Experiments}

The four LEP experiments and SLD operate on the $Z^0$ resonance. At this
energy, the cross section for $b\bar{b}$ production  is about
6.6 nb and the signal-to-noise ratio for hadronic events is 1:5, comparable
to the $\Upsilon(4S)$ resonance. 
Compared with $e^+e^-$ annihilation, the $b\bar{b}$ production cross section
at hadron colliders is enormous, about $\, 50 \mu b$ at 1.8 TeV.
However, a signal-to-background ratio of about 1:1000 makes it difficult to
extract $b$ quark signals and to fully reconstruct $B$ mesons.

The kinematic constraints available
on the $\Upsilon(4S)$ cannot be used on the $Z^0$.
However, due to
the large boost the 
$b$ quarks travel $\approx$ 2.5 mm before they decay and the decay products of 
the two $b$-hadrons are clearly separated in the detector.
The large boost makes precise lifetime measurements possible.

\subsection{Averaging Experimental Results}

To extract $B$ meson branching fractions, the detection efficiencies are
determined from a Monte Carlo simulation and the yields are corrected
for the charmed meson branching fractions. 
In order to determine world
average branching fractions for $B$ and $D$ meson
decays the results from individual experiments must be normalized
with respect to a common set of absolute branching fractions of
charm mesons and baryons. The branching fractions for
the $D^0$ and $D^+$ modes used to calculate the $B$ branching fractions
are given in Table~\ref{Cabfavd}. 
For the $D^0 \to K^- \pi^+$ branching fraction 
we have chosen an average of values recently reported by
the CLEO~II, ARGUS and ALEPH experiments \cite{DKpi}. 
The value 
${\cal B}(D^+\to K^-\pi^+\pi^+)=8.9\pm 0.7\%$ is used in this review
to normalize branching fractions for $D^+$ modes.
Our value for ${\cal{B}}(D^0 \to K^- \pi^+\pi^0)$ is calculated using
a recent result from CLEO II \cite{CLEO-d0kpipi0}, 
${\cal{B}}(D^0 \to K^- \pi^+\pi^0)/{\cal{B}}(D^0 \to K^- \pi^+)\, =
\, (3.67 \pm 0.08 \pm 0.23)$, averaged with an older measurement from
ARGUS \cite{ARGUS-d0kpipi0}.
The branching ratios of other $D^0$
decay modes relative to $D^0 \to K^- \pi^+$ are taken from the PDG
compilation\cite{PDG}. The $D^+$ branching ratios are also taken
from the PDG compilation \cite{PDG}. The CLEO II results for
$D^+ \to K^- \pi^+\pi^+$, however, has been re-scaled to account for the new
$D^0 \to K^-\pi^+$ branching fraction. For older measurements of $B$
decays involving $D^*$ mesons, the branching fractions have been rescaled
to account for improved measurements of the $D^*$ branching fractions.

Branching ratios for all $D_s$ decay modes are normalized relative to
${\cal{B}}(D_s^+\rightarrow \phi \pi^+)$. 
Two model-independent
measurements of the absolute branching fraction for
$D_s^+\rightarrow \phi \pi^+$ have been published by BES \cite{BES-Ds}
and CLEO \cite{CLEO-Ds}. These have been averaged to determine the value
used here (Table \ref{Cabfavds}). 
Branching ratios involving $D_s^*$ modes are also
re-scaled to account for  the isospin violating decay $D_s^* \to D_s \pi^0$
recently observed by CLEO \cite{Ds-isospin}. 

The determination of branching fractions for $B$ decays to charmed baryons 
requires knowledge of 
${\cal{B}}(\Lambda_c^+ \to pK^-\pi^+)$. The uncertainty in this
quantity is still large 
as it can only be determined by indirect and somewhat
model dependent methods. 
In this review we use 
${\cal{B}}(\Lambda_c^+ \to pK^-\pi^+)\, =\, 4.4 \pm 0.6$\%
as determined by the particle data group \cite{PDG}.
However, recent studies of baryon production in B decay indicate
that the production model that 
is assumed in many of the determinations of ${\cal B}(\Lambda_c\to p K^-
\pi^+)$ is flawed (see Section VC for a detailed discussion).
An alternate method used by CLEO \cite{cleo_lambdac} is based on a measurement of the
relative semileptonic rate, $\Gamma(\Lambda_c \to pK^-\pi^+)/\Gamma(\Lambda_c \to
\Lambda \ell^+ \nu_{\ell})$. With additional assumptions this leads to 
${\cal{B}}(\Lambda_c^+ \to pK^-\pi^+)\, =\, 5.9 \pm 0.3 \pm 0.14$\%

Statistical errors are recalculated in the same way as the branching ratios.
For results from individual 
experiments on $B$ decays to final states with $D$
mesons two systematic errors are quoted.
The second systematic error contains the contribution 
due to the uncertainties in the $D^0\to K^-\pi^+$, $D^+\to K^-\pi^+\pi^+$
or $D_s^+ \to \phi \pi^+$
branching fractions. This will allow easier rescaling at a time when these
branching fractions  are measured more precisely.
The first systematic error includes the experimental uncertainties and when 
relevant the uncertainties in the
ratios of charm branching ratios, e.g. $\Gamma(D^0 \rightarrow
K^- \pi^+ \pi^+ \pi^-)/\Gamma(D^0 \rightarrow K^- \pi^+)$ and the error in the
$D^*$ branching fractions. For modes involving $D_s^+$ mesons, the first 
systematic error also includes the uncertainties
due to the $D^0$ and $D^+$ branching ratios.
For all other modes only one systematic error is given.
For the world averages, the
 statistical and the first systematic error are combined
in quadrature while the errors due to the 
$D^0$, $D^+$ and $D_s$ branching ratio scales 
are still listed separately.

With the improvement in the precision
of the $D^0$ and $D^*$ branching fractions these are no longer the
dominant source of systematic error in the study of hadronic $B$
meson decay. The errors on the $D_s^+$ and $\Lambda_c^+$ 
branching ratio scales remain large.

\section{Lifetime Measurements}

\subsection{Theoretical Expectations
for Lifetimes of Hadrons with Heavy Quarks} 
In the naive spectator model the external spectator amplitude is the only
weak decay mechanism and thus the lifetimes of 
all mesons and baryons containing heavy quarks should be equal.
Differences in hadronic decay channels and interference between
contributing amplitudes modify this simple picture
and give rise to a hierarchy of lifetimes.
Experimentally, we find the measured
lifetimes to be significantly different. For example,
the $D^+$ lifetime is $\sim 2.5$ times longer than the $D^0$ lifetime.

 The decay width of charmed hadrons 
($\Gamma_{tot} = \Gamma_l + \Gamma_{sl} + 
\Gamma_{had}$) is dominated by the hadronic component.
For example, for the $D^+$ meson one finds that
the semileptonic component, 
$\Gamma_{sl} = (16.3 \pm 1.8) \times 10^{10} s^{-1}$, is a small 
fraction of the total width $\Gamma = (94.6 \pm 1.4) \times 10^{10} s^{-1}$.
The contribution from purely leptonic decays can be neglected.

 Measurements of the lifetime ratio 
$\tau(D^+)/\tau(D^0) = 2.547\pm 0.044$\cite{PDG} and of the inclusive 
semileptonic branching ratios, $D^+\rightarrow eX = (17.2\pm1.9)\%$\cite{PDG}
and $D^0\rightarrow eX = (6.64\pm 0.18\pm 0.29)\%$ (using a recent
result from CLEO\cite{CLEO_4}), show that the 
$D^0$ and $D^+$ semileptonic decay widths are nearly equal.
\begin{equation}
 \frac{\Gamma(D^0\rightarrow eX)}{\Gamma(D^+\rightarrow eX)} =
 \frac{{\cal{B}}(D^0\rightarrow eX)}{{\cal{B}}(D^+\rightarrow eX)} \times 
 \frac{\tau(D^+)}{\tau(D^0)} = 0.98 \pm 0.11
\end{equation}
This implies that differences in the total decay widths must be due to 
differences in the hadronic  decay amplitudes. 

 In the past, it was suggested that the large
difference in the charm meson lifetimes was
due to the presence of the exchange (annihilation) diagram for the
$D^0$ ($D_s$). A more reliable explanation invokes the destructive 
interference of the external and internal spectator diagrams, which decrease 
the hadronic width of the $D^+$. 
The external and internal spectator diagrams can give the 
same final states only for the $D^+$ meson 
and not for the $D^0$ or $D_s^+$ mesons 
(see Figs 1(a) and (b)). The two diagrams will interfere destructively. This 
effect reduces the total width of the $D^+$ and enhances its lifetime
(see Section \ref{a1-a2}). As a consequence we expect 
\hbox{$\Gamma_{had}(D^{+}) < \Gamma_{had}(D^{0}) \sim 
\Gamma_{had}(D_{s})$}, or the following hierarchy of lifetimes
\begin{equation}
\tau(D^0)\sim\tau(D_s)<\tau(D^+)
\end{equation}
It is important to note that the difference in the hadronic decay 
width should be understandable at the level of two-body decays, since
 three-body and four-body decays are experimentally found
to be dominated by quasi two-body channels.

The baryon sector is more complex.
The exchange mechanism is no longer helicity suppressed and can be 
comparable to  the spectator diagram. In addition, color suppression is 
operative only for particular decay channels.

 There are three large effects that modify hadronic 
widths in charm baryon decay\cite{Guberina}:
\begin{enumerate}
\item 
 destructive interference between 
the external spectator (Fig. ~2(a)) and the internal 
spectator (Fig.~2(c)) if a spectator quark is a u-quark (as in $\Lambda_{c}$
and $\Xi_{c}^{+}$), analogous to the effect in $D^+$ decay. 
\item 
constructive interference between two internal
spectator diagrams (Figs. 2(b) and 2(c)) if the spectator quark is an s-quark
(as in the $\Xi_{c}^{+}$, $\Xi_{c}^{0}$ and $\Omega_c$ baryons) 
\item 
 W-exchange contributions (Fig.~2(d)) which can be large 
 if the baryon  contains a $d$-quark(as in the $\Lambda_{c}$ and 
$\Xi_{c}^{0}$ baryons).  
\end{enumerate}
 Neglecting mass differences and Cabibbo-suppressed 
decays, the nonleptonic decay rates for charm baryons are qualitatively
given by:
\begin{eqnarray}
\Gamma(\Lambda_{c})&=&\Gamma_{spec}+\Gamma_{des.int.}+\Gamma_{exch.} \\
\Gamma(\Xi_{c}^{+})&=&\Gamma_{spec}+\Gamma_{des.int.}+\Gamma_{con.int.}
\nonumber \\ 
\Gamma(\Xi_{c}^{0})&=&\Gamma_{spec}+\Gamma_{con.int.}+\Gamma_{exch.}
\nonumber \\
\Gamma(\Omega_{c})&=&\Gamma_{spec}+\Gamma_{con.int.}
\nonumber 
\end{eqnarray}

where spec denotes the spectator component, exch denotes the $W$-exchange
component, con.int denotes the componet from constructive interference and 
des.int denotes the destructive interference component.
 Models with different relative weights for these non-spectator
effects lead to different predictions. 
There are two models, 
one by Guberina, R\"uckl, and Trampetic\cite{Guberina} 
and the  other by 
Voloshin and Shifman\cite{VS} that predict a baryon lifetime hierarchy
\begin{eqnarray}
\tau(\Omega_{c})\sim\tau(\Xi_{c}^{0})<\tau(\Lambda_{c})<\tau(\Xi_{c}^{+}) &
\qquad \hbox{Guberina, R\"uckl, and Trampetic} \\
\tau(\Omega_{c})<\tau(\Xi_{c}^{0})<\tau(\Lambda_{c})\sim\tau(\Xi_{c}^{+}) &
\qquad \hbox{Voloshin and Shifman} \nonumber
\end{eqnarray}

Since the ground state hadrons containing $b$
quarks decay weakly, their lifetimes should be typical of the weak
interaction scale, in the range of 0.1--2 ps. 
Ten years ago, before the MAC \cite{macblife} and MARK II \cite{mkiiblife}
collaborations presented the first measurements of the $b$ lifetime, the only
phenomenological guide to the strength of the coupling between the quark
generations was the Cabibbo angle. If the coupling between the third and
second generations ($|V_{cb}|$) 
had the same strength as the coupling between the second and
first($|V_{cs}|$), the $b$ lifetime would be about 0.1 ps. The measurements of
lifetimes from the PEP experiments that indicated a value
longer than 1 ps were not anticipated and it was then deduced 
that the CKM matrix element $|V_{cb}|$ was very small.

As in the charm sector we expect a lifetime hierarchy for $b$-flavored
hadrons. However, since the lifetime differences are expected to
scale as $1/m_Q^2$, where $m_{Q}$ is the mass
of the heavy quark, 
the variation in the $b$ system should be significantly smaller, on the
order of $10 \%$ or less \cite{bigimarch}. 
For the $b$ system we expect
\begin{equation}
\tau(B^-)\; \geq \; \tau(\bar{B}^0)\; \approx \;
 \tau(B_s)\; >\; \tau(\Lambda_b^0)
\end{equation}
Measurements of lifetimes for the various $b$-flavored hadrons 
thus provide a means to determine
the importance of non-spectator mechanisms in the $b$ sector.

\subsection{Techniques for Charm Lifetime Measurements}

 The measurements of the charm hadron lifetimes are dominated by fixed 
target experiments using silicon vertex detectors. The 
measurement of the lifetime is, in principle, very simple. One measures 
the decay length $L=\beta\gamma c t$ to extract the proper 
time $t$. The typical proper time for a c-hadron decay is in the range
$10^{-12}-10^{-13} s$, so that high precision vertex detectors are necessary. 
The lifetimes are determined using a binned maximum likelihood fit to the
distribution of reduced proper time, which is
defined as $t^{'} = t - N\sigma / \beta\gamma c$, where 
$\sigma$ is the error on the longitudinal
displacement (L) between the primary and the secondary vertex
(typically about $400~\mu m$).
The value of N 
varies depending on the analysis (typically $N=3$). 
The reduced proper time avoids the use of large corrections at short 
$t$ and is equivalent to starting the clock at a later time.
Corrections for acceptance and hadronic
absorption at long times and
resolution at short times are included in the fitting function.
Events from the mass sidebands are used to 
model the background lifetime distribution.

This technique must be modified slightly for  measurements
of the short lived charmed hyperons, for example,
the $\Omega_c$ lifetime is comparable to the E687 lifetime resolution,
{\it i.e.}, around 0.05 ps \cite{e687-nim}.
In E687, the fit is performed for all observed 
times greater than $-0.05$ ps in order 
to retain sufficient statistics. The effect of resolution is significant;
it is included in the analysis
 by convoluting the exponential decay and the resolution
function\cite{E687_4}.

\subsection{Techniques for Beauty Lifetime Measurements}

The lifetime of a particle is related to its decay length, $L_b$, by
\begin{equation}
\tau_b \; = \; \frac{L_b}{\gamma \beta c}
\label{tau_eq}
\end{equation}
At LEP energies, for example, the average $b$ momentum is about 30 GeV
which results in an average decay length of $2.5$ mm for $<\tau_b> \, = \,
1.5$ ps. Similarly, at CDF the mean vertex displacement in the  plane
transverse to the beam is about $0.9$ mm.

A variety of methods has been developed to measure the decay length and to
determine the $b$ lifetime. They all follow the same basic steps.
A purified sample is selected and the decay length is either measured 
directly or determined indirectly
 by using the impact parameter. The resulting decay length is
then corrected for the Lorentz boost. An additional correction for
background contamination is applied as well.

To
determine the lifetime of a specific $b$ hadron, as in charm hadron
lifetime measurements,
one would like to have a sample of fully reconstructed decays.
The $b$ vertex could then be
reconstructed allowing a measurement of the decay length. The momentum of
the $b$ hadron gives the $\gamma\beta$ factor in equation (\ref{tau_eq}) 
without any further assumptions. The resulting proper time distribution would
be an exponential function convoluted with a Gaussian 
resolution function representing the
measurement errors.
Although currently limited by statistics
this procedure will ultimately yield the most precise measurements of
individual $b$ hadron lifetimes.

The best statistical precision 
in the determination of lifetimes of hadrons 
containing $b$ quarks is currently obtained
from measurements using partial reconstruction of semileptonic
decays. These decays 
represent about 21\% of the total $b$ decay rate and have
the experimental advantage 
that both electrons and muons can be efficiently identified
with low background. 
The purity of the sample can be enhanced by kinematical cuts that
take advantage of the large mass of the $b$ quark e.g.
selecting leptons with large transverse momentum
with respect to the $b$ direction. 
Event samples with purities above 90\% have been obtained
at LEP. However, in such 
semileptonic decays the neutrino is not detected so the
$b$ hadron is not completely reconstructed. 
One then has to rely on Monte Carlo simulations
to estimate the $b$ momentum 
and to extract the proper time distribution from the
decay length measurements.

For inclusive lifetime measurements, 
the presence of a high $p_{\perp}$ lepton  or a $\psi$
meson is usually sufficient to demonstrate the presence of
a b quark,
while for exclusive measurements of individual $b$ hadron
lifetimes an additional decay particle 
has to be reconstructed in order to establish a signature
characteristic for the decaying $b$ hadron (Fig \ref{impact}(b)). 
The $\Lambda_b$ lifetime, for example, is measured
using a sample of events 
containing $\Lambda_c^+\ell^-$ or $\Lambda \ell^-$ combinations.

In early experiments 
the vertexing precision was not adequate to measure the
decay length, $l\, = \, \gamma \beta c \tau$, directly. 
The impact parameter method
shown schematically in Fig. \ref{impact}(a)
was developed as alternative. 
Because of  the finite lifetime of the $b$ hadron, a
lepton from the semileptonic decay of the heavy quark
will miss the primary
vertex where the $b$ hadron was produced.  
The miss distance or impact parameter,
$\delta$, is given by
\begin{equation}
\delta \; = \; \gamma \beta c \tau_b \sin{\alpha} \sin{\theta}.
\end{equation}
where $\alpha$ is the angle 
between the lepton and the $b$ directions and $\theta$ is the polar angle.
The $b$ direction is usually approximated by the axis of the
hadronic jet. A negative sign is assigned
to the impact parameter 
if the lepton track crosses the jet axis behind the beam spot
indicating a mismeasured
lepton or a background event. The main advantage of the
impact parameter method is that it is 
rather insensitive to the unknown boost of the parent;
as $\gamma\beta$ increases 
with the $b$ momentum, $\sin{\alpha}$ decreases approximately as
$1/\gamma\beta$ for $\beta \approx 1$.

In experiments 
with sufficient statistics and vertex resolution, the decay length for
the $b$ hadron vertex is reconstructed by using the 
lepton track and the direction of the reconstructed
charm meson as shown in Fig \ref{impact}(b).
The momentum of the $b$ hadron 
is estimated by using the observed decay 
products, the missing momentum and a correction factor 
determined from a Monte Carlo simulation. The proper
time distribution is then given by 
an exponential convoluted with a Gaussian resolution
function and the momentum 
correction factor. A maximum likelihood fit is used to extract
the lifetime \cite{aleph_eps_bs}.

\subsubsection{Averaging $B$ hadron Lifetime Measurements}
\label{aver_life}

To obtain 
the most precise
value for inclusive and exclusive $b$ lifetimes the results of
lifetime measurements from different experiments have been combined. 
Using the
conventional approach of weighting 
the measurements according to their error does not
take into account the 
underlying exponential decay-time distribution. If a measurement
fluctuates low then its weight in the average 
will increase, leading to a bias towards
low values. This is particularly 
relevant for low statistics measurements such as
the $B_s$ lifetime. According to a study by Forty\cite{forty}, 
this bias can be
avoided
if the weight is calculated using the 
relative error $\sigma_i/\tau_i$.
\footnote{This procedure assumes good vertex resolution, 
{\it {i.e.}} $\sigma<\tau/10$.}
We find a 1-3\% difference in the average lifetimes computed,   
with the second method giving the larger
value. A slight bias of the latter method  towards 
higher lifetime values could be avoided
by taking into account asymmetric errors. 
This effect has been found empirically to be rather small and
we omit this additional complication in the calculation
of our lifetime averages.

\subsection{Results on Lifetimes of Hadrons That  Contain $c$ Quarks}

 The experimental results are summarized in Fig.\ref{clife} where updated 
world averages for the c-hadron lifetimes are 
given\cite{PDG}\cite{E687_4}\cite{WA89}. From these results,
the full lifetime hierarchy can be studied.

The measurements of the charm hadron lifetimes
are now extremely precise. Systematic effects will soon
become the largest
component of the error for some measurements e.g. the $D^0$ 
and $D^+$ lifetimes. These systematic effects are due to the
uncertainty in the $D$ momentum distribution, to the nuclear absorption of
the $D$ meson or its decay products in the target, and to the lifetime
of the background.

In the baryon sector the measurements are
still statistics limited. There are now results 
for the $\Omega_c$ lifetime from E687\cite{E687_4} and WA89\cite{WA89} 
which complete the baryon hierarchy. It is quite remarkable
that the lifetime of this rare and short lived baryon is now 
being measured.

 The world averages for lifetime measurements
 are now dominated by results from E687,
 which is the only single experiment which has  
measured all the charmed hadron lifetimes\cite{E687_3}\cite{E687_4}. 
The results are internally consistent and the 
ratios of lifetimes,
 which characterize the hierarchy, are to a large extent 
unbiased by systematic effects\cite{Malvezzi}.
 For the charm mesons lifetimes a clear pattern emerges, in agreement with the 
theoretical predictions
\begin{equation}
\tau(D^0)<\tau(D_s)<\tau(D^+)
\end{equation}
 The meson lifetimes are now measured at the level of few percent, probably 
beyond the ability to compute them. The near equality of 
$\tau(D_s)$ and $\tau(D^0)$ is direct evidence for the 
reduced weight of the non-spectator (W-exchange and W-annihilation) in 
charm meson decays\cite{Bigi_3}. 

 The agreement between the measurements of 
 charm baryon lifetimes and theoretical expectations
 is remarkable, since in addition to the exchange diagram, 
there are constructive as well as destructive contributions to the decay 
rate. The experimental results lead to the following baryon 
lifetime hierarchy 
\begin{equation}
\tau(\Omega_{c})\leq\tau(\Xi_{c}^{0})<\tau(\Lambda_{c})<\tau(\Xi_{c}^{+})
\end{equation}
 Although statistically limited the present values tend to favor the model
of Guberina, R\"uckl, and Trampetic \cite{Guberina}.

\subsection{Results on Lifetimes of Hadrons That Contain $b$ Quarks}

%
%
%
Inclusive measurements of the $b$ lifetime were important historically 
to establish the long $b$ lifetime. In addition, they
provided the first evidence that the 
coupling between the second and third quark generation is quite small.
They are still needed for some electroweak studies such as the determination
of the forward-backward asymmetry in $Z \to b\bar{b}$ 
where the different hadrons containing $b$ quarks are not distinguished.
For $B$ physics, {\it {i.e.}} the study
of $B$ meson decays, exclusive measurements of individual $b$ hadron
lifetimes are preferable. For example,
to extract the  value of the CKM matrix element $|V_{cb}|$ 
from measurements of semileptonic $B$ decays
the average of the $B^+$ and $\bar{B}^0$ lifetimes should be used rather than 
the inclusive $b$ lifetime which contains additional
contributions from $B_s$ mesons and $b$ baryons.

The current world average for the inclusive $b$ lifetime
which includes many measurements
is \cite{kroll},
\begin{center}
\begin{tabular}{ll}
$<\tau_b> \; = \; 1.563 \pm 0.019$ &ps.
\end{tabular}
\end{center}
The  world average for this quantity in 1992 was $(1.29\pm 0.05)$ ps. 
The substantial
change in the value has been attributed to several improvements:
the use of neutral energy when calculating the b jet direction,
and better knowledge of the resolution function as a result of the
use of silicon vertex detectors\cite{forty},\cite{Sharma}.


Precise measurements of exclusive lifetimes for b-flavored hadrons 
have been carried out by CDF \cite{cdf_eps}, \cite{cdf_bs_life},
by some of the LEP experiments
\cite{aleph_eps_b0} -- \cite{delphi_xi}
and by SLD \cite{sld_eps}.
The most recent results and the techniques used are given in Table
\ref{Tblife}.

\subsubsection{$B^-$ and $\bar{B}^0$ Lifetime Measurements}

The best statistical precision in the determination of exclusive lifetimes
is obtained from measurements using
lepton-particle correlations. For example, a sample of 
$B^0$ candidates can be
obtained from events with lepton-$D^{*+}$ correlations of the correct
sign; these events originate from the decay $\bar{B^0}\to D^{*+} \ell^-\nu$,
$D^{*+}\to D^0 \pi^+$ and $D^0 \to K^- \pi^+$ (see Fig. \ref{impact} (b) for
the method and Fig. \ref{cdf_dstarlnu} for the CDF results).
The pion from the strong decay and the lepton
form a detached vertex. 
This information combined with the direction of the reconstructed 
$D^0$ meson determines the location of the $B$ decay vertex so that
the decay length can be measured. 
To obtain the lifetime from the decay length, requires knowledge
of $\gamma\beta$ which is estimated from the
momenta of the observed decay products. Since the neutrino is not
observed, a correction is made to determine the boost factor. The uncertainty
in the size of this correction is included in the systematic error
and is typically on the order of $3\%$.
Another systematic problem is the contamination from decays
$B^-\to D^{**}~l^- \nu$, followed by $D^{**}\to D^{*+}\pi^-$ 
where the $\pi^-$
from the strong decay of the $D^{**}$ (p-wave) meson is not detected.
These backgrounds will lead to a $B^-$ meson contamination
in the $\bar{B}^0$ lifetime sample (and vice-versa).
Since the branching fractions for such decays are poorly measured, this
is another important systematic limitation and gives a contribution
of order 5\% to the systematic error. Significant contributions
to the systematic error
also result from the uncertainty in the level of background and its
lifetime spectrum. More detailed discussions of exclusive lifetime measurements
can be found in recent reviews  by Sharma and Weber 
\cite{Sharma} and Kroll \cite{kroll}.

The systematic problems associated with the boost correction and
the contamination from poorly measured backgrounds can be avoided
by using fully reconstructed decays such as $\bar{B^0}\to D^+ \pi^-$
or $B^- \to \psi K^-$.
However, since exclusive $B$ branching ratios are 
small, this method has much poorer statistical precision.
In hadron collider experiments, this approach has been successfully
used to determine the $\bar{B^0}$, $B^-$, and $B_s$ lifetimes from
exclusive modes with $\psi$ mesons e.g. $\bar{B^0}\to \psi K^{*0}$,
$B^-\to \psi K^-$\cite{cdflife} and $B_s\to \psi \phi$ \cite{cdf_bs_life}.

A topological vertexing method has been used by 
the DELPHI and SLD experiments. Candidate $\bar{B^0}$ and
$B^+$ mesons
are distinguished on the basis of the net charge of the tracks at
the decay vertex. This method has small statistical errors however
care must be taken to assure that systematic uncertainties from tracking 
and incorrect assignments of decay
vertices are controlled. The neutral $B$ lifetime that is extracted 
represents an average over the lifetimes over all neutral $b$ flavored hadrons
including $B_d^0$, $B_s^0$, and $\Lambda_b^0$. With good knowledge
of the production fractions, the exclusive $B^0$ lifetime can be extracted. 
In the case of SLD, the excellent resolution of their CCD vertex detector
compensates to some degree for their low statistics.

Using the procedure
for averaging measurements described in Section \ref{aver_life}, we combine
the individual $B^-$ and $\bar{B}^0$ lifetime measurements and obtain
\begin{center}
\begin{tabular}{ll}
$\tau_{B^-} \; = \; 1.62 \, \pm \, 0.04 $ &ps \cr
$ \tau_{\bar{B}^0} \; = \; 1.57 \, \pm \, 0.04 $ & ps 
\end{tabular}
\end{center}
When  averaging the results obtained by studying $D^{(*)}-\ell$ correlations
a common systematic error of 3\% has been assumed.

\subsubsection{$B_s$ Lifetime Measurements}

The $B_s$ lifetime was measured by CDF \cite{cdf_bs_life}
and the LEP experiments using
partial reconstruction of the 
semileptonic decay $\bar{B}_s^0 \to D_s^- \ell^+ \nu$.
Candidate $D_s^-$ mesons were reconstructed in the  $\phi \pi^-$ or
$K^{*0}K^-$ final states. 
Fig. \ref{aleph_bs_life}(a) shows the $K^-K^+\pi^+$ invariant mass spectrum
obtained by ALEPH \cite{aleph_eps_bs}
for right-sign and wrong-sign $D_s\ell$ combinations.
The $B_s$ decay length was measured and converted to the $B_s$ proper
time using a $B_s$ momentum estimator based on the reconstructed lepton
and the $D_s$ momentum as well as on an estimated neutrino energy obtained by
using a missing mass technique. The $B_s$ lifetime
was extracted from the proper time distribution using a maximum
likelihood fit. The result of such a procedure is shown in Fig. 
\ref{aleph_bs_life}(b).

The uncertainty in the $B_s$ lifetime is still dominated by the statistical
error. Assuming a common systematic error of 2\% \cite{Sharma} for the
uncertainty in the vertex resolution and the neutrino energy estimate we
obtain
\begin{center}
\begin{tabular}{ll}
$\tau_{B_s} \; = \; 1.55 \, \pm \, 0.09 $ & ps
\end{tabular}
\end{center}

For the $B_s$ meson, there are two weak eigenstates
with different lifetimes that can be distinguished by their
CP quantum number. The decay $\bar{B}_s^0 \to D_s^- \ell^+ \nu$
contains an equal mixture of the two eigenstates. 
An appreciable lifetime difference $\Delta\Gamma$
is expected for the $B_s$ (O(10\%))
and should be measurable at future experiments. 
Measurements of the $B_s$ lifetime difference 
may be used to constrain $|V_{ts}|/|V_{td}|$\cite{Browpak}.

\subsubsection{$b$ Baryon Lifetime Measurements}

Studies of $\Lambda_c^+ \ell^-$ and $\Lambda \ell^-$ correlations at LEP
are used to determine the lifetime of the $\Lambda_b^0$
baryon. For example, using the decay chain
$$\Lambda_b  \to  \Lambda_c^+ \ell^- \bar{\nu},
~\Lambda_c \to \Lambda X \to  p \pi^- X$$
the $p\pi^-$ invariant mass distribution shown in Fig. 
\ref{opal_lambdab_life}(a) was obtained by OPAL\cite{opal_eps_lambda}.
Although the composition of the $b$ baryon sample is not known, it is
expected that the $\Lambda_b$ baryon is the most copiously produced. 
Both impact parameter and decay length measurements have been 
used to determine $\tau_{\Lambda_b}$. Since the $\Lambda_c^+$ lifetime
is short, the $\Lambda_b$ decay length can be estimated by using the 
displacement of the $\Lambda \ell^-$ vertex. The  
time distribution from the OPAL analysis which uses this technique is
shown in Fig. \ref{opal_lambdab_life}(b).

A better estimate of the $\Lambda_b^0$ decay point is obtained from fully
reconstructing the $\Lambda_c^+$ baryon and finding the $\Lambda_c^+ \ell^-$
vertex. However, the sample sizes become somewhat smaller. Using this method,
CDF finds $\tau_{\Lambda_b} \, = \, 1.33 \pm 0.16 \pm 0.07$ ps.
Combining the results listed in Table \ref{Tblife} the
world average $\Lambda_b$ lifetime is found to be 
\begin{center}
\begin{tabular}{ll}
$\tau_{\Lambda_b} \; = \; 1.21 \, \pm \, 0.07 $ & ps.
\end{tabular}
\end{center}
This confirms the original indications that the lifetime of
$\Lambda_b$ is very short, a fact
that is  difficult to accommodate theoretically.

DELPHI and ALEPH have observed
small signals in $\Xi^- \ell^-$ correlations.
 These are expected to come from
$\Xi_b^- \to \Xi_c^0 \ell^- \bar{\nu}X$ and
$\Xi_b^0 \to \Xi_c^+ \ell^- \bar{\nu}X$ followed by
$\Xi_x \to \Xi^- X'$. These samples have been used to measure
the lifetime of $\Xi_b^-$ \cite{aleph_xi}, \cite{delphi_xi}.

\subsubsection{Measurements of Lifetime Ratios}
\label{liferat}

The ratio of the $B^-$ and $\bar{B^0}$ lifetimes has been measured
by a number of experiments. These measurements are performed
either by using
correlations between $D$ mesons and leptons
or  by using exclusive final states such as
$B^-\to \psi K^-$ and $\bar{B^0}\to \psi K^{*0}$. The CLEO~II experiment has
measured ${\cal B}(B^0\to X~l^-\nu)$ and ${\cal B}(B^-\to X~l^-\nu)$
using the yield of leptons found opposite fully
and partially reconstructed B decays \cite{cleoiitptz}.
From isospin invariance, the ratio of the two branching fractions
is the ratio of the lifetimes.

Averaging the results listed in Table \ref{Tbratio} we obtain
$$
\frac{\tau_{B^-}}{\tau_{\bar{B}^0}} \; = \; 1.00 \, \pm \, 0.05
$$
Note that this value is not exactly equal to the ratio 
of the world averages for the
$B^-$ and $\bar{B}^0$ lifetimes since the average value of
$\tau_{B^-}/\tau_{\bar{B}^0}$ is calculated directly 
from the ratios reported by the experiments.

\subsection{Lifetime Summary}

The experimental results on lifetimes for
hadrons with $c$ quarks are shown in Fig.\ref{clife}.
For the $D^0$ and $D^+$ mesons, the lifetimes 
measurements will soon
become systematics dominated.
It is clear that from the observed
lifetime hierarchy that non-spectator effects are important
in the charm sector.

A summary of the
measurements of $b$ hadron lifetimes 
is shown in Fig.~\ref{blifetime}.
The pattern of measured lifetimes follows the 
theoretical expectations outlined above and non-spectator effects
are observed to be small.
However, the $\Lambda_b$ baryon lifetime is unexpectedly short.
As has been noted by several authors,
the observed value of the $\Lambda_b$ lifetime is quite difficult
to accommodate theoretically\cite{neubertlb},\cite{rosnerlb}.

Assuming that the relative production ratios
of $B^-,\,\bar{B}^0,\,B_s,\,\Lambda_b$ at the $Z^0$ are
$0.39\,:\,0.39\,:\,0.12\,:\,0.10$ the exclusive lifetime
measurements can be averaged to give
$<\tau_{excl.}>\; = \; 1.551 \pm  0.025 $ ps; this is 
consistent with the world average for the
inclusive $b$ lifetime, $\tau_b \; = \; 1.563\pm 0.019$ ps.

\section{Nonleptonic Decays of c-quark Hadrons}
\subsection{Introduction}

 In the past few years there has been an impressive increase 
in the size of charm particle data samples.  
$D$ mesons are the only heavy quark systems in which 
Cabibbo-allowed decays,  
single Cabibbo-suppressed decays, and double Cabibbo-suppressed
decays (DCSD)  have all been
measured. 

 The high statistics now 
available allow for isospin analyses of related decay modes. 
The effect of elastic FSI can then
be taken into account when making comparisons with model predictions. 
 A further improvement are sophisticated 
amplitude analyses for three-body and four-body final states, 
from which the resonant substructure 
of multibody final states can be determined.

A systematic investigation of charm baryon decay modes has also begun.
This is complementary to investigations of the meson sector. In the case
of charm baryons, the W-exchange mechanism is no longer helicity
suppressed and can be studied in detail.

\subsection{Double Cabibbo Suppressed Decays}

 The decay mode $D^{+}\rightarrow K^{+}\pi^{+}\pi^{-}$, recently 
measured by the E687 and E791 collaborations\cite{E687_DCSD},\cite{E791_DCSD} 
has an unambiguous interpretation 
as a doubly Cabibbo suppressed spectator decay.
The decays $D^{0}\rightarrow K^{+}\pi^{-}$
and $D^{0}\rightarrow K^{+}\pi^{+}\pi^{-}\pi^{-}$ may occur
either by DCSD  or by mixing  (see Fig.1(f)).
 To observe these decay modes,
 experiments use the decay chain $D^{*+} \rightarrow D^{0}\pi^+$ to 
tag the flavor of the neutral $D$. The mixing
and DCSD contribution are separated by using 
their different time dependences. The DCSD component
follows the usual $e^{-t/\tau}$ time evolution, while the rate for the 
mixing events is proportional to $t^{2}e^{-t/\tau}$ in the limit of small 
mixing\cite{Bigi_1}. In the Standard Model $D^{0}-\overline{D^{0}}$ 
mixing is expected to be small: The ratio of mixed to unmixed decays,
$r_m$, is less than $10^{-8}$\cite{Burdman}. The best limit on $r_m$ 
comes from the E691 experiment which found $r_{m}<0.0037$\cite{E691_1}.
A measurement of the corresponding DCSD parameter, $r_{DC}$, expected 
to be on the order of $\tan^{4}\theta_{c} \sim 0.0026$, can be obtained
 from the time integrated measurement of
CLEO if no mixing is assumed\cite{CLEO_1}. 
Since CLEO does not measure time evolution,
their measurement is sensitive to  mixing and to DCSD as well as to possible
interference between the two mechanisms.

 In Table~\ref{Doublyd} we summarize the status of the DCSD 
measurements. Since there
is no Pauli interference for DCSD in $D^+$ decays, the ratio of a DCSD decay
to a Cabbibo favored decay, which has destructive interference,
is expected to be greater than $\tan^{4}\theta_{c}$.  The rate for
DCSD in $D^0$ decay is somewhat greater than was 
expected from $SU(3)$ breaking.
However, the errors are still too large to draw any firm conclusions.

\subsection{Amplitude Analyses of Hadronic Charm Decays}

 Dalitz plot analyses of nonleptonic decays have become an important
source of information on the dynamics of
charm hadron decay. Resonant substructure analyses of 
three-body and four-body final states of $D$ mesons, which correctly
 take into account interference effects, 
allow for meaningful comparisons of
experimental data and theoretical models.

For example, Fig.~\ref{Kkpi_dalitz}
shows the $K^{\mp}K^{\pm}\pi^{\pm}$ invariant mass distribution,
and Dalitz plots for the $D^+$ and $D_s$ mass region 
from experiment E687\cite{E687_1}. 
One notes that the $\overline{K^\star}$ 
and $\phi$ bands dominate both decays. A Dalitz plot analysis shows
that these modes are saturated by
quasi two-body processes: $\phi\pi^+$ and $\overline{K^{\star 0}}(892)K^+$
for $D_s^+$; and $\phi\pi^+$, $\overline{K^{\star 0}}(892)K^+$ and
$\overline{K^{\star 0}}(1430)K^+$ for $D^+$. Amplitude analyses 
have also been performed for several
$D\rightarrow K\pi\pi$ and $D\rightarrow K\pi\pi\pi$ modes. 
These analyses support the hypothesis that all $D$ and $D_s$ 
nonleptonic decays are dominated by two-body modes. The one exception 
is the decay mode $D^+\rightarrow K^-\pi^+\pi^+$, which cannot be 
fitted  without including
a large non-resonant three-body component\cite{Wiss}.

 It is important to study the decay $D_s \rightarrow \pi^-\pi^+\pi^+$,
which is observed with a branching fraction of $0.31\pm 0.06 \%$\cite{bfd},
in order to determine the importance of the W-annihilation diagram. 
In this decay mode,
none of the initial quarks is present in the final state and the 
decay is Cabibbo-favoured. However, a Dalitz plot analysis 
is crucial, as the presence of resonant submodes that contain 
a meson with $s\overline{s}$ quark content, such as 
$D_s^+\to f_{0}(980)\pi^+$\cite{comment_f} occur
 through a spectator process rather than through W-annihilation.

 Preliminary results from the E687 experiment on this decay have been 
presented recently\cite{Moroni}. Their Dalitz plot analysis is appreciably 
different from the previously 
accepted scenario\cite{PDG}. No significant
 nonresonant $D_s^+\to 3\pi$ is observed. Two new decay modes 
($D_s^+\to f_{2}(1270)\pi^+$ and $D_s^+\to f_{0}(1300)\pi^+$) have been found. 
The presence of a sizeable $D_s^+\to f_{0}(980)\pi^+$ component
has been confirmed. The absence of the $D_s^+\to \rho^0\pi^+$ mode is
also confirmed with higher statistics. 
We note that all the resonant submodes
observed in the $D_s^+\to\pi^+\pi^-\pi^+$ decay 
have a meson with $s\overline{s}$ quark 
content and thus can be attributed to the spectator process.

 The observation of a nonresonant contribution might be interpreted
as evidence for W-annihilation; however it is experimentally
difficult to distinguish 
this possibility from  a coherent sum of wide resonances that could 
easily mimic a flat distribution. The observation of the decay mode
$D_s^+\to \rho^0\pi^+$ would be a 
clear indication of the presence of the W-annihilation mechanism.
However, the absence of the $D_s^+\to \rho^0\pi^+$ channel may have other 
explanations and does not preclude a significant
 contribution from the annihilation 
diagram. 
A sizeable branching fraction 
for the decay mode $D_s^+ \rightarrow \omega(782)\pi^+$ could be a strong 
signature for the existence of the W-annihilation 
diagram. E691 finds \hbox{$D_s^+\rightarrow\omega(980)\pi^+/
D_s^+\rightarrow\phi\pi^+<0.5$\cite{E691_2}} which is not restrictive enough
to rule out this possibility.
 More data, therefore, are necessary to demonstrate the presence 
of non-spectator contributions in charm meson decay.

\subsection{Hadronic Decays of Charmed Baryons}

 In the baryon sector 
only charmed baryons with one $c$-quark have been observed so far. 
Impressive progress in the study
of charmed baryons has been made in the last few years. 
The existence 
of the $\Omega_c$ has been established\cite{Omegac}. 
Experimental sensitivity has progressed to the level that
Cabibbo-suppressed decay modes of $\Lambda_c$ have been 
observed\cite{CLEO_2}\cite{E687_2}.

The study of charm baryons gives information complementary 
to that gained from the study of the charm mesons. Due to the 
presence of a diquark, the exchange diagram is 
no longer helicity suppressed.
 Predictions for decay rates of charmed baryons into two-body 
final states are now available\cite{Baryon_models}. 
The agreement with experimental results is mixed. 

 To date, the decays $\Lambda_c^+\to\Lambda (n \pi)^+$, $\Sigma^0(n\pi)^+$,
$\Sigma^- (n\pi)^+$, $\Sigma^+ (n\pi)^0$, $p K^- (n\pi)^+$, 
and $p K_s (n\pi)^0$ with $n\le 3$ and including up to 1 $\pi^0$ have 
been reconstructed. Recently some decay modes of $\Lambda_c$ with an
$\eta$ meson in the final states have been observed\cite{Cleolceta}; 
these decays are expected to proceed entirely through nonfactorizable 
internal W-emission and W-exchange diagrams. 

 The observation of certain decay modes such as $\Lambda_c^+\rightarrow 
\Xi^{\star 0}K^+$\cite{ARGUS} or 
$\Lambda_c^+\rightarrow\Sigma^+\phi$\cite{CLEO_3} provides strong evidence 
for the importance of W-exchange in charmed baryon decays. The simplest 
way for these decays to proceed is through the W-exchange 
diagram, although it is hard to completely rule out contributions from FSI. 
Table~\ref{Lambdacwex} gives branching fractions for this class of 
$\Lambda_c$ decay modes. Evidence for color suppressed decay modes
such as $\Lambda_c\to p \phi$ has also been found by CLEO\cite{CLEO_2}.

\section{Inclusive B Decay}

\subsection{Motivation}
Because of the large mass of the $b$ quark $B$ meson decays
give rise to a large
number of secondary decay products. For instance,
CLEO finds that the charged and photon multiplicities at the
$\Upsilon (4 S)$ are:
$n_{\rm charged}=10.99 \pm 0.06 \pm 0.29$ and
$n_{ \gamma}=10.00\pm 0.53 \pm 0.50$, respectively
\cite{multi},\cite{multiARG}.
The high multiplicity of final state particles leads to a large
number of possible exclusive final states. Even with a
detector that has a large acceptance 
for both charged tracks and photons, it is difficult to reconstruct
many exclusive final states because of combinatorial backgrounds.
Furthermore the detection efficiency drops for high multiplicity final states.
Thus, to get a complete picture of $B$ meson decay, it is important to
study inclusive decay rates.

A number of theoretical calculations of inclusive $B$ decay rates have been
made using the parton model. It is believed that measurements
of such inclusive rates can be more reliably compared to the theoretical
calculations than can measurements of exclusive decays 
While this is sufficient motivation for studying the inclusive rates,
there is also a need for accurate measurements in order to model the
decays of $B$ mesons both for high energy collider experiments, and for
experiments at the $\Upsilon (4S)$. As a specific example, the inclusive
rate for $B\to\psi$ has been used to determine the $B$ meson production
cross-section at the Tevatron \cite{pppsi}.

The branching ratios for inclusive $B$ decays to particular final state
particles are determined by measuring the inclusive yields of these
particles in data taken
on the $\Upsilon (4S)$ resonance, and subtracting the non-resonant background
using data taken at energies below the
$\Upsilon (4S)$ resonance. The off-resonance data are scaled to correct
for the energy dependence of the continuum cross-section.
Results on inclusive production at the $\Upsilon (4S)$ are usually presented
as a function of the variable $x$, which is the fraction of the maximum
possible momentum carried by the particle, $p_{max}=\sqrt{E_{beam}^2 - M^2}$.
The endpoint for production in $B$ decays is at $x=0.5$.

The results reported by the
different experiments have been rescaled to accommodate 
the new charm branching
fraction. The world averages for inclusive $B \to$~meson  decays are given in 
Table~\ref{khinc}.

\subsection{Inclusive $B$ Decay to Mesons}

CLEO~1.5 \cite{CLEOK} has measured 
the branching fractions of inclusive $B$ decays
to light mesons, while
ARGUS has determined the average multiplicities of light mesons
in $B$ decay. 
If more than one meson of the particle type under study is 
produced in a $B\bar{B}$ decay, 
then the branching fraction and the multiplicity
will differ. Unless otherwise noted, 
the results reported in Table \ref{Tbmulti} are averaged over $B$ and
$\bar{B}$ decay.

In the decay $b \to c \to s$ the 
charge of the kaon can be used to determine the
flavor of the $b$ quark. A first attempt to measure the tagging efficiency and
misidentification probability for this method
was made by ARGUS \cite{ARGUSK}.
With the large sample of 
reconstructed $B^0$ and $B^+$ decays from CLEO~II it should
be possible to measure these quantities directly.
The experiments also measure the momentum spectra for the particles listed
in Table \ref{Tbmulti}. 
These results provide important information needed
to improve Monte Carlo generators and to determine tagging efficiencies for 
future $B$ experiments\cite{dunietztag}.
The inclusive production of $D^0, D^+, D_{s}^+$ and $D^{*+}$ mesons
in $B$ decay has been measured 
by ARGUS \cite{ARGUSD} and CLEO~1.5 \cite{CLEOD}. 
Preliminary measurements of several of these inclusive
branching fractions from CLEO~II have also
become available\cite{cleodds},\cite{dpfd0}.
To improve signal to background and obtain low systematic errors, 
only the $D^0 \to K^- \pi^+$,
$D^+ \to K^- \pi^+ \pi^+$ and $D_{s}^{+} \to \phi \pi^+$ decay modes are used.
The results are given in Table~\ref{khinc}.

Analyses of the shape of the $D_s$ momentum
spectrum (Fig. \ref{Fdsmomdata})
indicates that there is a substantial two body component. 
In model dependent fits the ARGUS and CLEO~1.5 
experiments find two body fractions of $(58 \pm 7 \pm 9)$\%
\cite{ARGUSD} and $(56 \pm 10)$\% \cite{CLEOD}, respectively. 
CLEO~II finds a somewhat smaller 
two body fraction, $45.7\pm 1.9\pm 3.7\pm 0.6\%$
where the last error accounts for the uncertainty due to model
dependence in the predictions for the rates of
two-body modes\cite{cleodds}.  This result does not include
additional uncertainty from the $D_s^+ \to \phi \pi^+$ branching fraction.
Averaging the results from the three experiments we find a two body
component of $(48.3 \pm 3.6)\%$ which leads to ${\cal{B}}(B\to D_sX\;
{\rm (two~body)})\; = \; (4.9\pm 1.3)\%$
It is important to determine
what mechanisms are responsible for the production
of the remainder, the lower momentum $D_s$ mesons. 
Two possibilities are external
$W^-$ emission with $W^-\to \bar{c} s$ or $W^-\to \bar{u} d$ with 
$s \bar{s}$ quark popping at the lower vertex. A limit on the latter possibility
($<31\%$ of $D_s$ mesons are produced by this mechanism) is obtained
from the absence of wrong sign $D_s^+-\ell^+$ correlations \cite{CLEO_lowerds}.

Results on inclusive $B$ decay to final states with $\psi$ and $\psi '$ mesons
have been reported by CLEO~1.5, ARGUS, and
CLEO~II\cite{CLEOpsiinc} and are given in Table \ref{khinc}.
In the most recent high statistics analysis from CLEO~II, the effect of final
state radiation has been taken into account.
The resulting invariant dielectron and dimuon mass distributions are
shown in Fig. \ref{Fpsi}. The theoretical predictions for the
production of charmonium states in $B$ decay\cite{bodwin,kuehn_psi,palmstech}
is discussed further in Section~\ref{fac-color}.

The momentum spectrum  for $B \to \psi, \psi^{'}$ transitions 
has been measured (Fig.~\ref{Fpsimomdata}).
The two body component due to $B\to \psi K$ and $B\to \psi K^*$
saturates the spectrum in the momentum range between 1.4 and 2.0 GeV. 
By subtracting the contributions from $\psi$'s
originating in $\psi$' and $\chi_c$ decays 
CLEO and ARGUS measured the momentum
distribution of the direct component shown in  Fig.~\ref{Fpsimomdata}(b).
The average branching ratio for  direct $\psi$ production is found to be
${\cal{B}}(B \to \psi$, where $\psi$ not from $\psi ')
\; = \; (0.82 \pm 0.08)\%$. 
The two body component constitutes about 1/3 of direct $\psi$ production.
The composition of the remainder is not yet determined.

Results on 
inclusive $B \to \chi_c X, \chi_c \to \gamma \psi$
decays have been reported by ARGUS \cite{arguschi} 
and CLEO~II \cite{CLEOpsiinc,fastpsi}. 
ARGUS assumes there is no $\chi_{c2}$ production.
CLEO~II has significantly better $\chi_c$ mass resolution 
than ARGUS and allows for both possibilities. 
The branching ratio for $\chi_{c0} \to \gamma \psi$ is 
$(6.6\pm 1.8) \times 10^{-3}$ so the contribution
of the $\chi_{c0}$ meson to the $\psi\gamma$ 
final state can be neglected.
CLEO finds evidence at the 2.5 standard deviation level
for a $B\to \chi_{c2}$ contribution which would
indicate either non-factorizable contributions or higher order processes
$O(\alpha_s^2)$ in $b\to c \bar{c} s$\cite{bodwin}.

The decay of $B$ mesons to the lightest charmonium state, 
the $\eta_c$, has not
yet been observed. A recent CLEO~II search  placed an
upper limit of 0.9\% on the process $B \to \eta_c X$ 
at the 90\% confidence level \cite{CLEOpsiinc}.

By using the results in Table~\ref{khinc}, it is possible to isolate the
component of $B\to \psi$ production which is due to production of higher
charmonium states in B decay and the direct component. 
Similarly, the direct
$B\to \chi_{c1}$ component can be determined by removing the contribution
from $B\to \psi^{'}$, $\psi^{'}\to \chi_{c1} \gamma$.
It is assumed, that all $\psi$' mesons are directly produced.

\subsection{Inclusive $B$ Decay to Baryons }

ARGUS\cite{argusbary}
and CLEO~1.5\cite{crawbary} have observed inclusive production of $\bar{p}$,
$\Lambda$, $\Xi$ and the charmed $\Lambda_c^+$ baryon. Recently
CLEO~II has reported the observation 
of $B \to \Sigma_c X$\cite{sigmamz}, $B\to \Xi_c^{0} X$ and
$B\to \Xi_c^{+} X$\cite{cleocascade}.
The measured branching ratios for these decays and the world
averages can be found in Table \ref{khinc}.


The momentum spectrum of $B\to \Lambda_c$ transitions has been measured
by  CLEO \cite{sigmamz}.
The spectrum is rather soft indicating
$\Xi_c$ production or the presence of a significant multibody component.
Similarly,
CLEO~II has found that $B\to \Sigma_c^0 X$ and $B\to \Sigma_c^{++} X$ 
decays have no two body contribution. 

In addition to the inclusive branching ratios given above, the experimental
data has been used in  attempts to disentangle which of the baryon
production mechanisms shown in  Fig.~\ref{btobaryon} dominates.
CLEO~1.5 \cite{crawbary}
and ARGUS \cite{argusbary} have investigated 
baryon correlations in  $B$ decay in order
to elucidate the underlying decay process.
We follow the notation of Reference\cite{crawbary}. Let N
denote baryons with $S=C=0$ (e.g. p, n, $\Delta$, $N^*$). Let Y
refer to baryons with $S=-1, C=0$ (e.g. $\Lambda$, $\Sigma^0$, $\Sigma^+$).
Let $Y_c$ refer to baryons with $S=0, C=1$ [e.g. $\Lambda_{c}^{+}$,
$\Sigma_{c}^{(+,0,++)}$] . Then the following final states can be used
to distinguish possible mechanisms for baryon production in $B$ decay 
(Fig. \ref{btobaryon}).

\begin{enumerate}
\item {$\bar{B} \to Y_c \bar{N} X$, $\bar{B} \to \Xi_c \bar{Y} X$}\\
These final states are produced by the usual $b \to c W^-$ coupling
in a spectator or exchange diagram in conjunction with the popping of
two quark pairs from the vacuum (as shown in Figs.~\ref{btobaryon}(a),(b)).
 It should be noted that the two
mechanisms can be distinguished by examination of the $Y_c$ momentum
spectrum,  since the exchange diagram will
produce two body final states (e.g. $\Lambda_c \bar{p}$ or
$\Sigma_c^{++} \bar{\Delta}^{--}$).

\item {$\bar{B}\to D N \bar{N} X$, $\bar{B} \to D Y \bar{Y} X$}\\
 The non-charmed baryon-antibaryon pair is produced from W fragmentation
after hadronization with two quark-antiquark pairs popped from
the vacuum (as shown in Figs.~\ref{btobaryon}(c),(d)). 
The $D$ meson is formed from the charm spectator quark system.
If this mechanism is significant, inclusive production of
charmless baryon-antibaryon pairs should be
observed in $B$ decay.

\item{$\bar{B} \to Y_c \bar{Y} X$,  $\bar{B} \to \Xi_c \bar{Y_c} X$}\\
 These states are produced by the internal spectator graph
with $W^- \to \bar{c} s$ in conjunction with the popping of two quark
antiquark pairs. 

\item {$\bar{B}\to D_{s}^{-} Y_c \bar{N} X$,
 $\bar{B}\to D_{s}^{-} \Xi_c \bar{Y} X$}\\
This is the same as mechanism (1) with $W^- \to \bar{c} s$.
Since the minimum mass of the final state system is 5.2 GeV, this mechanism is
highly suppressed by phase space.
\end{enumerate}

The low rates for $B\to \Lambda \bar{\Lambda} X$, $\Lambda \bar{p} X$ and
$D^* p \bar{p} X$
suggest that mechanism (2) is small. 
The absence of a two body component in the momentum spectra
 of $B\to \Lambda_c X$, $\Sigma_c X$ indicates that the W-exchange
mechanism is small. Thus it was thought
reasonable to assume that $\bar{B}\to Y_c \bar{N} X$ with
an external spectator  $b\to c W^-$
coupling (Fig.~\ref{btobaryon}(a)) 
 is the principal mechanism in $B$ to baryon transitions.

If $B$ decays to baryons are dominated by $\bar{B} \to \Lambda_c \bar{p} X$
and $\bar{B} \to \Lambda_c \bar{n} X$ then
measurements of the branching fractions for
 $B \to \bar{p} X$,
$B \to p \bar{p} X$ can be used to extract the absolute $\Lambda_c^+ \to
p K^- \pi^+$ branching fraction.  The CLEO~1.5 measurements give
$B (\Lambda_c \to p K^- \pi^+) = 4.3 \pm 1.0 \pm 0.8 \%$ which can be
used to normalize all other measured $\Lambda_c^+$ branching fractions. 
In a similar
fashion, ARGUS finds $(4.1\pm 2.4)$\% for this branching fraction.

An alternate explanation for the absence of a two body component
in $B$ decays to baryons was recently proposed by Dunietz, Falk and
Wise\cite{dunietzbary}. These authors suggested that the primary
mechanism in such decays is 
the internal W-emission process $b\to c \bar{c} s$.
This might lead to two body final states such as 
$\bar{B}\to \bar{\Lambda_c} \Xi_c$
which would account for the softness of the $\Lambda_c^+$ momentum spectrum.  
CLEO has searched for the mechanism suggested by Dunietz \etal~in 
a variety of ways.
By examining $\Lambda_c$-lepton correlations, 
it is possible to constrain the size of the $b\to c \bar{c} s$ component
in $B\to {\rm baryon}$ decays. The $b\to c \bar{c} s$ component gives
rise to opposite sign $\Lambda_c^+ \ell^-$ 
correlations (Fig.~\ref{lambdalep}(b)).
whereas the internal
process W-emission process
$b\to c u {\bar d}$ gives same sign $\Lambda_c^+ \ell^+$ correlations
(Fig.~\ref{lambdalep}(a)).
From the ratio of same sign to opposite sign $\Lambda_c$-lepton 
yields, CLEO finds $b\to c \bar{c} s/
b\to c \bar{u} d = (20\pm 12 \pm 4) \%$ for internal W-emission processes. 
This shows that $b\to c \bar{c} s$, although present at a modest
level, is not
the dominant mechanism operating in B decays to baryons. 

Since the $b\to c \bar{c} s$ mechanism is present, $\Xi_c^+$
and $\Xi_c^0$ baryons should be produced in $B\to$baryon transitions.
However, $\Xi_c^0$ baryons can also be produced from $b\to c\bar{u}d$
transitions with $s\bar{s}$ popping. CLEO~II has observed signals
$B\to \Xi_c^+$ and $B\to \Xi_c^0$. The observed rates are consistent with
what is expected from the measurements of $\Lambda_c$-lepton correlation
and quark popping.
%

To verify whether the dominant mechanism for baryon production in
B decays is the external spectator mechanism 
with $b\to c \bar{u} d$ as was previously assumed 
by the CLEO and ARGUS analyses, 
CLEO~II has searched for evidence of $B\to \Lambda_c
\bar{N} \ell \nu$. This should give rise to several 
distinctive experimental signatures:
$\Lambda$-lepton correlations, $\Lambda_c$-lepton correlations,
and semi-exclusive $B\to \Lambda_c^+ \bar{p} \ell^- \nu$ production having a 
massing mass consistent with a B decay. No significant signals were
observed in these correlations\cite{glasbary}. 
This indicates that
the conventional and previously
accepted picture of baryon production in $B$ decay is incorrect.

A possible explanation of all the existing data requires the
simultaneous presence
 of several production mechanisms. The internal spectator
process $b\to c \bar{u} d$ followed by $u \bar{u}$ 
or $d \bar{d}$ quark popping is
dominant. This leads to production of
a high mass excited anti-nucleon in conjunction
with a charmed baryon and accounts for the soft momentum spectrum
of charmed baryons produced in B decay as well as the absence of
$B\to \Lambda_c \bar{N} X \ell \nu$. The internal spectator process
$b\to c \bar{c} s$ with quark popping as well as the internal spectator
process $b\to c \bar{u} d$ with $s \bar{s}$ quark popping are also
operative at the 10-20\% level. The latter two mechanisms appear to
account for the production of $\Xi_c$ baryons in $B$ decay.

\subsection{Charm Production in $B$ Decay}
\label{charmpro}

The measurements of inclusive decay rates can be used to test the parton level
expectation that most $B$ decays proceed via a $b\to c$ transition.
If we neglect the small contributions from $b\to u$
and penguin transitions, we expect about
1.15 charm quarks to be produced per  $B$ decay.
The additional $15\%$ is due
to the fact that the virtual W forms a $s \bar{c}$ quark pair with
a probability of approximately $0.15 $. 
To verify this expectation we use the experimental results listed in 
Table~\ref{khinc} and determine the charm yield, denoted $n_c$, to be
\begin{eqnarray*}
 n_c & = & {\cal B}(B \to D^0 X) + {\cal B}(B \to D^+ X) +
{\cal B}(B \to D_s X)  \\
 & +&  {\cal B}(B \to \Lambda_c X) +  {\cal B}(B \to \Xi^+_c X) + 
 {\cal B}(B \to \Xi^0_c X) \\
 & + &2\times{\cal B}(B\to \psi X)
 + 2\times{\cal B}(B\to \psi{\rm '} X)
 + 2\times{\cal B}(B\to \chi_{c1} X)  \\
 &+& 2\times{\cal B}(B\to \chi_{c2} X)
 + 2\times{\cal B}(B\to \eta_c X~({\rm incl.~other ~c\bar{c}}))\\
 & =  &  1.15 \pm 0.05  \\ 
\end{eqnarray*}

The factor of 2 that multiplies ${\cal{B}}(B\to c\bar{c}X)$ accounts for the 
two charm quarks produced in $b \to c\bar{c}s$ transitions. Wherever possible
the branching fractions for direct production are used.
The contribution of $B\to \eta_c X$ and
other charmonium states is generously taken to be at the CLEO 90\% confidence
level upper limit for the process $B\to \eta_c X$.
This value of $n_c$ is slightly larger than the
values reported in the 1996 conferences due to the use of
a smaller world average for the absolute branching fraction
${\cal B}(D^0\to K^-\pi^+)$.

Another interesting quantity is the fraction of $B$ decays in which two charm
quarks are produced which is naively expected to be about 15\%.
This expectation can be 
compared to the sum of the experimental measurements
\begin{eqnarray*}
{\cal{B}}(B \to X_{c \bar{c}}) & = & 
{\cal{B}}(B \to D_s X) + {\cal{B}}( B \to \psi X) + {\cal{B}}(B\to \psi' X) \\
 & +& {\cal {B}}(B\to \chi_{c1} X) + {\cal {B}}(B\to \chi_{c2} X) +
{\cal {B}}(B\to \Xi_c X)  \\
 & +& {\cal {B}}(B\to \eta_c X~({\rm incl.~other ~\bar{c}})) \\
 &= & (15.8 \pm 2.8)\% 
\end{eqnarray*}
where the direct $B \to \psi$ and  $B \to \chi_{c1}$ branching fraction have
been used. 
The contribution from $B\to \Xi_c^0 X$ is reduced by 1/3 to take
into account the fraction that is produced not by the 
$b\to c \bar{c} s$ subprocess but by
$b\to c\bar{u}d \, + \, s \bar{s}$ quark popping. 
The measured value of
${\cal{B}}(B \to X_{c \bar{c}})$ calculated in this way is far below $30\%$.


The possibility of an additional contribution
from $B\to D \bar{D} K X$ decays to the hadronic B width
was suggested by Buchalla \etal. These decays proceed
via the quark level process $b\to \bar{c} c s$ with light
quark popping at the upper $\bar{c} s$ vertex.
Such decays give
wrong sign D-lepton correlations from the $\bar{D}$ mesons
that hadronize from the virtual $W$.

Note that such decays would increase the calculated
${\cal B}(b\to c \bar{c} s)$  but do not modify
the determination of $n_c$ (the number of charm quarks
produced per $B$ decay). 

Preliminary evidence for the presence of this decay mechanism
has been presented by CLEO from the observation
of $D-\ell^-$ correlation in $B\bar{B}$ events \cite{kwon}.
An energetic lepton above $1.4$ GeV is chosen in the same hemisphere
as the $D$ meson in order to tag the flavor of the other $\bar{B}$ meson.
After subtracting backgrounds from mixing and lepton misidentification,
${\Gamma}(B\to D X)/{\Gamma}(B\to \bar{D} X)=0.107\pm 0.029\pm 0.018$
which gives ${\cal B}(B\to D X)=8.1\pm 2.6 \%$ for the branching fraction
of the new mechanism.  The adddition of this new decay mechanism increases
${\cal{B}}(B\to c \bar{c} s)$ to $23.9\pm 3.8\%$.
Attempts to reconstruct exclusive modes
such as $B\to D \bar{D} K^{(*)} $ are in progress.

With the addition of recent
experimental results the understanding of baryon production in B decay is 
improving. In contrast
to meson production in $B$ decay, $B \to {\rm baryon}$
transitions proceed predominantly through the internal W-emission process
$b\to c\bar{u}d$ followed by light quark pair popping.
In a parton level calculation with diquark correlation taken into
account, Palmer and Stech \cite{palmstech}
have performed a calculation of the total rate
for inclusive $B$ decay to charmed baryons.
They find ${\cal{B}}(B \to$ charmed baryons) $\approx 6\%$.
In order to compare this prediction with experimental data,
we will assume most $B$ to charmed 
baryon decays proceed through a $\Lambda_c$
baryon but correct for the small fraction of $\Xi_c$ baryons produced
by $b \to c\bar{u}d$ transitions combined with $s\bar{s}$-popping.
This gives
\begin{center}
\begin{tabular}{rcl}
${\cal B}(B \to {\rm charmed~baryons})$ & = & 
${\cal{B}}(B\to \Lambda_c X) + 1/3 \times {\cal{B}}(B\to \Xi_c^0)$\\
 & =  & $(7.1 \pm 1.6)\% $
\end{tabular}
\end{center}

The experimental result for the charm yield per $B$ decay
is consistent with the naive expectation
that $1.15$ charm quarks are produced per $b$ decay. However,
it is not consistent with proposals
that suggest that $1.3$ charm quarks should
be produced per $b$ decay. Such a high charm yield 
is required by some recent theoretical efforts that explain
the discrepancy between 
theoretical calculations and experimental measurements of
the inclusive semileptonic rate by an enhancement of the
$b\to c \bar{c} s$ mechanism (see Section~\ref{baffle}
for a more detailed discussion).

\section{Exclusive Hadronic B Decay}
\label{BDpiDrho}
The experimental branching ratios for $B$ meson decay to
exclusive final states containing $D$ mesons 
are given in Tables ~\ref{kh3}  and \ref{kh4}.

\subsection{Measurements of $D (n \pi)^-$ Final States}

To date, final states containing a $D$ meson and one or two 
pions have been observed. 

To select $ \bar{B} \to D \rho^-$ candidates additional requirements are
imposed on the $\pi^-\pi^0$ invariant mass and the $\rho$ helicity angle
in $ \bar{B} \to D \pi^-\pi^0$ decays.
By fitting the $\pi^- \pi^0$ mass spectrum and the helicity angle distribution,
CLEO~II finds that
at least 97.5\% of the $B \to D \pi^-\pi^0$ rate in the $\rho$ mass region
can be attributed to the decay $B \to D \rho^-$\cite{mcdd}.

\subsection{Measurements of $D^*(n\pi)^-$ Final States}

Final states containing a $D^*$ meson and one, two or
three pions have also been observed. 
These include the $B \to D^* \pi^-$ ,
$B \to D^* \rho^-$, and $B \to D^* a_1^-$ decay channels.
The results for the decays $\bar{B^0} \to D^{*+} \pi^-$,
$\bar{B^0} \to D^{*+} \rho^-$ and $\bar{B^0} \to D^{*+} \pi^-\pi^-\pi^+$
are listed in Table~\ref{kh4}, and the results for
$B^- \to D^{*0} \pi^-$, $B^- \to D^{*0} \rho^-$ and
$B^- \to D^{*0} \pi^-\pi^-\pi^+$ are given in Table ~\ref{kh3}.

The  $B^-$ and $\bar{B}^0$ signals in the $D^* \pi$ and $D^* \rho$
decay channels from the CLEO II experiment are shown in Fig. \ref{dspi}.
It is found that $B \to D^* \pi^-\pi^0$ in the $\pi^-\pi^0$ mass
region near the $\rho$ meson is saturated by the
decay $B \to D^* \rho^-$ (Fig. \ref{subs}) and
a tight upper limit of $<9$\% at
90\% C.L. is set on a possible non-resonant contribution \cite{mcdrho}.

The CLEO~II data also suggest that the signal in
$B\to D^{*}\pi^-\pi^-\pi^+$ arises predominantly from $B\to D^{*} a_1^-$. 
Taking into account 
the $a_1 \to \pi^-\pi^-\pi^+$ branching fractions, it follows
that ${\cal{B}}(B\to D^{*} a_1^-) = 2 \times
{\cal{B}}(B\to D^{*}\pi^-\pi^-\pi^+)$. 
A fit to the
$\pi^-\pi^-\pi^+$ mass distributions with contributions from
$B \to D^{*+} a_1^-$ and a $B \to D^{*+}\pi^- \rho^0$ non-resonant background.
gives an upper limit of 13\% on
the non-resonant component in this decay.

The Cabibbo suppressed decay modes such as 
$B\to D K$ should also be observed and studied in the near future. 
These modes, in particular, 
$B^+\to D^0 K^+$ and $B^+ \to \bar{D}^0 K^+$ with
$D^0\to |f_{CP}>$ (where $|f_{CP}>$ denotes a CP eigenstate) 
will be used at B factories to constrain one of the three angles
of the unitary triangle.

\subsection{Polarization in $B \to D^{*+}\rho^-$ Decays}
\label{pol-D*-rho}

By comparing the measured polarization in $\bar{B^0} \to D^{*+}\rho^-$
with the expectation from the corresponding semileptonic
B decay a test of the factorization hypothesis
can be performed (see Sec.~\ref{fac-ang-cor}).
The polarization is obtained from the distributions of the helicity angles
$\Theta_{\rho}$ and $\Theta_{D^*}$. The $D^{*+}$ helicity angle,
$\Theta_{D^*}$, is the angle between the $D^0$ direction
in the $D^{*+}$ rest frame and the $D^{*+}$ direction
in the rest frame of the $B$ meson.
After integration over $\chi$, the angle between the
normals to the $D^{*+}$ and the
$\rho^-$ decay planes, the helicity angle distribution can be expressed
as
\begin{equation}
{d^2\Gamma\over{d\cos\Theta_{D^*}d\cos\Theta_{\rho}}}
\propto
{1\over{4}}\sin^2\Theta_{D^*}\sin^2\Theta_{\rho}(|H_{+1}|^2+|H_{-1}|^2)
+\cos^2\Theta_{D^*}\cos^2\Theta_{\rho}|H_{0}|^2 \label{polar3d}
\end{equation}
where $H_{i}$ are the amplitudes for the various possible
$D^*$ helicity states.
The fraction of  longitudinal polarization is defined by
\begin{equation}
 {{\Gamma_L}\over{\Gamma}}
 ~ = ~ {{|H_0|^2}\over{|H_{+1}|^2 + |H_{-1}|^2 + |H_{0}|^2}} \label{ratiohel}
\end{equation}
If $\Gamma_L$ is large both
the $D^{*+}$ and the $\rho^{-}$ helicity angles will
follow a $\cos^{2}\Theta$ distribution, whereas a large transverse
polarization, $\Gamma_T$, gives a $\sin^2\Theta$ distribution for both
helicity angles.
An unbinned two dimensional likelihood fit
to the joint $(\cos\Theta_{D^*}, \cos\Theta_{\rho})$ distribution gives
\begin{equation}
(\Gamma_{L}/\Gamma)_{\bar{B^0} \to D^{*+} \rho^-}\; =\; 93 \pm 5 \pm 5 \%
\end{equation}

The same procedure has been applied to a sample of exclusively reconstructed
$B^- \to D^{*0}\rho^-$ decays. While $\bar{B}^0 \to D^{*+}\rho^-$ is an
external spectator decay, $B^- \to D^{*0} \rho^-$ can proceed via both
the external and the internal spectator mechanisms. 
The interference between the two
amplitudes can modify the polarization \cite{keum-pol}. CLEO II
finds \cite{kh-pisa}
\begin{equation}
(\Gamma_{L}/\Gamma)_{B^- \to D^{*0} \rho^-}\; =\; 84.2 \pm 5.1\%
\end{equation}

\subsection{Measurements of $D^{**}$ Final States}
\label{B->D**}

In addition to the production of $D$ and $D^*$ mesons,
the charm quark and spectator antiquark can hadronize as a $D^{**}$ meson.
The $D^{**0}(2460)$ has been observed experimentally and identified
as the J$^P=2^+$ state, while the
$D^{**0}(2420)$ has been identified as the $1^+$ state. These states have
full widths of approximately 20 MeV. Two other states, a $0^+$ and another
$1^+$ are predicted but have not yet been observed, presumably because of their
large intrinsic widths.
There is evidence for $D^{**}$ production in semileptonic $B$
decays\cite{Dssin}, and $D^{**}$ mesons have also been seen in hadronic
decays. However, early experiments did not have sufficient data to
separate the two narrow $D^{**}$ states and hence reported branching
ratios only for the combination of the two (see results listed under
$B \to D_J^{(*)0}$ in Tables~\ref{kh3} and \ref{kh4}).

In order to search for $D^{**}$ mesons from $B$ decays the
final states $B^- \to D^{*+} \pi^- \pi^-$ and
$B^- \to D^{*+} \pi^- \pi^- \pi^0$ are studied.
These decay modes are not expected to occur via
a spectator diagram in which the $c$ quark and the spectator
antiquark form a $D^*$ rather than a $D^{**}$ meson.
The $D^{*+}$ is combined with a $\pi ^-$ to form a $D^{**}$ candidate.
CLEO~II has also looked for $D^{**}$ production in the channels
$B^-\to D^+ \pi^- \pi^-$ and $\bar{B^0}\to D^0 \pi^- \pi^+$.
Since $D^{**0}(2420)\to D \pi$ is forbidden, only the
$D^{**0}(2460)$ is searched for in the $D \pi \pi$ final state.

CLEO II has reported a significant signal in the
$D^{**0}(2420) \pi^-$ mode.
ARGUS has also found evidence for $B \to D^{**}(2420) \pi^-$ using
a partial reconstruction technique in which they observe a fast and slow pion
from the $D^{**}$ decay but 
do not reconstruct the $D^0$ meson\cite{Krieger}.

Other final states with higher pion multiplicities should be systematically
studied in the future. 

\subsection{Exclusive Decays to $D$ and $D_s$ Mesons}
\label{doubledees}

Another important class of modes are decays to two charmed mesons.
As shown in Fig. ~1~(a)
the production of an isolated pair of charmed mesons
($D_s^{(*)}$ and $D^{(*)}$) proceeds through a Cabibbo favored
spectator diagram in which
the $s\overline{c}$ pair from the virtual $W^-$ hadronizes into a
$D_s^-$ or a $D_s^{*-}$ meson and the remaining spectator quark and the
$c$ quark form a $D^{(*)}$ meson. 
These modes have been observed by the CLEO~1.5,
ARGUS and CLEO~II\cite{cleodds} experiments.
B mesons are reconstructed in eight decay modes:
$D_s^-D^+$, $D_s^-D^0$,
$D_s^{*-}D^+$, $D_s^{*-}D^0$,
$D_s^-D^{*+}$, $D_s^-D^{*0}$,
$D_s^{*-}D^{*+}$, and $D_s^{*-}D^{*0}$.
The sum of the branching fractions for the
exclusive modes, averaged over $B^-$ and $\bar{B}^0$
decays, is $5.50 \pm 0.81  \%$. This can be compared
to the branching fraction of the 
two body component found in the fit to the inclusive $D_s$ momentum spectrum
of $4.9\pm 1.3\%$. The error is dominated by the uncertainty in
${\cal{B}}(D_s \to \phi \pi)$. The remaining contribution to 
the inclusive production of $D_s$ mesons must be due to the decay modes
$B\to D_s^{**} D^{(*)}$, $B\to D_s^{(*)} D^{(*)} (n\pi)$ or
$D_s^{(*)} D \pi$.

Partial reconstruction techniques have also been 
used to improve the size of the signals in
$B\to D^{(*)} D_s^{(*)+}$. Larger samples not only  reduce the
statistical error in the branching ratio measurements, they
also allow the polarization in $B\to D^* D_s^{*+}$ decays to be determined.
Comparison of the yield in partially reconstructed 
and fully reconstructed $B\to D^* D_s^{(*)+}$
events gives a model independent 
measurement  of ${\cal B}(D_s\to \phi\pi^+)$ which sets the
scale for the $D_s$ branching fractions.
Branching fractions and background levels for CP eigenstates
such as $\bar{B}^0\to D^{(*)+} D^{(*)-}$ will also be studied.

Since the internal spectator mechanism cannot contribute
to the $B \to D^{(*)}D_s^{(*)}$ decay modes, 
in the absence of higher order 
processes the $B^-$ and $\bar{B}^0$ decay widths will be equal
$$
\frac{\Gamma(\bar{B}^0 \to D^{(*)}D_s^{(*)})}{\Gamma(B^- \to D^{(*)}D_s^{(*)})}
= \frac{{\cal{B}}((\bar{B}^0 \to D^{(*)}D_s^{(*)})}
{{\cal{B}}(B^- \to D^{(*)}D_s^{(*)})}
\times \frac{\tau_{B^-}}{\tau_{\bar{B}^0}}
= 1
$$
Using the world average for the lifetime ratio we find
$$
\frac{\Gamma(\bar{B}^0 \to D^{(*)}D_s^{(*)})}{\Gamma(B^- \to D^{(*)}D_s^{(*)})}
= 0.78 \pm 0.23
$$
which is consistent with this expectation.

\subsection{Exclusive B Decay to Baryons}
The first exclusive $B\to$baryon decay has been observed by
CLEO~II\cite{exclbaryon}. A small number of decays
were reconstructed in the
modes $\bar{B}^0\to \Lambda_c^+ \bar{p} \pi^+ \pi^-$ and 
$\bar{B}^0\to \Lambda_c^+ \bar{p} \pi^-$ corresponding to
branching fractions
 of $0.162^{+0.019}_{-0.016}\pm 0.038\pm 0.026\% $ and
  $0.63^{+0.023}_{-0.020}\pm 0.012\pm 0.010\% $, respectively.
In addition, CLEO ~II has 
set limits on other higher multiplicity 
exclusive modes with baryons in the final state.

\subsection{Color Suppressed B decay}
\label{B->psi-K(*)}

\subsubsection{Exclusive $B$ Decays to Charmonium}
\label{intro-B->psi-K(*)}

In $B$ decays to charmonium mesons, the $c$ quark from the
$b$ decay combines with a $\bar{c}$ quark from the virtual $W^-$ decay to
form a charmonium state. This process is described by the color suppressed
diagram shown in Fig.~1~(b).
The branching fractions for these modes are listed in Tables~\ref{kh3}
and \ref{kh4}. 
Comparing $B$ meson decays to different final states containing
charmonium mesons allows to investigate the dynamics of the underlying
decay mechanism.

The decay modes $\bar{B^0} \to \psi K^0$ and  $\bar{B^0} \to \psi' K^0$ are
of special interest since the final states are
CP eigenstates. These decays are of great importance
for the investigation of
one of the three CP violating angles accessible to study in $B$ decays.
It is also possible to use the decay
$\bar{B^0} \to \psi K^{*0}$, $K^{*0} \to K^0 \pi^0$ which has a
somewhat higher branching ratio, but this final state 
consists of a mixture of CP eigenstates.
It has even CP if the
orbital angular momentum L is 0 or 2 and odd CP for L=1.
If both CP states are present the CP asymmetry will be diluted.
A measurement of CP violation in this channel is only possible if one of the
CP states dominates, or if a detailed moments analysis 
is performed \cite{Idunit}.
Measurements of the polarization in the decay $\bar{B^0}
\to\psi \bar{K^{*0}}$  can be used to determine the
fractions of the two CP states.

To form $B$ meson candidates a $c\bar{c}$ meson is combined with 
a strange meson candidate. 
Decay modes of this type have been reconstructed by
CLEO~1.5, ARGUS, and CLEO~II.
The CDF Collaboration  \cite{cdfpsik} has also reported signals
for $B\to \psi K^{*0}$ and $B\to \psi K^-$ 
and measurements of polarization in $B\to \psi K^*$ decays\cite{cdfpolar}.
Because of the large uncertainties associated with the $b$-quark production 
cross section at the Tevatron the results are given as ratios
of branching fractions,
\begin{eqnarray*}
\frac{{\cal{B}}(B^0 \to \psi K^0)}{{\cal{B}}(B^+ \to \psi K^+)} & = &
1.13 \pm 0.22 \pm 0.06 \% \\
\frac{{\cal{B}}(B^0 \to \psi K^{*0})}{{\cal{B}}(B^+ \to \psi K^+)} & = &
1.33 \pm 0.27 \pm 0.11 \% \\
\frac{{\cal{B}}(B^+ \to \psi K^{*+})}{{\cal{B}}(B^+ \to \psi K^+)} & = &
1.55 \pm 0.46 \pm 0.16 \% \\
\end{eqnarray*}
Assuming equal production of $B^+$ and $B^0$ mesons the measurements can be combined
to determine the vector to pseudoscalar ratio in $B \to \psi$ decay
\begin{eqnarray}
\frac{{\cal{B}}(B \to \psi K^*)}{{\cal{B}}(B \to \psi K)} & = &
1.32 \pm 0.23 \pm 0.16 \\
\label{eq:cdfpsiratio}
\end{eqnarray}

Using the world average branching fractions
from Tables \ref{kh3} and \ref{kh4}
and combining $B^-$ and $\bar{B}^0$ decays
we determine the sum of the  exclusive two-body decays to ${\cal{B}}(B\to \psi
\:K(K^*,\:\pi)) \; = \; 0.258 \pm 0.030\%$ and 
${\cal{B}}(B\to \psi$'$\:K(K^*,\:\pi)) \; = \; 0.22 \pm 0.09\%$. Thus
about 1/4 of the inclusive rate for direct
$B\to \psi$ production can be accounted for by exclusive modes. 
The experimental investigation of the remaining
fraction is important, since any additional 
quasi-two body channel open to $B\to \psi$ transitions
could be useful for future studies of CP violation.
$\psi$  mesons of lower momentum could originate from 
multibody final states or from two-body decays involving heavier
$K^{(*)}$ resonances.

Evidence for the decay mode $B\to \chi_{c1} K$ has been reported by CLEO~II
\cite{fastpsi,SixthB}
and  ARGUS \cite{FifthB}. The average branching fraction is 
${\cal B}(B^-\to \chi_c K^-) = (0.104\pm 0.040) \%$.
The CLEO~II collaboration has also placed upper limits on $\chi_{c1}K^0$
and $\chi_{c1}K^*$ production in $B$ decay.

Signals for Cabibbo suppressed $B$ decays with charmonium states
have been found by CLEO~II and CDF in the
decay mode $B^+\to \psi \pi^+$\cite{cleopsipi},\cite{cdfpsipi}.

\subsubsection{Polarization in $B\to\psi K^*$ }

The polarization in $B\to\psi K^*$ is studied by using the methods described
for the 
$\bar{B^0}\to D^{*+}\rho^-$ polarization measurement
in Section \ref{pol-D*-rho}.
After integration over the azimuthal angle between the $\psi$ and the
$K^*$ decay planes, the angular distribution in $B \to \psi K^*$ decays
can be written as
\begin{equation}
 {d^2\Gamma\over{d\cos\Theta_{\psi}d\cos\Theta_{K^*}}}
\propto {1\over{4}}\sin^2\Theta_{K^*}
(1+\cos^2\Theta_{\psi})(|H_{+1}|^2+|H_{-1}|^2)
+\cos^2\Theta_{K^*}\sin^2\Theta_{\psi}|H_{0}|^2 , \label{psipolar}
\end{equation}
where the $K^*$ helicity angle $\Theta_{K^*}$ is the angle between
the kaon direction in the $K^*$ rest frame and the $K^*$ direction in the
$B$ rest frame, $\Theta_{\psi}$ is the corresponding $\psi$ helicity angle,
and $H_{\pm1,0}$ are the helicity amplitudes.
The fraction of longitudinal polarization in $B \to \psi K^*$
is determined by an unbinned fit to the $\psi$ and $K^*$ helicity angle
distributions. The results obtained by the CLEO~II, ARGUS and
CDF collaborations are given in Table \ref{Tpsipolex}.
The efficiency corrected distributions in
each of the helicity angles $\cos\Theta_{\psi}$ and $\cos\Theta_{K^*}$
are shown in Fig.~\ref{expol} (CDF).
Assuming that the systematic errors from the various
experiments are uncorrelated, these three results can be averaged to obtain
\begin{equation}
{\Gamma_L\over \Gamma} = 0.78 \pm 0.07  \label{psikstavg}
\end{equation}
In addition, CDF has reported the first measurement of polarization
for the $B_s\to \psi\phi$ mode,
$${\Gamma_L}/{\Gamma}= 0.56\pm{0.21}^{+0.02}_{-0.04}.$$

Although the decay mode $B \to \psi K^*$  may not be completely
polarized,  it is still dominated by a single CP eigenstate and,
therefore, will be useful for measurements of CP violation.

\subsubsection{Exclusive Decays to a $D^{0 (*)}$ and a Neutral Meson.}
\label{color-supress}

$B$ decays that can occur
via an internal W-emission graph but that
do not yield charmonium mesons in the final
state are expected to
be suppressed relative to decays  that
occur via the external W-emission graph.
For the internal graph, in the absence
of gluons, the colors of the quarks from the virtual $W$ must
match the colors of the $c$ quark
and the accompanying spectator antiquark.
In this simple picture, one expects that the suppression
factor should be  $1/18$ in rate for decays involving $\pi^0$, $\rho^0$
and $\omega$ mesons.
In heavy quark decays the effects
of gluons cannot be neglected, and QCD based calculations
 predict larger suppression factors on the order of $1/50$\cite{Neubie}.
If color suppression is much less than expected,
as is the case for some charm meson decays, then these
$B$ decay modes 
modes could also be useful for CP violation studies\cite{Dunietz}.

CLEO~II has searched for color suppressed decay modes of $B$ mesons that
contain a single $D^0$ or $D^{*0}$ meson in the final state\cite{wex}.
The relevant color suppressed modes are listed in Table~\ref{Tbrcolcomp}.
No signals were observed.
Upper limits \cite{PDGul} on the
branching ratios for color suppressed modes are given in Table~\ref{Tbrcolcomp}.
These limits indicate that color suppression is present in $B$ decay.

\section{\bf Hadronic Decays: Theoretical Interpretation}
\subsection{The Effective Hamiltonian}
The Hamiltonian for weak hadronic charm (or beauty) 
decays is modified by gluon
exchange between the quark lines in two ways. Hard gluon 
exchanges can be accounted for by perturbative methods and renormalization 
group techniques\cite{AG}. There are also
 long distance or non-perturbative 
interactions are responsible for the binding of 
quarks inside the asymptotic hadron states.
 It is possible to separate the two regimes by means of 
the operator product expansion\cite{Wilson}, which 
incorporates all long range 
QCD effects into the hadronic matrix element of local four-quark 
operators\cite{Neubie}.
 The effective Hamiltonian\cite{GW}, for example in the case of the charm 
decays, can be expressed as,
\begin{equation}
H_{eff} = \frac{G_F}{\sqrt{2}} V_{cs}^{*} V_{ud}  
[ \frac{c_{+} + c_{-}}{2} (\overline{u}d)(\overline{s}c) + 
  \frac{c_{+} - c_{-}}{2} (\overline{s}d)(\overline{u}c) ]
\label{Heff}
\end{equation}
where Cabibbo-suppressed transitions and penguin diagrams are neglected. 
Here, $(\overline{q_i}q_j)$ denotes 
$\overline{q_i}\gamma^{\mu}(1-\gamma^5)q_j$, $G_F$ is the Fermi coupling 
constant and $c_{\pm}$ are the Wilson coefficients. Gluon 
exchange has the effect of generating the second term which is
an effective neutral current.
Without QCD corrections,
$c_{+} = c_{-} = 1$ and the usual weak Hamiltonian is recovered. 
The Wilson coefficients $c_{\pm}(\mu)$ can be evaluated from QCD in
the leading logarithmic approximation\cite{Neubie}.
There is a large uncertainty in the calculation from the
choice of the scale $\mu$. 
The usual scale is taken to be $\mu \sim m_Q$, so that for the c-quark 
\hbox{($m_c=1.5 \, {\rm GeV},
\Lambda^{(4)}=234 \, {\rm MeV}$)} and b-quark 
\hbox{($m_b=5 \, {\rm GeV}, \Lambda^{(5)}=200 \, {\rm MeV}$)} we obtain,
\begin{eqnarray}
\label{c1c2}
 {\rm c-quark} \quad : \quad c_{1} = \frac{c_{+}+c_{-}}{2} = +1.25 \qquad
                       c_{2} = \frac{c_{+}-c_{-}}{2} = -0.49 \\
 {\rm b-quark} \quad : \quad c_{1} = \frac{c_{+}+c_{-}}{2} = +1.12 \qquad
                       c_{2} = \frac{c_{+}-c_{-}}{2} = -0.27 \nonumber
\end{eqnarray}

\subsection{Factorization}
 The hypothesis that the decay amplitude can be expressed
as the product of two single current matrix elements is called
factorization. This hypothesis is taken 
by  analogy to semileptonic decays where 
the amplitude can be decomposed into a leptonic and a hadronic current. 
A qualitative justification for the factorization hypothesis based on 
color transparency was suggested by Bjorken\cite{Bjorken}. 
For example, in a $B^-$ decay, a $\overline{u}d$ pair, 
which is produced as a color singlet from the virtual $W^-$, could travel 
fast enough to leave the interaction region without influencing the second 
hadron formed from the c quark and the spectator antiquark. 
Buras, Gerard, and Ruckl\cite{BGR} show that factorization is valid in the 
limit $1/N_c \rightarrow 0$ and have considered leading $1/N_c$ corrections to 
this limit. Dugan and Grinstein\cite{DG} have suggested that 
factorization follows from perturbative QCD in certain kinematic regions.
 It is expected that the factorization hypothesis will be more reliable in $B$ 
hadronic decays than in the equivalent $D$ 
hadronic decays because of the larger energy transfers.

 There are several phenomenological models of the nonleptonic two-body decays 
of heavy flavors\cite{Models}. The
model of Bauer, Stech, and Wirbel (BSW) is widely used\cite{BSW}. 
In addition to factorization,
the BSW model uses hadronic currents instead of quark currents and 
allows the coefficients $a_{1},a_{2}$  
of the products of currents to be free 
parameters determined by experimental data. 
The effective Hamiltonian becomes
\begin{equation}
H = \frac{G_F}{\sqrt{2}} V_{cs}^{*} V_{ud} 
[ a_{1} (\overline{u}d)_{H} (\overline{s}c)_{H} + 
  a_{2} (\overline{s}d)_{H} (\overline{u}c)_{H} ]
\label{Heffa1a2}
\end{equation}
The relation between $a_{1}$, $a_{2}$ and the QCD coefficients $c_{1}$,
$c_{2}$ is:
\begin{eqnarray}
\label{a1a2c1c2}
 a_{1} & = c_{1} + \xi c_{2} \\
 a_{2} & = c_{2} + \xi c_{1} \nonumber
\end{eqnarray}
where the factor $\xi (=1/N_c)$ is the color matching factor.
 
 Three classes of decay can be distinguished: decays determined by $a_1$ 
(class I) e.g. $D^{0}\rightarrow K^{-}\pi^{+}$ (Fig.1a); decays determined by 
$a_2$ (class II), e.g. $D^{0}\rightarrow \overline{K^{0}}\pi^{0}$ (Fig.1b) and 
those where both
the $a_1$ and $a_2$ contributions are present
and interfere (class III) e.g. the decay 
$D^{+}\rightarrow \overline{K^{0}}\pi^{+}$.
In this model, the 
rate for any exclusive two-body decay can be calculated once the 
parameters $a_1$ and $a_2$ are given. 

  For example, the amplitude for the decay 
$D^{0} \rightarrow K^{-}\pi^{+}$ (neglecting the exchange diagram term)
is given by,
\begin{equation}
A = \frac{G_F}{\sqrt{2}}V_{cs}^{*}V_{ud}
a_{1}<\pi^{+}\mid(\overline{u}d)_{H}\mid 0> 
<K^{-}\mid(\overline{s}c)_{H}\mid D^0>
\end{equation}
where the first matrix element is the amplitude for creating a pion from the 
vacuum via the axial current, proportional to the pion decay constant $f_\pi$;
the second term is the matrix element for the transition $D^{0}\to K^-$ 
that can be expressed in terms of form factors extracted from the data 
on semileptonic decays\cite{Review_2}.

\subsection{Final State Interactions}
 FSI can dramatically modify observed decay rates. 
These interactions occur in a space-time region where the final state 
particles have already been formed by the combined action of weak and strong 
forces, but are still strongly interacting while recoiling from each 
other\cite{Neubie}.
 
 The relation between the decay amplitudes $A_i$, corresponding to final state
{\it i}, and the bare amplitude $A^{0}_{j}$ (without FSI) is
\begin{equation}
 A_{i} = \sum_{j} S^{1/2}_{ij} A^{0}_{j} \label{Fsiai}
\end{equation}
where S denotes the strong interaction S-matrix for hadron-hadron scattering.
 As a result, there is mixing between
 channels having the same quantum numbers and relative phases will be induced.
It is worthwhile to note that, in general, the final state
{\it i} might not be directly accessible through weak decay diagrams.
For example, the observed rate for a decay mode
that is small in a short-distance quark level calculation
 can be dramatically enhanced by rescattering 
from modes with larger branching fractions.
 The factorization approximation can only be used to determine the 
bare decay amplitudes $A^{0}_{j}$.
One way to eliminate the uncertainties associated with the S-matrix is 
to sum over all decay channels with the same conserved quantum numbers. From
Equation~(\ref{Fsiai}) and the unitarity of the S-matrix, we have 
\begin{equation}
 \sum _{i} \mid A_{i} \mid^{2} = \sum_{j} \mid A^{0}_{j} \mid^{2}
\end{equation}
 that is, the sum of related decay rates remains unaffected by FSI.

 It is customary to distinguish elastic and inelastic FSI. 
For example, for two coupled channels Equation (~\ref{Fsiai})
gives,
\begin{equation}
 \left( \begin{array}{c}
     A_1 \\ 
     A_2 \\ 
        \end{array} 
 \right) = 
 \left( \begin{array}{c}
  \quad \eta e^{2i\delta_1} \qquad\qquad\qquad i\sqrt{1-\eta^2} 
 e^{i(\delta_1+\delta_2)} \quad \\
  \quad  i\sqrt{1-\eta^2} e^{i(\delta_1+\delta_2)} \qquad\qquad\qquad 
 \eta e^{2i\delta_2} \\
        \end{array} 
 \right) 
 \left( \begin{array}{c}
     A^0_1 \\ 
     A^0_2 \\ 
        \end{array} 
 \right)_{bare} 
\end{equation}
where $\delta_1,\delta_2$ are the strong interaction phase shifts and 
$\eta$ is the elasticity parameter\cite{BSW}\cite{Kamal_1}. 
Inelastic FSI $(\eta < 1)$  alter
the observed amplitudes compared with the factorization predictions. 
It should be noted that 
elastic FSI $(\eta=1)$ may also change the observed width of 
coupled channels by modifying the interference between 
two isospin amplitudes.

\subsection{Heavy Quark Effective Theory}

It has recently been appreciated that there is a symmetry of QCD
that is useful in understanding systems containing one heavy quark.
This symmetry arises when the quark becomes sufficiently heavy
to make its mass irrelevant to the nonperturbative dynamics of the
light quarks. This allows the heavy quark degrees of freedom to
be treated in isolation from the light quark degrees of freedom.
This is analogous to the canonical treatment of a hydrogen atom,
in which the spin and other properties of the nucleus can be neglected.
The behavior and structure of the atom 
are determined by the electron degrees of freedom.
Heavy quark effective theory (HQET) was developed by
Isgur and Wise \cite{ISGW} who define a single universal form factor,
$\xi(v\cdot v^{'})$, known as the Isgur-Wise function. In this function $v$ and
$v^{'}$ are the four-vector 
velocities of the initial and final state heavy quarks.
In the heavy quark limit all the
form factors for hadronic matrix elements such as $B\to D^*$ and
$B\to D$ can be related to this single
function. The value of this function can then be determined from a measurement
of the $B\to D^* \ell \nu$ rate as a function of $q^2$ \cite{ISGW}.
The theory also provides a framework for systematic calculations
of corrections to the heavy quark limit.

The evaluation of amplitudes for hadronic decays requires not only the
assumption of factorization, but also the input of hadronic form factors and
meson decay constants. As a result of the development of HQET it is now
believed that many of the hadronic form factors for $b \to c $ transitions
can be calculated quite
well in an essentially model-independent way. This has been done by
several groups \cite{Neubie},\cite{Bari}. The comparison of these theoretical
predictions with the experimental results can be used to test the range of
validity of HQET, and the extent to which $1/M_Q$ corrections to the heavy
quark symmetry are needed. It is not yet clear whether
HQET can also be correctly applied to the calculation of form factors for
charm quark decays.
\subsection{FSI in Charm Decay}

 The presence of FSI often complicates the comparison 
between experimental data and theoretical predictions. 
In charm
decay FSI are particularly problematic because there are 
several resonances in the mass region of charm hadrons. 
From the measurements now
available, it is possible to disentangle these
contributions.

 An isospin analysis gives quantitative results about the FSI. 
For example, consider the decays 
$D^{0}\rightarrow \pi^{-}\pi^{+}$, $D^{0}\rightarrow \pi^0\pi^0$
and $D^{+}\rightarrow\pi^0\pi^+$ which represent all possible two-body 
$D\rightarrow \pi\pi$ transitions. The amplitudes for these
final states can be expressed in terms of 
amplitudes for the isospin 0 and 2 eigenstates.
 The isospin decomposition gives\cite{GL}:
\begin{eqnarray}
 A(D^{0}\rightarrow \pi^{-}\pi^{+}) & = & \frac{1}{\sqrt{3}}( \sqrt{2}A_{0} +
 A_{2} ) \\
 A(D^{0}\rightarrow \pi^{0}\pi^{0}) & = & \frac{1}{\sqrt{3}}( -A_{0} +
 \sqrt{2}A_{2} ) \nonumber \\
 A(D^{+}\rightarrow \pi^{0}\pi^{+}) & = & \sqrt{3/2}A_{2}  \nonumber
\end{eqnarray}
 where $A_{I} =  A_{I}~ e^{i\delta_{i}}$ is the complex amplitude 
for isospin I and $\delta$ is the phase shift from FSI. 
These expressions lead to the following phase independent relations,
\begin{eqnarray}
\label{Phaseind}
 |A(D^{0}\rightarrow \pi^{-}\pi^{+})|^2 + 
 |A(D^{0}\rightarrow \pi^{0}\pi^{0})|^2 & = & |A_{0}|^2 + |A_{2}|^2 \\
 |A(D^{+}\rightarrow \pi^{0}\pi^{+})|^2 & = & \frac{3}{2}|A_{2}|^2  \nonumber
\end{eqnarray}
 Using the $D$ branching fractions from this review,\cite{bfd}
we have calculated world averages for 
 phase shifts and isospin amplitudes (Table~\ref{Isospind}). 
The results in Table~\ref{Isospind} show that several phase shifts between
different isospin amplitudes are close to $90^{\circ}$ 
indicating large contributions from FSI. Moreover, the 
lower isospin amplitudes are always larger than the higher ones.

If inelastic FSI can be neglected, it is possible
to extract branching fractions 
corrected for FSI\cite{BSW}. The prescription consists in 
adding the isospin amplitudes with zero phase shift. 

 In Table~\ref{Cabfavd} and ~\ref{Cabfavds} branching fractions for 
a number of Cabibbo-favoured and 
Cabibbo-suppressed decays are compared to predictions of the BSW model 
using updated values for $a_1$ and $a_2$. 
The values in parentheses are BSW model predictions corrected for 
isospin phase shifts (taken from Table~\ref{Isospind}) due to FSI.
 These corrections generally improve the agreement with the data.
However, some serious discrepancies still 
remain in the decays to vector particles 
such as $K^*$ and $\rho$. These discrepancies could be due 
to either an incorrect 
determination of these form factors or to mixing between 
$\overline{K}\rho$ and $K^\star\pi$ due to inelastic FSI.

While the BSW model agrees reasonably well with the experimental measurements 
of branching fractions
for two-body decays of $D^{0}$ and $D^+$, it fails 
to predict the observed pattern of $D_s^+$ decays. 
For example, the ratio of hadronic branching fractions
\hbox{$\Gamma(D_s\rightarrow \eta\pi^+)/\Gamma(D_s\rightarrow\eta^\prime
\pi^+)=0.39\pm 0.13$}\cite{PDG} is considerably different from the
corresponding semileptonic ratio \hbox{$\Gamma(D_s\rightarrow \eta\ell^+\nu)/
/\Gamma(D_s\rightarrow\eta^\prime\ell^+\nu)=2.45\pm0.94$}\cite{Batt94}.
These two ratios are expected
to be nearly equal if factorization holds. A possible 
explanation could be the interference between spectator and annihilation 
diagrams\cite{Lipkin}; a relatively small annihilation amplitude 
could have a large effect via an interference term.
Alternately, there could be a gluonium component in the $\eta$' meson
that is responsible for the enhancement of the hadronic current.

  Another puzzling problem is the anomalous value of the ratio
of the Cabibbo-suppressed decay of $D^0$ into $K^+ K^-$ and 
$\pi^+ \pi^-$\cite{PDG}:
\begin{equation}
\frac{\Gamma(D^{0}\rightarrow K^{-}K^{+})}
 {\Gamma(D^{0}\rightarrow\pi^{-}\pi^{+})} = 2.85 \pm 0.20 
\end{equation}
Models predict a substantially lower number, $\sim 1.4$ from
SU(3) breaking  in the decay constants.
The suggestion that penguin contributions could explain 
such a high value\cite{Sanda} 
seems to be ruled out by a recent calculation\cite{Kamal_2}. 
 FSI seem to be responsible for the sizeable 
branching fraction 
${\cal B}(D^{0}\rightarrow K^{0}\overline{K}^{0})$ as the quark
level contributions from 
two W-exchange diagrams are small.
 A better way to look at this problem would rather be to consider 
the ratio\cite{Bigi_2}:
\begin{equation}
\frac{\Gamma(D^{0}\rightarrow K^{-}K^{+}) + \Gamma(D^{0}\rightarrow 
 K^{0}\overline{K}^{0})}
 {\Gamma(D^{0}\rightarrow \pi^{-}\pi^{+}) + \Gamma(D^{0}\rightarrow 
 \pi^{0}\pi^{0})} = 2.3 \pm 0.4
\end{equation}
 This ratio should not be affected by elastic FSI because the sum of $D^0$ 
decay modes is independent of strong interaction phases. 
The measured value is still above  the expected level 
of SU(3) breaking (1.4). 
Inelastic final state interactions may explain this 
ratio\cite{Kamal_1}\cite{Kamal_2}. A recent calculation
that takes into account both non--spectator diagrams and rescattering 
effects can also accommodate this result\cite{BLMPS}.

Inelastic FSI are probably responsible for the decay mode 
$D^0\rightarrow\phi\overline{K}^0$, which is observed at the level of 
$0.8\pm 0.1 \%$\cite{bfd}. Initially this
decay mode was called the {\it smoking
gun} for W-exchange in charm decay, 
since it cannot occur at the quark level
in any other way. However, Donoghue noted that the large rate for 
the $\phi\overline{K}^0$ channel could be the result of rescattering from 
other decay modes such as $\overline{K}^{\star0} \eta$\cite{Donoghue}. 
This explanation requires the branching fraction for
$D^0\rightarrow\overline{K}^{\star0} \eta$
be large enough  O($2 \%$) to allow for significant 
rescattering. The observed value, $\Gamma(D^0\rightarrow 
\overline{K}^{\star0} \eta) = 1.7 \pm 0.5 \%$\cite{bfd}, supports
this interpretation. In this sense the decay
$D^0\rightarrow\phi\overline{K}^0$ should now be considered 
the {\it smoking gun} for inelastic FSI in charm decay.

Observation of the decay $D^+\to \phi K^+$ was
reported by the E691 collaboration\cite{e691phik}
with a branching ratio 
${\cal B}(D^+\to \phi K^+)/
{\cal B}(D^+\to \phi \pi^+)= (5.8^{+3.2}_{-2.6} \pm 0.7) \%$.
This decay mode is quite unusual and intriguing. At the quark level,
it is doubly Cabibbo suppressed and requires annihilation.
Rescattering may contribute but the rescattering must proceed
from an initial state which is doubly Cabibbo suppressed.
Observation of a similiar non-resonant decay $D^+\to K^+ K^- \pi^+$
was reported by the WA82 collaboration\cite{wa82phik}. 
These signals are not confirmed by E687 which finds\cite{e687phik}
${\cal B}(D^+\to \phi K^+)/
{\cal B}(D^+\to \phi \pi^+)<2.1 \%$ at the 90\% confidence level
${\cal B}(D^+\to K^+ K^- K^+)/
{\cal B}(D^+\to K^- \pi^+\pi^+)<2.5 \%$ at the 90\% confidence level,
which are marginally consistent with the original observations.

\subsection{Tests of the Factorization Hypothesis }
\label{test-factor}

\subsubsection{Tests of Factorization with Branching Fractions}

 The factorization hypothesis can be tested by comparing hadronic 
exclusive decays to the corresponding semileptonic mode. These tests
can be performed for exclusive hadronic decays of
either $D$ or $B$ mesons\cite{Stone}.

As an example, we 
consider the specific case of $\bar{B^0}\to D^{*+}\pi^-$.
The amplitude for this reaction is
\begin{equation}
A ={G_F\over \sqrt 2}V_{cb}V_{ud}^*{c_1}
\langle \pi^- | (\bar{d} u) | 0 \rangle
\langle D^{*+} | (\bar c b) | \bar{B^0} \rangle.\label{EHeffDP}
\end{equation}
The CKM factor $|V_{ud}|$ arises from the $W^-\to\bar u d$ vertex. The first
hadron current that creates the $\pi^-$ from the vacuum is related to
the pion decay constant, $f_{\pi}$, by:
\begin{equation}
\langle \pi^-(p) | (\bar d u) | 0 \rangle = -if_{\pi}p_{\mu}.\label{Efpi}
\end{equation}
The other hadron current can be determined from the
semileptonic decay $\bar{B^0}\to D^{*+}\ell^- \bar{\nu_{\ell}}$.
Here the amplitude is the product of a lepton current and the hadron
current that we seek to insert in Equation~(\ref{EHeffDP}).

Factorization can be tested experimentally by
verifying the relation
\begin{equation} {\Gamma\left(\bar{B^0}\to
D^{*+}\pi^-\right)\over\displaystyle{d\Gamma\over
\displaystyle dq^2}
\left(\bar{B^0}\to D^{*+} \ell ^- \bar{\nu_{l}}
\right)\biggr|_{q^2=m^2_{\pi}}} =
6\pi^2{ c_1^2}
f_{\pi}^2|V_{ud}|^2 ,\label{Efact}
\end{equation}
is satisfied. Here
$q^2$ is the four momentum transfer from the $B$ meson to the $D^*$
meson. Since $q^2$
is also the mass of the lepton-neutrino system, by
setting $q^2 = m_{\pi}^2=0.019 ~ GeV^2$
we are requiring that the lepton-neutrino system has
the same kinematic properties as does the
pion in the hadronic decay. 
For the coefficient $c_1$ we will use the value $1.12\pm 0.1$ 
at $\mu=m_b$ deduced from
perturbative QCD for the factorization tests involving
hadronic $B$ decays\cite{qcd}. 
The error in $c_1$ reflects the uncertainty
in the mass scale at which the coefficient $c_1$ should be evaluated.
In the original test of Equation~(\ref{Efact}),  Bortoletto and Stone
\cite{Bort} found that the equation was satisfied for $c_1$=1.
In the following discussion we will denote the left hand side of
Equation~(\ref{Efact}) by
$R_{Exp}$ and the right hand side by $R_{Theo}$.

We now consider the channels $D^0\rightarrow K^-\pi^+$
and $D^0\rightarrow K^{\star-}\pi^+$ which are
examples of $P\rightarrow PP$ 
and $P\rightarrow VP$ decay modes, respectively. The semileptonic modes
which should be compared are $D\to K\ell\nu$ and $D\to K^*\ell\nu$,
respectively. 
An updated value for $a_1$ has been determined in 
this review and is used in place of $c_1$ in the factorization test, 
while the semileptonic quantities are extracted from a recent 
review\cite{Review_2}. In the following, each form factor 
is assumed to have a pole form for the $q^2$ dependence, with $M_p=2.1$
GeV/$c^2$ for the vector and $M_p=2.5$
GeV/$c^2$ for the axial vector. The other ingredients for the 
factorization test are collected in Table~\ref{Ingred}. 

We emphasize that the branching fraction 
with elastic FSI removed is the quantity
which should be compared to the semileptonic rate for a factorization test.  
Therefore, we correct the measured branching fractions using the values of 
the isospin amplitudes and phase shifts from Table~\ref{Isospind}.
For example, the $D^0\rightarrow K^-\pi^+$ branching fraction without 
elastic FSI is
$\Gamma(D^0\rightarrow K^-\pi^+)_{no~ FSI}=(1.3\pm 0.1) \times 
\Gamma(D^0\rightarrow K^-\pi^+)_{measured}$. 
The results of factorization tests for charm decays
are given in Table~\ref{Restestd}.

 There is excellent agreement for the pseudoscalar decay mode, while we 
note a serious discrepancy for the vector mode. On the other hand we have 
already observed that models based on
factorization give a poor description of the observed rates for charm
decay modes involving vector particles, such as $K^{\star}$ and $\rho$.

 An alternative way to test factorization has been proposed by Kamal 
and Pham\cite{Kamal_3}. The main feature of their proposal is to compare
quantities that are independent of the strong interaction phases (see 
for example Equation~(\ref{Phaseind})), 
focusing on the isospin amplitudes instead 
of the decay amplitudes. The breakdown of factorization in channels
involving vector particles is attributed to
an inelastic coupling between $\overline{K}\rho$ and
$\overline{K}^{\star} \pi$ channels in the I=3/2 state, which could feed   
the $\overline{K}^{\star} \pi$ final state at the expense of the 
$\overline{K}\rho$ channel.

The large samples of reconstructed hadronic $B$ decays
have been used to obtain precise measurements of branching fractions
as discussed in section \ref{BDpiDrho}. 
These results can also  be used to test the factorization hypothesis. 
The factorization  tests 
can be extended to $B$ decays by using the modes
$\bar{B^0}\to D^{*+} X^-$ decays, e.g. $X=\pi^-$, $ X^- =\rho^-$ or
$a_1^-$.

To obtain numerical predictions for $R_{theo}$,
we must interpolate the observed differential
$q^2$ distribution \cite{width} for
$B \to D^* \ell ~\nu$ to $q^2=m_\pi^2$, $m_\rho^2$, and $m_{a_1}^2$,
respectively. 
Until this distribution is measured more precisely we have to use
theoretical models to perform this interpolation.
The differences between the extrapolations using
models for $B \to D^* \ell ~ \nu$ are 
small, on the order of 10-20\%.
The measurement of this differential distribution recently published
by CLEO~II can be combined 
with the earlier results from the ARGUS and
CLEO 1.5 experiments\cite{Bort,bhp}.
The values of $d\Gamma/dq^2(B\to D^*\ell \nu)$  used for the
factorization test are given in Table~\ref{TFactst}.
Using the information listed in Table~\ref{TFactst}
we obtain from Equation~(\ref{Efact})the results (which are
similar for $D^*\rho$) 
given in Table ~\ref{Tfactc}.

At the present level of precision, there is good
agreement between the experimental results and the expectation from
factorization for hadronic $B$ decays in
the $ q^2$ range $ 0 < q^2 < m_{a_1}^2$.
Note that it is  possible that factorization will be a poorer
approximation for decays with smaller energy release or larger $q^2$.
Factorization tests can be extended to higher $q^2$  using
$B\to D^{*} D_s^{(*)}$ 
decays as is discussed in section \ref{facapply} .

\subsubsection{Factorization and Angular Correlations}
\label{fac-ang-cor}

More subtle tests of the factorization hypothesis can be performed by examining
the polarization in $B$ (or $D$)
meson decays into two vector mesons\cite{Kg}.
Again, the underlying principle is to compare the hadronic decays to the
appropriate semileptonic decays evaluated at a fixed value in $q^2$.
For instance, the ratio of longitudinal to transverse polarization
($\Gamma_{L}/\Gamma_{T}$)
in $\bar{B^0} \to D^{*+} \rho^{-}$ should be equal to the corresponding ratio
for $B\to D^{*}\ell \nu$
 evaluated at $ q^2={m_\rho}^2=0.6~ \rm{GeV}^2$.
\begin{equation}
 {{\Gamma_{L}}\over{\Gamma_{T}}} ({\bar{B^0} \to D^{*+} \rho^{-}})
= {{\Gamma_{L}}\over{\Gamma_{T}}}
{(B\to D^*\ell\nu)|}_{q^2=m_{\rho}^2}
\label{Galgatpol}
\end{equation}
The advantage of this method is that it is not affected by QCD
corrections \cite{lepage}.

For $B \to D^*\ell\nu$ decay (or $D\to K^*\ell\nu$), 
longitudinal polarization dominates at low $q^2$, whereas near
$ q^2= q^2_{\rm max}$ transverse polarization dominates. There is a simple
physical argument for the behaviour of the form factors
near these two kinematic limits. Near $ q^2=q^2_{\rm max}$,
the $D^*$ is almost at rest and its small velocity is uncorrelated with the
$D^*$ spin, so all three $D^*$ helicities
are equally likely and we expect $\Gamma_T / \Gamma_L$ = 2.
At $q^2=0$, the $D^*$ has the maximum possible momentum, while the lepton and
neutrino are collinear and travel in the direction opposite to the $D^*$.
The lepton and neutrino helicities are aligned to give
$S_z= 0$, so near $q^2=0$ longitudinal polarization is dominant.

 Factorization breaks down in the charm sector as a result of the presence of 
FSI. From MARK III results\cite{Coffman} on the decay mode 
$D^0\to K^{\star0}\rho^-$, and the measured form factors for 
the semileptonic decay mode, we can evaluate Equation~(\ref{Galgatpol})
for this vector-vector decay mode,
\begin{eqnarray}
 {{\Gamma_{L}}\over{\Gamma_{T}}} ({D^0 \to K^{*+} \rho^{-}}) 
 = 0.90 \pm 0.65 \\ 
 {{\Gamma_{L}}\over{\Gamma_{T}}} {(D^0 \to K^*\ell\nu)|}_{q^2=m_{\rho}^2}
 = 0.78 \pm 0.07 \nonumber
\end{eqnarray}
This result supports the factorization hypothesis with large errors.
The observation of large transverse polarization and a D-wave 
component in the color suppressed decay $D^0\to K^{*0}\rho^0$ 
indicates the presence of large non-factorizable 
contributions\cite{Cheng_1},\cite{DeJongh}. In the future, polarization
in the vector-vector mode $D_s^+\to \phi\rho^+$ will also be measured.

For $\bar{B^0} \to D^{*+} \rho^-$,
we expect $88\%$ longitudinal polarization from the argument
described above \cite{Rosfac}. Similar results have been obtained by
Neubert\cite{neub}, Rieckert\cite{ricky}, and Kramer \etal \cite{Kramfac}.
Using the measured $q^2$ distribution for $B \to D^* \ell \nu$ 
Neubert \cite{neub} calculates the transverse and
longitudinal polarization in $B \to D^*\ell\nu$ decays.
Using his result we find $\Gamma_L /\Gamma$ to be
85\% at $q^2={m_\rho}^2=0.6$.
The agreement  between these predictions and the
experimental result (Sec.~\ref{pol-D*-rho})
\begin{equation}
\Gamma_L /\Gamma \; = \;  93 \pm 5 \pm 5 \%
\end{equation}
supports the factorization hypothesis in hadronic $B$ meson decay
for $q^2$ values up to $m_{\rho}^2$.

 The strength of FSI in $B$ decay can be determined 
by performing an isospin analysis of related decay
channels such as
$B^- \to D^0\pi^-$,
$\bar{B}^0 \to D^0\pi^0$, and
$\bar{B}^0 \to D^+\pi^-$ as was done
for the $D\to K \pi$ and $D \to K^* \pi$
systems. At the present level of experimental precision and in 
contrast to $D$ decay, there
is no evidence for non-zero isospin phase shifts in B decay.
From a maximum likelihood fit to the observed branching
fractions, Yamamoto found
that $\cos\delta^* > 0.82$ at the 90\% confidence level, 
where $\delta^*$ is the phase shift
 for the $B\to D \pi$ system
and comparable constraints, and
$\cos\delta^* > 0.57 (0.92)$, for the $B\to D^* \pi$ ($B\to D\rho$) isospin
multiplets\cite{hitoshi}. 
In $B$ (and $D$)
decays to two vector mesons, such as $B \to D^*\rho$, the  presence
of final state interaction could also be probed by studying the
angle $\chi$ between the $D^*$ and $\rho$ decay planes. FSI would cause
a phase shift between the helicity amplitudes and break 
the symmetry of the $\chi$ distribution. The
presence of FSI would lead to a angular distribution
proportional to $\sin\chi$ or $\sin 2\chi$\cite{hitoshichi}.

Until the $D_s$ decay constant, $f_{D_s}$, 
is measured more precisely, in $D_s \to
\mu\nu$, tests of the factorization hypothesis based on  
branching fractions 
cannot be applied to $B\to D^* D_s$ decays.
As data samples increase, it will become
possible to measure the polarization in $\bar{B}^0 \to D^{*+}D_s^{*-}$
decay modes and to investigate
whether factorization is still valid  at $q^2=m^2_{D_s}$.

\subsubsection{Applications of Factorization}
\label{facapply}

If factorization holds, hadronic $B$ decays can be used to extract information
about semileptonic
decays. For example, we can determine the poorly measured  
rate $B\to D^{**}(2420)~\ell~\nu$ from
the branching ratio of $B\to D^{**}(2420)\pi$.
By  assuming that
the rate for $B\to D^{**}(2420)\pi$ is related to
$d\Gamma/dq^2 (B \to D^{**}(2420) \ell \nu)$ evaluated at $q^2 = m_{\pi}^2$.
Using the model of Colangelo \etal \cite{Bari}
to determine the shape of 
the form factors we obtain the ratio
$$\frac{\Gamma(B \to D^{**}(2420) ~\ell ~\nu)}
{\Gamma(B \to D^{**}(2420)\pi)}= 3.2$$
Combining this 
with the experimental result,
${\cal {B}}(B^- \to D^{**0}(2420)\pi^- )\, = \, 0.16 \pm 0.05\, \%$,
(Table~\ref{kh3}) we predict
${\cal B} (D^{**}(2420) \ell \nu ) = 0.51 \pm 0.16 \%$. This
is not inconsistent with the average of 
recent direct measurements \cite{Review_1}
${\cal B} (D^{**}(2420) \ell \nu ) = 1.17 \pm 0.24\%$.

A second application of factorization is the determination of $f_{D_s}$
using the decays $B \to D^*D_s$.
The rate for $\bar{B^0}\to D^{*+}D_s$ is related
to the differential rate for
$\bar{B^0}\to D^{*+}\ell^-\nu$ at $q^2 = m_{D_s}^2$ if factorization continues
to be valid at larger values of $q^2$:
\begin{equation} {\Gamma\left(\bar{B^0}\to
D^{*+} D_{s}^{-}\right)\over\displaystyle{d\Gamma\over
\displaystyle dq^2}
\left(\bar{B^0}\to D^{*+}\ell^-\nu\right)\biggr|_{q^2=m^2_{D_s}}} =
6\pi^2 \delta ~{ c_1^2}
f_{D_s}^2|V_{cs}|^2 ,\label{Efacts}
\end{equation}
The factor $\delta = 0.37$ accounts for the different form factors
which enter in  $B \to D^* D_s$ and $B\to D^*\ell\nu$ \cite{Neubie}.

Using the value listed in Table \ref{TFactst}
for $d\Gamma/dq^2(B\to D^*\ell \nu)$
at $q^2\: = \: m_{D_s}^2$ and the average branching ratio
for ${\cal B}(B\to D^{*} D_{s}^-)=1.02 \pm 0.27 \%$, we obtain
$$
f_{D_s} = (277 \pm 77) \sqrt{3.6\%/{\cal{B}}(D_s \to \phi \pi^+)} ~\rm{MeV}
$$
and with ${\cal B}(B\to D^{*} D_{s}^{*-})=2.23 \pm 0.60 \%$, we find 
($\delta = 1$)
$$ 
f_{D_s^*} = (243 \pm 70) \sqrt{3.6\%/{\cal{B}}(D_s \to \phi \pi^+)} ~\rm{MeV}
$$
This result can be compared to the value
$$ f_{D_s} = (288\pm 30\pm 30 \pm 24) 
\sqrt{B(D_s \to \phi \pi^+)/3.6\%} ~\rm{MeV}$$
that was obtained from a
direct measurement of $D_s\to \mu \nu$ decays in continuum charm events
\cite{dsmnew}. 
Both values of $f_{D_s}$ are entirely consistent with 
theoretical predictions that are in the range $f_{D_s}=200-290$~MeV
\cite{fds_expect}.
If both the $D_s^+ \to \phi \pi^+$ branching fraction and $f_{D_s}$
are measured more precisely, then measurements of the
branching ratios of $B\to D^* D_s$ decays can be used
to test factorization in $B$ decay at $q^2 = m_{D_s}^{2}$. 
As noted earlier,
it will also be possible to test factorization in this $q^2$ range
by measuring $\Gamma_{L}/\Gamma$ in $B \to D^* D_{s}^*$ decays.

\subsubsection{Factorization in Color Suppressed Decay}
\label{fac-color}

It is not obvious whether the factorization hypothesis 
will be satisfied in decays that proceed via  internal W-emission
e.g $B\to \psi K^{(*)}$. Two observables have been compared
to phenomenological
models based on the factorization hypothesis: the ratio of vector
to pseudoscalar modes and the polarization in $B\to \psi K^*$ decays.

Using the results listed in Tables \ref{kh3} and \ref{kh4}
we can determine the ratio of vector to pseudoscalar meson production
\begin{equation}
{{\cal B}(B \to \psi K^*)\over{{\cal B}
(B \to \psi K)}} = 1.69 \pm 0.33
\nonumber
\end{equation}
Combined with the CDF measurement (Equation \ref{eq:cdfpsiratio}) we obtain
\begin{equation}
{{\cal B}(B \to \psi K^*)\over{{\cal B}
(B \to \psi K)}} = 1.47 \pm 0.21
\end{equation}

This quantity
can be calculated using factorization and the ratio of the
$B\to K^*$ and $B\to K$ form factors. 
The revised BSW model of Neubert \etal\cite{Neubie}
predicts a value of 1.61 for this ratio, which is close to the
experimental result. Another test is the corresponding ratio
for $\psi^{'}$ decays:
\begin{equation}
{{\cal B}(B \to \psi' K^*)\over{{\cal B}
(B \to \psi' K)}} = 2.1 \pm 1.5
\end{equation}
This measurement 
can be compared to the revised BSW model, which predicts 1.85
for this ratio. 
Gourdin \etal \cite{gkpeta}
argue, that the ratio 
${{\cal B}(B \to \eta_c  K^*)/{{\cal B}
(B \to \eta_c K)}} $ would provide a good test of the
factorization hypothesis in internal spectator decays. However,
it will require a significantly larger data sample than is
available at present before this ratio can be measured with sufficient
precision. Other ratios of decay rates in modes with charmonium mesons
may also be used to test factorization\cite{gkpother}.

The experimental results on $\psi K^*$ polarization
 can be compared to the theoretical predictions of
Kramer and Palmer\cite{Krampalm}
which depend on the assumption of factorization and on the
unmeasured $B\to K^*$ form factor. Using the BSW model to estimate the form
factors, they find $\Gamma_{L}/\Gamma= 0.57$. 
Using HQET to extrapolate
from the E691 measurements of the $D\to K^*$ form factor, they obtain
$\Gamma_{L}/\Gamma=0.73$. 
The group of Gourdin, Kamal and Pham as well as the collaboration of 
Aleksan, Le Yauoanc, Oliver, P\`ene, and Raynal 
have noted that there is no set of experimental
or theoretical form factors that can  simultaneously reproduce the
measured values of $\Gamma_{L}/\Gamma$ and ${\cal B}(B\to \psi K^*)/
{\cal B}(B\to \psi K)$ \cite{gkp},\cite{ayopr}. They conclude that
there is either a
fundamental problem in heavy to light form factors or a
breakdown of factorization for this class of decay modes.
Kamal and Santra have suggested that all the measured
observables in exclusive $B\to \psi$ can be accommodated with a
single non-factorizable amplitude\cite{kamalpsi}.

CLEO also finds evidence at the 2.5 standard deviation level
for $B\to \chi_{c2}$ transitions at a branching ratio
of $ 0.25\pm 0.10\pm 0.03\%$. If confirmed, this
would indicate the presence of either
non-factorizable color octet
contributions that are neglected in the usual treatment of
hadronic $B$ decays
or higher order processes
$O(\alpha_s^2)$ in $b\to c \bar{c} s$ decays\cite{bodwin}.

\subsection{Determination of the Color Suppressed Amplitude}
\label{eff-color-supp}

\subsubsection{Color Suppression in $B$ and $D$ Decay}

In the decays of charm mesons, the effect of color suppression
is obscured by the effects of FSI or reduced by nonfactorizable 
effects. The nonfactorizable contribution arises from the soft gluon
exchange between color currents\cite{Cheng_1}.
Table~\ref{Tcolsupd} gives ratios of several charm meson decay modes with 
approximately equal phase space factors where the mode in the numerator is 
color suppressed while the mode in the denominator is  an external spectator 
decay. With the exception of the 
decay $D^0\to \bar{K}^0\rho^0$ it is clear that 
the color suppressed decays do not have significantly smaller branching ratios.

 The data on charm decays supports the {\it new factorization} scheme
\cite{BGR}, that is $N_c \to \infty$ in Equation~\ref{a1a2c1c2}.
 This scheme gives values of $a_1\sim 1.25$ and $a_2\sim -0.49$ for 
nonleptonic charm decays. Assuming that the values of the coefficients can be
extrapolated from $\mu = m_{c}^2$ to $\mu = m_{b}^2$ taking into account the 
evolution of the strong coupling constant $\alpha_s$, we obtain the 
predictions $a_1\sim 1.12$ and $a_2\sim -0.27$ for $B$ decays.

The smaller magnitude of $a_2$ means that in contrast to the charm sector
one expects to find a more consistent pattern of color suppression in $B$ meson
decays.
In Section ~\ref{color-supress} we obtained upper limits for color suppressed
$B$ decays with a $D^0$ or $D^{*0}$ meson in the final state.
In Table~\ref{Tbrcolcomp} these results are compared to the predictions
of the BSW and the RI models\cite{ri}.

In contrast to charm decay, color suppression seems to be operative
in hadronic decays of $B$ mesons. The limits on the color suppressed
modes with $D^{0(*)}$ and neutral mesons are still above the level
expected by the two models, but we can already exclude a prediction
by Terasaki \cite{tera} that
${\cal{B}}(\bar{B^0} \to D^0 \pi^0) 
\approx 1.8 {\cal{B}}(\bar{B^0} \to D^+\pi^-)$.
To date, the only color suppressed $B$ meson decay modes
that have been observed are final states
that contain charmonium mesons e.g. $B\to \psi K$ and 
$B\to \psi K^*$\cite{psicomment}.

\subsubsection{Determination of $|a_1|$, $|a_2|$ and
the Relative Sign of ($a_2/a_1$)}
\label{a1-a2}

 We have determined the free parameters $a_1$ and $a_2$ of the BSW model 
for $D$ decays taking into account 
the isospin phase shifts due to FSI.
Using updated world averages of
the branching fraction\cite{bfd} for the decay
$D\rightarrow K\pi$, where no inelastic effects are expected, 
gives
\begin{eqnarray}
\label{a1a2d}
 a_{1} & = & +1.10 \pm 0.03 \\
 a_{2} & = & -0.50 \pm 0.03 \nonumber 
\end{eqnarray}
A comparison with the QCD Wilson coefficients (see Equation~\ref{c1c2})
shows that $a_{1} \simeq c_1$ and $a_{2} \simeq c_2$, that is $\xi \sim 0$. 
This result is anticipated in the $1/N_c$ expansion 
by Buras, Gerard, and R\"uckl\cite{BGR} and 
implies that quarks belonging to different color singlet currents do not 
easily combine to form a single meson\cite{BauerS}. 

 If instead we use the perturbative QCD result,
equation~(\ref{a1a2c1c2}) with $N_c = 3$, we obtain 
the following values of $a_{1}$ and $a_{2}$,
\begin{eqnarray}
 c-quark \quad : \quad a_{1} = +1.08 \qquad
                       a_{2} = -0.07 \\
 b-quark \quad : \quad a_{1} = +1.03 \qquad
                       a_{2} = +0.11 \nonumber
\end{eqnarray}
 The value of $a_{2}$ from the QCD calculation is inconsistent 
with the experimental results for hadronic charm decay. 
This discrepancy suggests that 
non-factorizable contributions and FSI play an important role. 
Nonperturbative soft gluon effects become more important in decays with 
smaller energy release, allowing more time for FSI. This may explain why 
$a_2$ is class dependent in charm decay, whereas it appears to be fairly 
stable in B decays\cite{Cheng_1}. 

 In a recent paper\cite{Kamal_4} Kamal and coauthors (see also 
Cheng\cite{Cheng_2}) have argued that $a_{1}$ and $a_{2}$ (in the factorized 
amplitude) should be replaced by an effective and unitarized 
parameters $a^{U,eff}_{1}$ and $a^{U,eff}_{2}$. These quantities receive 
contributions from annihilation and nonfactorizable processes as well as 
FSI. Since these effective parameters are process dependent, 
factorization tests (comparing hadronic to semileptonic rate) should be 
used as a tool to determine the moduli of these 
quantities\cite{Kamal_4}. In this way, much of the predictive 
power of the phenomenological models (based on factorization) is lost, 
because the $a_{1}$ and $a_{2}$ parameters are now dependent on the 
particular decay channel.

In the BSW model \cite{Neubie}, the branching
fractions of the $\bar{B}^0$ normalization 
modes are proportional to $a_1^2$ while
the branching fractions of the
$B\to\psi$ decay modes depend only on $a_2^2$. A fit to the
branching ratios for the $B$ decay modes
$\bar{B^0}\to D^+\pi^-$, $D^+\rho^-$, $D^{*+}\pi^-$ and $D^{*+}\rho^-$
using the model of Neubert \etal\ yields
\begin{equation}
|a_1| = 1.07 \pm 0.04  \pm  0.06
\label{normal_a1}
\end{equation}
and a fit to
the modes with $\psi$ mesons in the final state gives
\begin{equation}
|a_2| = 0.23 \pm 0.01 \pm 0.01
\label{psi_a2}
\end{equation}
The first error
on $|a_1|$ and $|a_2|$ includes the
uncertainties from the charm or charmonium branching ratios,
 the experimental systematics associated with detection
efficiencies and background subtractions as well as the statistical
errors from the branching ratios.
The second  error quoted is the uncertainty due to
the $B$ meson production fractions and lifetimes.
 We have assumed that the ratio of $B^+ B^-$ and $B^0 \bar{B^0}$ production
at the $\Upsilon(4S)$ is one \cite{dstlnu}, 
and assigned an uncertainty of 10\% to it. 

The magnitude of the amplitude for external spectator processes,
$|a_1|$ can also be determined from $B\to D^{(*)}D_s^{(*)}$ decays. 
Since these transitions are not subject to interference with the internal
spectator amplitude we can combine $B^-$ and $\bar{B}^0$ decays to reduce
the statistical error. Using the average branching fractions given in
Tables~\ref{kh3},~\ref{kh4} we obtain
\begin{equation}
|a_1|_{DD_s} = 0.98 \pm 0.06 \pm 0.04
\label{dds_a1}
\end{equation}
It is interesting to note that this value of $|a_1|$ agrees with the
result of the fit to the $B\to D^{(*)} \pi$ and $B\to D^{(*)}\rho$ modes
(see Equation~\ref{normal_a1}).
In general, $|a_1|$ could be different
for exclusive $b\to c \bar{u} d$ and $b\to c \bar{c} s$ processes.

By comparing branching ratios of $B^-$ and $\bar{B^0}$ decay modes it is
possible to determine the sign of $a_2$ relative to $a_1$.
The BSW model,~\cite{Neubie} predicts the following ratios:
\begin{equation}
R_1 = {{\cal B}(B^- \to D^0 \pi^-) \over {\cal B}(\bar{B^0}\to D^+ \pi^-)}
                = (1 + 1.23 a_2/a_1)^2  \label{colrate1}
\end{equation}
\begin{equation}
R_2 = {{\cal B}(B^- \to D^0 \rho^-)
\over {\cal B}(\bar{B^0} \to D^+ \rho^-)}
                = (1 + 0.66 a_2 /a_1)^2  \label{colrate2}
\end{equation}
\begin{equation}
R_3 = {{\cal B}(B^- \to D^{*0} \pi^-)
         \over {\cal B}(\bar{B^0} \to D^{*+} \pi^-)}
                     =(1 + 1.29 a_2/a_1)^2  \label{colrate3}
\end{equation}
\begin{equation}
R_4 = {{\cal B}(B^- \to D^{*0} \rho^-)
          \over{\cal B}(\bar{B^0} \to D^{*+} \rho^-)}
                     \approx (1 + 0.75 a_2/a_1)^2   \label{colrate4}
\end{equation}

Table~\ref{Tbswexpc} shows a comparison between the
experimental results and
the two allowed solutions in the BSW model.
The systematic errors due to
detection efficiencies partly cancel each other out.
In the ratios $R_3$ and $R_4$ the $D$ meson branching ratio uncertainties
do not contribute to the systematic error.

A least squares fit to the ratios $R_1$ - $R_3$ gives
\begin{equation}
a_2/a_1 = 0.26 \pm 0.07 \pm 0.05
\label{a2a1_ratio}
\end{equation}
where we have ignored uncertainties in the
theoretical predictions.
$R_4$ is not included in the fit since
the model prediction in this case is not thought to be reliable \cite{volkie}.
The second error is due to the uncertainty in
the $B$ meson production fractions and lifetimes
that enter into the determination of $a_1/a_2$ in the combination
$(f_+  \tau_{+}/ f_{0} \tau_{0})$. 
 As this ratio increases,
the value of $a_2/a_1$ decreases. 
The allowed range of $(f_+  \tau_{+}/ f_{0} \tau_{0})$ 
excludes a negative value of $a_2/a_1$.

Other uncertainties in the magnitude\cite{fdvari}
 of $f_D$, $f_{D^*}$ and in the hadronic form
factors can change the magnitude of $a_2/a_1$ but not its sign.
The numerical factors that
multiply $a_2/a_1$ include the ratios of $B \to \pi$($B\to\rho$)
to $B\to D$ ($B\to D^*$) form
factors, as well as the ratios of the meson decay constants. We
assume values of 220~MeV for $f_D$ and $f_{D^*}$ \cite{rosfd}.
To investigate the model dependence of the result we have recalculated
$|a_1|$, $|a_2|$, and $a_2/a_1$ in the model of Deandrea \etal\ We find
$|a_1| = 1.00 \pm 0.04 \pm 0.06$, 
$|a_2| = 0.24 \pm 0.01 \pm 0.01$, and
$a_2/a_1 = 0.25 \pm 0.07 \pm 0.05$, consistent with the results discussed
above. A different set of $B \to \pi$ form factors can be calculated using
QCD sum rules. By using the form factors determined by Belyaev, Khodjamirian 
and R\"uckl \cite{brueckl} and 
by Ball \cite{Ballff}, $a_2/a_1$ changes by 0.04. Kamal and Pham
have also considered the effect of uncertainties in form factors,
the effects of final state interactions, and annihilation terms. They
conclude that these may change the magnitude of $a_2/a_1$ but 
not its sign \cite{KPham}. Systematic uncertainties in the ratio 
of $D$ branching fractions could also modify its magnitude.

The magnitude of $a_2$ determined from this fit 
to the ratio of $B^-$ and $B^0$ modes is consistent with the value
of $a_2$ determined from the fit to the $B\to\psi$ decay modes.
The sign of $a_2$ disagrees with the theoretical
extrapolation from the fit to charm meson decays using
the BSW model\cite{oldfit}. It is also disagrees with the 
expectation from the $1/N_{c}$ rule\cite{BS},\cite{halperin}.
The result may be consistent with the expectation of perturbative QCD
\cite{Burasa1a2}.

\subsection{The Sign of a$_2$/a$_1$ and 
the Anomalous Semileptonic Branching Ratio}
\label{baffle}

The observation 
that the coefficients $a_1$ and $a_2$ have the same relative sign
in $B^+$ decay, came as a surprise 
since destructive interference was observed in hadronic charm
decay. 
Although constructive interference has been observed in all the
$B^+$ modes studied so far these comprise only a small fraction of the
total hadronic rate. 
If the constructive interference which is observed in $B^+$ decay
is present at the same level in the remainder of hadronic $B^+$ decays,
then we would expect a lifetime ratio $\tau_{B^+}/\tau_{B^0}\sim 0.83$
unless there is a large compensating 
contribution from W-exchange to $B^0$ decay \cite{hsw}. 
It is also possible that there is no interference
in the higher multiplicity $B$ decays that have not yet been 
reconstructed. It, therefore, is important  
to measure $a_1$ and $a_2$ for a large variety of decay modes.

It is intriguing that $a_1$ determined from $B\to D^{(*)}\pi , \,
D^{(*)} \rho$
modes agrees well with the value of $a_1$ extracted from 
$B \to DD_s$ decays. 
The observation of  
color suppressed decays such as $\bar{B}^0 \to D^0 \pi^0$
would give another measure of $|a_2|$ complementary to that obtained
from $B \to $ charmonium decays .

Keum \cite{keum-pol} has suggested that the relative sign of $a_1$ and $a_2$
could be determined from a measurement of the polarization in 
$B^- \to D^{*0} \rho^-$ decays. For $a_2/a_1 > 0 $ the amount of
longitudinal polarization should be less than 88\% while for $a_2/a_1<0$
the converse will hold.
At the present level of precision, both possibilities are consistent
with the data on polarization.

The experimentally measured semileptonic branching ratio is determined to be
$(10.35\pm 0.17 \pm 0.35)$\% in the model independent dilepton analysis
\cite{Review_1}.
Comparable but more precise rates are also obtained from the analysis of the
single lepton spectrum.
These measurements are significantly below the theoretical lower
bound  ${\cal{B}}_{sl}>12.5 $\% from  QCD calculations within the
parton model\cite{bbsl}.

It is possible to 
understand the origin of the theoretical limit in a simple way.
In the absence of QCD corrections, the virtual $W$ emitted by the
b quark can decay into either a lepton-antineutrino pair, a $\bar{u}-d$ quark
pair, or a $\bar{c}-s$ quark pair. For the decay into a quark pair,
there are three possible color states that are equally probable.
In addition, corrections must be made for the reduction in phase space
in the $W\to \tau \nu$ and $W\to \bar{c} s$ decays.
Then the semileptonic fraction, ${\cal B}_{SL}$ is given by
\begin{equation}
 {\cal B}_{SL} = {{f_c} \over {5 f_c + 3 f_{\bar{c} s} +  f_{c \tau}} }
\end{equation}
Using the phase space factors, $f_c=0.45$, $f_{\bar{c} s} \approx f_{c \tau}
=0.12$ gives ${\cal B}_{SL} = 16.5\%$. QCD corrections modifiy
the hadronic contributions to the width and give ${\cal B}_{SL} = 14.3\%$.
The theoretical lower limit of $12.5\%$
is obtained by varying the quark masses and
QCD scale to their lower limits.

Several explanations of this discrepancy, awaiting experimental
confirmation have been proposed.
\begin{enumerate}
\item There might be an increased $b\to c\bar{c}s$ component of the 
$B$ meson hadronic width 
\cite{bbsl}, \cite{palmstech},\cite{dunietz}. However, recent
experimental data rule out 
the mechanism suggested by Dunietz \etal \cite{dunietz}
 as a major contributor to $B \to$ baryon decays.
\item Higher order contributions might reduce the theoretical expectation
or the assumption of duality may not hold for b quark decay\cite{falk}.
The former
has been advocated by Bagan, Ball, Braun, and Gosdzinsky
who find results that are marginally consistent 
with experiment\cite{bagan1},\cite{bagan2} but
also predict $N_{c}=1.24\pm 0.06$ for the number of charm quarks
produced per $b$ decay again due to higher order QCD enhancements
of the $b\to c \bar{c} s$ channel\cite{bagan2}.
\item Constructive interference in $B^-$ decays would reduce the theoretical
expectation for the semileptonic branching ratio.
A small contribution from W exchange to $\bar{B}^0$ decays would
keep the lifetime ratio close to unity and satisfy the experimental
constraints on this quantity\cite{hsw}.
\item A combination of a larger $b\to c \bar{c} s$
component and a non-perturbative 
enhancement of $b\to c \bar{u} d$ as in (3)
could also suffice to explain the discrepancy.
\item There could also
be a large contribution to the inclusive rate that has not been measured.
 It has been suggested by Palmer and Stech\cite{palmstech},
that $b \to c \bar{c} s$ followed by $c \bar{c} \to \rm{gluons}$,
 which in turn hadronize into a final state with no charm, has a large
branching ratio. The charm content for this mechanism would not be properly
taken into account.
\item It is possible that the rate for the hadronic penguin
diagram $b\to sg$ is larger than expected\cite{kaganbsg}.
This possibility will lead to significant production of high
multiplicity charmless final states which are quite
difficult to distinguish experimentally.
\item In addition, there 
could be a  systematic experimental
flaw in the computation of the yield of charm quarks from $b$ decay.
\end{enumerate}
Increasing the $b\to c\bar{c}s$ component would increase the average number of
$c$ quarks produced per $b$ quark 
decay and would lead to another interesting problem:
The predicted number of charm quarks per $b$ decay would increase 
to 1.3 while the current experimental world average  
for this number is $1.15\pm 0.05$ (see section~\ref{charmpro}).
Moreover, ${\cal B}(b\to c\bar{c}s)=15.8\pm 2.8$ which is far below 30\%.
With the recent observation of $B\to D \bar{D} K X$ transitions, the
branching fraction for $(b\to c \bar{c} s)$ has increased from
$15.8\pm 2.8\%$ to 
$23.9 \pm 3.8\%$ which is now consistent with the QCD calculations
of Ball et al. However,
the new source of $b\to c \bar{c} s$ decays does not modify
the charm yield and was already included in the determination of $n_c$.
This suggests that the problem of reconciling the 
semileptonic branching fraction and $n_c$ has not yet
been completely resolved.

The experimental result for the charm yield per $B$ decay
$$1.15 \pm 0.05$$ is consistent with the naive expectation
that $1.15$ charm quarks are produced per $b$ decay. However,
it does not support a number of  proposals
that suggest that at least $1.3$ quarks should
be produced per $b$ decay. 
In some recent theoretical efforts, large charm quark yields 
are a consequence of enhancements of $b\to c \bar{c} s$ which
are required in order to explain
the discrepancy between 
theoretical calculations and experimental measurements of
the inclusive 
semileptonic rate, ${\cal B}(B \to X\ell \nu)$ \cite{falk_baffle}.
Other explanations of this discrepancy such as a non-perturbative 
enhancement of $b\to c\bar{u} d$ or $b\to s ~g$  do not require
a larger charm yield.

The data are not yet sufficiently precise 
to convincingly rule out the possibility of a larger
charm yield. In addition, there
are several possible systematic experimental
flaws in the computation of the yield of charm quarks.
The charm meson absolute branching
fractions can contribute a systematic uncertainty, although the 
errors from this source
 have been significantly reduced by the precise determinations
of ${\cal B}(D^0\to K^-\pi^+)$\cite{DKpi}
and ${\cal B}(D^+\to K^-\pi^+\pi^+)$. 
The effect of a small change in the branching fractions for charm
meson is demonstrated by the following example:
decreasing ${\cal B}(D^0\to K^-\pi^+)$
and ${\cal B}(D^+\to K^-\pi^+\pi^+)$ by 7\% increases the total charm
yield in $B$ decay to $125\pm 6$\% (CLEO II measurements). 
Note that the value for $n_c$ reported here is slightly higher
than the value given at the 1995 conferences due to the smaller
world average for the absolute branching fraction
${\cal B}(D^0\to K^-\pi^+)$ used in this review.
The absolute branching
fraction scales for the $D_s$ meson and $\Lambda_c$ baryons are
still quite uncertain. 
Since the inclusive branching ratios to
these particles are small, a substantial change to the branching ratio
scale would be required to significantly modify the charm yield.

A systematic study of inclusive hadronic $B$ decays
to mesons and baryons and a concerted effort
to obtain more precise measurements 
of charm meson absolute branching fractions
will be required to fully resolve this problem.

\section{~Conclusions}

 Significant progress in
the physics of $B$ and $D$
mesons has been made in the last several years.

Fixed target experiments with silicon vertex detectors
such as E691 and E687 led to precise measurements
of charm meson and baryon lifetimes. The observed hierarchy
of lifetimes can be compared to theoretical models and is used to assess
the size of non--spectator effects. The $D^+$/$D^0$ lifetime
difference is attributed to constructive interference in
$D^+$ decays. This conclusion is supported by the observation
of destructive interference in many exclusive $D^+$ decay modes.

There has also been rapid progress in the measurement of
lifetimes of b-flavored hadrons from the LEP experiments, SLD, and
CDF. These results now clearly show that to a good approximation
$\tau_{B^+}\approx \tau_{B^0}\approx \tau_{B_s}$ while the $\Lambda_b$ lifetime
is significantly shorter. The small value of $\tau(\Lambda_b)$ is unexpected
and cannot be easily accommodated in most theoretical frameworks given
the observed size of non-spectator effects in charm decay\cite{neubertlb},
\cite{rosnerlb}.

The fixed target experiments and CLEO have reported many new measurements
of hadronic charm decay modes. There are now sufficiently precise data
to isolate the effects of FSI, and to solve
for the isospin amplitudes and relative phases in a number of quasi two-body
decay modes. Unambiguous evidence for DCSD in $D^+$ decay has been observed.
There is no compelling evidence for W-exchange or W-annihilation
in charm meson decay from measurements of
either hadronic decays or from the lifetime hierarchy.
Charm baryon decay shows strong evidence for W-exchange contributions. In
this case, however, there is no helicity suppression. 
Comparison of the observed
rates for hadronic charm meson decays and models based on factorization
show a number of discrepancies in $D^0$, $D^+$, and $D_s$ decays. 
The most dramatic of these
are in exclusive
$D_s$ decays to final states with $\eta$ and $\eta^{'}$ mesons.
These discrepancies may 
indicate the breakdown of factorization in hadronic charm decay .

Results from CLEO~II have significantly changed
our understanding of hadronic B decay.  
A complete experimental picture of inclusive $B$ decay is now
emerging. A new contribution to $b\to c \bar{c} s$ decays has been 
observed. However, the problem of 
simultaneously accomodating the low
value of $n_c$ and the $B$ semileptonic branching fraction remains.
The data and measurements of exclusive hadronic branching fractions
are now of sufficient quality to perform non-trivial
tests of the factorization hypothesis including comparisons
of rates for $\bar{B}^0\to D^{*+} X^-$
(where $X^-=\pi^- ,\rho^-$, or $a_1^-$)
with rates for  $D^{*+} \ell ^- \bar{\nu}$
at $q^2=M_X^2$, as well as comparisons of  the polarizations in 
$\bar{B}^0\to D^{*+}\rho^-$ with 
$\bar{B}^0 \to D^{*+} \ell^-\bar{\nu}_\ell$. In all cases, the
factorization hypothesis is consistent with the data at the present
level of experimental precision and for $q^2 < m_{a_1}^2$ in contrast
to the situation in charm decay.
No evidence for FSI is observed in $B$ decay. Limits
on the strong interaction phase shift in $B\to D\pi$, $B\to D\rho$ have
been obtained.

Improved measurements of branching ratios of two-body
decays with a $\psi$ meson in the final state
have been reported from ARGUS, CDF, and CLEO~II. 
The decay $B \to \psi K^*$ is polarized with 
$\Gamma_L / \Gamma = (78 \pm 7)$ \%. Therefore, this mode
will be useful for measuring CP violation. However, it is difficult
to simultaneously accommodate these results on polarization 
and the ratio of ${\cal B}(B\to \psi K^*)/{\cal B}(B\to \psi K)$
branching fractions in models that assume factorization.

Color suppression
appears to operate in hadronic $B$ decays in contrast to charm decays.
There is no experimental evidence for 
color suppressed decays of neutral 
$B$ mesons to a charm meson and light neutral
hadron in the final state.
The most stringent limit, 
${\cal B}(\bar{B^0}\to D^0\pi^0) / {\cal B}(\bar{B^0}\to D^+\pi^-) < 0.07$
from CLEO~II, is
still above the level where these 
color suppressed $B$ decays are expected in most models. 
The observation of $B \to \psi$ modes
shows that color suppressed decays are present. 
The appearance of many  internal spectator decays
at levels comparable to external spectator decays in the charm sector
may be due to FSI. 

Using results on exclusive
$B \to \psi$ decays from CLEO~1.5, CLEO~II and ARGUS, 
 we obtain values of the BSW parameter
$|a_2|\; = \; 0.23 \pm 0.01 \pm  0.01$. We also report a new value for
the BSW parameter
$|a_1|\; = \; 1.03 \pm 0.04 \pm 0.06$.
By comparing the rates for $B^-$ and $\bar{B}^0$ modes, it is
 been shown that the sign of
$a_2/a_1$ is positive, in dramatic
contrast to what is found in charm decays. It is difficult to reconcile
this result with the near equality of the 
$B^+/B^0$ meson lifetimes  unless
the pattern is significantly different for higher multiplicity decay modes
or there is a large W-exchange contribution to $B^0$ decay.

 In the next few years the samples of reconstructed charm particles 
should increase by a factor of 10 as E791 complete their data analysis 
and as FOCUS, the upgrade of the E687 experiment, 
SELEX and CLEO III begin taking data. 
These large charm samples will allow for more sensitive
searches for $D^0-\bar{D^0}$ mixing, rare decays,
 and CP violation, and for
a systematic investigation of charm baryons and their lifetimes.

Large samples of reconstructed hadronic $B$ decays will be obtained
in the next few years by CLEO~II/CLEO~III as a result
of further improvements in the luminosity of CESR, and upgrades of
the detector. There will also
be significant increases in the size of data samples available from the
CDF experiment.
Accurate tests of the factorization hypothesis
over the full $q^2$ range will become feasible. 
The large tagged sample at CLEO can be used
to study inclusive properties of $B^+$ and $B^0$ decays.
Measurements of additional decays to final states with charmonium 
mesons will be performed and 
other color suppressed decays will be observed.

The ultimate goal of the study of heavy flavor mesons is to 
 measure the large CP asymmetries 
predicted by the Standard Model in decay modes such as
$\bar{B}^0\to \psi K^0$, $\bar{B}\to \pi^+ \pi^-$ and $B^-\to D^0 K^-$. 
In order to throughly test
the consistency of the Standard Model's description of
CP violation in these decays at future facilities, 
the mechanisms that operate in hadronic decays of heavy quarks 
must be well understood. This review shows that
rapid progress is being made in this program.

\acknowledgements
We thank our colleagues from the CLEO, E687, E691, E791,
ARGUS, CDF, SLD, ALEPH, OPAL, DELPHI and L3 experiments for their contributions
to the work discussed in this review.
We thank  the US Department of Energy, the Italian Istituto Nazionale di
Fisica Nucleare, the University of Hawaii and The Ohio State University for 
their unwavering support.

%
%
\tighten

%
%
\clearpage
\begin{table}[htb]
\caption{Charm event samples of $e^+e^-$ colliding beam experiments.}
\label{charmepem}
\begin{tabular}{lll}
 \qquad\qquad Experiment \qquad\qquad & $\sqrt{s}$ & Number of
charm events produced\\
\hline
\qquad\qquad Mark III \qquad & $3.77$ GeV & $28000\quad D^0\overline{D^0}$ 
\qquad \\
\qquad\qquad &            & $20000\quad D^+ D^-$ \qquad \\
\qquad\qquad & $4.14$ GeV & $3000\quad D_s\overline{D_s}$ \qquad \\
\qquad\qquad BES \qquad      & $4.03$ GeV & $6000\quad D_s\overline{D_s}$ 
\qquad \\
\qquad\qquad CLEO II \qquad & $\sim 10.5$ GeV & $4 \times 10^6 \quad 
c\overline{c}$ \qquad \\
\qquad\qquad ARGUS \qquad & $\sim 10.5$ GeV & $0.7 \times 10^6 \quad 
c\overline{c}$ \qquad \\
\qquad\qquad LEP \qquad & $91$ GeV & $220000 \quad c\overline{c}$ 
per experiment \\
\qquad\qquad SLD \qquad & $91$ GeV & $14000 \quad c\overline{c}$ \quad \\ 
\end{tabular}
\end{table}
\begin{table}[htb]
\caption{Fully reconstructed charm samples of fixed-target experiment}
\label{charmfix}
\begin{tabular}{lll}
 \qquad\qquad Experiment \qquad\qquad & Reaction & Fully reconstructed
Charm Decays\\
\hline
\qquad\qquad E691 \qquad  & $\gamma$ Be $170$ GeV & $10000$  \qquad \\
\qquad\qquad E687 \qquad  & $\gamma$ Be $220$ GeV & $100000$ \qquad \\
\qquad\qquad WA75 \qquad  & $\pi^-$  N  $350$ GeV & $350$    \qquad \\
\qquad\qquad NA32(ACCMOR) & $K^-$ and $\pi^-$  N  $200$ GeV & $1300$ \\
\qquad\qquad WA82 \qquad  & $\pi^-$  N  $340$ GeV & $3000$    \qquad \\
\qquad\qquad E653 \qquad  & $\pi^-$  N  $600$ GeV & $1000$    \qquad \\
\qquad\qquad E769 \qquad  & $\pi^-$  N  $250$ GeV & $4000$    \qquad \\
\qquad\qquad E791 \qquad  & $\pi^-$  N  $500$ GeV & $200000$  \qquad \\
\qquad\qquad EXCHARM      & $n$\quad N  $40$  GeV &           \qquad \\
\qquad\qquad WA89 \qquad  & $\Sigma^-$ N  $330$ GeV &          \qquad \\
\end{tabular}
\end{table}
\begin{table}[htb]
\let\tabbodyfont\scriptsize
\caption{Measurements of exclusive lifetimes for b flavored hadrons.}
\medskip
\label{Tblife}
\begin{tabular}{lllllll}
Particle & Method &CDF  & ALEPH & OPAL & DELPHI &SLD \\  \hline \medskip
$ \bar{B}^0$& $D^{*+}l^-$ & $1.57\pm0.08\pm0.07$& $1.61\pm 0.07 \pm 
0.05$& $1.53\pm 0.12\pm0.08$& $1.61^{+0.14}_{-0.13}\pm 0.08$&
$1.60^{+0.15}_{-0.14}\pm 0.10$    \\ 
$ \bar{B}^0$& excl. &$1.64\pm0.11\pm0.06$& $1.25^{+0.15}_{-0.13}\pm 0.05$
      &        &  & \\ 
$ \bar{B}^0$& topol. & & & & $1.63\pm 0.14\pm 0.13$ & $1.55\pm 0.07\pm 0.12$\\ 
$ \bar{B}^0$& $\pi\pi$ & &$1.49^{+0.17+0.08}_{-0.15-0.06}$
 & &  &\\ \hline\medskip 
$ B^-$& $D^0l^-$ & $1.51\pm 0.12\pm 0.08$ & $1.58\pm0.09\pm0.04$
 & $1.52\pm0.14\pm0.09$ & $1.61\pm 0.16\pm 0.12$ &
$1.49^{+0.11}_{-0.11}\pm 0.05$ \\
$ B^-$&  excl. & $1.68\pm 0.09\pm 0.06$ & 
$1.58^{+0.21}_{-0.18}\pm 0.04$ &   & & \\ 
$ B^-$&  topol. & &  &   & $1.72\pm0.08\pm0.06$& 
$1.67\pm 0.06\pm 0.09$\\ \hline \medskip
$ B_s^0$&$D_s -l$&$1.42^{+0.27}_{-0.23}\pm 0.11$& 
$1.64^{+0.16}_{-0.14}\pm0.04$& 
$1.54^{+0.25}_{-0.21}\pm 0.06$& $1.54^{+0.31}_{-0.27}\pm 0.06$ & \\ 
$ B_s^0$&    $D_s-h$ &      & $1.61^{+0.30+0.29}_{-0.18-0.16}$&
  & $1.57^{+0.45+0.15}_{-0.37-0.14}$  & \\ 
$ B_s^0$&    $D_s$ incl. &      &  &
  & $1.61^{+0.34+0.18}_{-0.29-0.13}$  & \\ 
$ B_s^0$&    $\phi-l$ &      &  &
  & $1.45^{+0.20+0.32}_{-0.23-0.16}$  & \\ 
$ B_s^0$&$\psi\phi$&$1.74^{+1.08}_{-0.69}\pm 0.07$ &&   & & \\ \hline\medskip 
$ \Lambda_b$  & $\Lambda - l$&      & $1.21\pm 0.09\pm 0.07$& $1.16
\pm 0.11 \pm 0.06$ &  $1.10^{+0.16+0.05}_{-0.14-0.08}$  &  \\
$ \Lambda_b$  & $\Lambda_c -l$& $1.33\pm 0.16\pm 0.07$     
 & $1.24^{+0.15}_{-0.14}\pm 0.05$& 
   $1.14^{+0.22}_{-0.19}\pm 0.07$  &  $1.26^{+0.26+0.03}_{-0.22-0.05}$   & \\ 
$ \Lambda_b$  & $p-l$ &      &                               &
                     &  $1.27^{+0.35+0.08}_{-0.29-0.09}$   & \\ \hline\medskip
$ \Xi_b^0$  & $\Xi-l$ &      &      $1.25^{+0.55}_{-0.35}\pm 0.20$
                         &
                     &  $1.5^{+0.7}_{-0.4}\pm0.3$   & \\ 
\end{tabular}
\let\tabbodyfont\small
\end{table}
\begin{table}[htb]
\let\tabbodyfont\scriptsize
\caption{Measurements of the $B^+/B^0$ lifetime ratio.}
\medskip
\label{Tbratio}
\begin{tabular}{clllll}
 Method &CDF  & ALEPH & OPAL & DELPHI&SLD  \\ \hline \medskip
 $D-l$& $0.96\pm0.10\pm0.05$ & $0.98\pm 0.08 \pm
0.02$ & $0.99\pm0.14^{+0.05}_{-0.04}$ & $1.00^{+0.17}_{-0.15}\pm 0.10$&
$0.94^{+0.14}_{-0.12}\pm 0.07$    \\ 
 $excl$& $1.02\pm 0.09\pm 0.15$& $1.27^{+0.23}_{-0.19}
\pm 0.03$& & & $1.08^{+0.09}_{-0.08}\pm 0.10$   \\
  topol. & & & &  $1.06^{+0.13}_{-0.11}\pm 0.10$ \\ \hline
  $\rm{B~tags}$  & & & & &   $0.93\pm 0.18\pm 0.12$\\
  (CLEO II)  & & & & &   \\
\end{tabular}
\let\tabbodyfont\small
\end{table}
\begin{table}
\caption{Measured ratios of Double Cabibbo suppressed to Cabibbo favoured 
$D$ decays}
\label{Doublyd}
\begin{tabular}{ll}
 \qquad\qquad Ratio \qquad\qquad\qquad & 
Branching ratio(\%) \\
\hline
\qquad\qquad $\Gamma(D^+\rightarrow K^+\pi^+\pi^-)/
\Gamma(D^+\rightarrow K^-\pi^+\pi^+)$ \qquad & 
$ 0.0072 \pm 0.0023 \pm 0.0017 $ \qquad \\
\qquad\qquad $\Gamma(D^0\rightarrow K^+\pi^-)/\Gamma(D^0\rightarrow K^-\pi^+)$
\qquad\qquad & 
$ 0.0077 \pm 0.0025 \pm 0.0025 $ \qquad \\
\qquad\qquad $\Gamma(D^0\rightarrow K^+\pi^+\pi^-\pi^-)/
\Gamma(D^0\rightarrow K^-\pi^+\pi^+\pi^-)$ \qquad & 
\quad $\le 0.018 \quad (CL=90\%)$ \qquad \\
\qquad\qquad $\Gamma(D^0\rightarrow K^+\pi^-\pi^0)/
\Gamma(D^0\rightarrow K^-\pi^+)$ \qquad & 
\quad $\le 0.0068 \quad (CL=90\%)$ \qquad \\
\end{tabular}
\end{table}
\begin{table}
\caption{Decay modes of $\Lambda_c$ which can occur through the W-exchange 
diagram}
\label{Lambdacwex}
\begin{tabular}{ll}
 \qquad\qquad Decay Mode \qquad\qquad & 
Branching fraction(\%)\cite{PDG} \\
\hline
\qquad\qquad $\Lambda_c\rightarrow \Delta^{++} K^-$ \qquad & 
$ 0.7 \pm 0.4 $  \qquad \\
\qquad\qquad $\Lambda_c\rightarrow \Sigma^+ \phi$ \qquad & 
$ 0.30 \pm 0.13 $ \qquad \\
\qquad\qquad $\Lambda_c\rightarrow \Xi^0 K^+$ \qquad & 
$ 0.34 \pm 0.09 $ \qquad \\
\qquad\qquad $\Lambda_c\rightarrow \Xi^{\star0} K^+$ \qquad & 
$ 0.23 \pm 0.09 $ \qquad \\
\end{tabular}
\end{table}
\begin{table}[htb]
\let\tabbodyfont\scriptsize
\caption{Multiplicities and branching fractions of light mesons in $B$ meson decay.}
\label{Tbmulti}
\begin{tabular}{lll}
Mode & CLEO 1.5 \cite{CLEOK} & ARGUS \cite{ARGUSK} \\ 
     & (Branching Ratio) & (Multiplicity) \\ \hline
$ B/\bar{B}\to \pi^{\pm} $ &      & $ 3.59\pm 0.03\pm0.07$   \\
(not from $K_s,\Lambda$) & & \\
$ B/\bar{B}\to \pi^{\pm} $ &      & $ 4.11\pm 0.03\pm0.08$   \\
(incl. $K_s,\Lambda$) & & \\
$ B/\bar{B}\to K^{\pm} $ & $ 0.85\pm 0.07\pm 0.09$ & $0.78\pm 0.02\pm 0.03$ \\
$ \bar{B}\to K^{-} $  & $ 0.66\pm 0.05\pm 0.07$ &  \\
$ \bar{B}\to K^{+} $  & $ 0.19\pm 0.05\pm 0.02$ &  \\
$ B/\bar{B}\to K^0/\bar{K}^0 $ & $ 0.63 \pm 0.06\pm0.06$ & $0.64\pm 0.01 \pm 0.04$   \\
$ B/\bar{B}\to K^{*0} $    &    & $0.146\pm 0.016\pm 0.020$ \\
$ B/\bar{B}\to K^{*+} $    &    & $0.182\pm 0.054\pm 0.024$  \\
$ B/\bar{B}\to \rho^0 $    &    & $0.209\pm 0.042 \pm 0.033$ \\
$ B/\bar{B}\to \omega $    &    & $< 0.41$ (90\% C.L.)       \\
$ B/\bar{B}\to f_0(975) $  &    & $<0.025$ (90\% C.L.) \\
$ B/\bar{B}\to \eta   $  & $0.176\pm 0.011 \pm 0.0124$ (CLEO II) &  \\
$ B/\bar{B}\to \eta ' $  &    & $<0.15$ (90\% C.L.) \\
$ B/\bar{B}\to \phi $ & $ 0.023 \pm 0.006  \pm 0.005 $ 
& $0.039\pm 0.003\pm 0.004 $ \\
\end{tabular}
\let\tabbodyfont\small
\end{table}
\begin{table}[htb] 
\let\tabbodyfont\scriptsize 
\caption{Branching fractions [\%] of inclusive $B$ decays }
\label{khinc} 
 \begin{tabular}{l|lll|l} 
\multicolumn{1}{l}{Particle} & 
\multicolumn{1}{l}{ARGUS} & 
\multicolumn{1}{l}{CLEO 1.5} & 
\multicolumn{1}{l}{CLEO} & 
\multicolumn{1}{l}{Average} \\  
\hline 
 $\bar{B} \rightarrow \bar{D}^0 X$ & $ 51.6 \pm 4.0  \pm 6.6  \pm 2.1  $ & 
$ 61.9 \pm 3.4  \pm 3.7  \pm 2.5  $ & 
$ 67.1 \pm 2.2  \pm 1.5  \pm 2.7  $ & 
$ 64.8 \pm 2.2  \pm 2.6  $ \\ 
 $\bar{B} \rightarrow D^- X$ & $ 23.5 \pm 3.0  \pm 4.5  \pm 1.8  $ & 
$ 25.5 \pm 3.4  \pm 2.0  \pm 2.0  $ & 
$ 24.0 \pm 1.2  \pm 0.8  \pm 1.9  $ & 
$ 24.2 \pm 1.3  \pm 1.9  $ \\ 
 $\bar{B} \rightarrow D^{*-} X$ & $ 27.7 \pm 2.3  \pm 4.7  \pm 1.1  $ & 
$ 23.6 \pm 1.3  \pm 2.3  \pm 0.9  $ & 
$ 25.3 \pm 1.7  \pm 1.0  \pm 1.0  $ & 
$ 24.9 \pm 1.5  \pm 1.0  $ \\ 
 $\bar{B} \rightarrow D^{*0} X$ &  & 
 & 
$ 28.1 \pm 1.5  \pm 1.9  \pm 1.1  $ & 
$ 28.1 \pm 2.4  \pm 1.1  $ \\ 
 $\bar{B} \rightarrow D_s^- X$ & $ 8.1 \pm 1.1  \pm 0.9  \pm 2.0  $ & 
$ 8.5 \pm 1.3  \pm 2.1  $ & 
$ 11.8 \pm 0.4  \pm 0.9  \pm 2.9  $ & 
$ 10.1 \pm 0.7  \pm 2.5  $ \\ 
 $\bar{B} \rightarrow \phi X$ & $ 3.9 \pm 0.3  \pm 0.4  $ & 
$ 2.3 \pm 0.6  \pm 0.5  $ & 
$ 3.7 \pm 0.1  \pm 0.3  $ & 
$ 3.6 \pm 0.3  $ \\ 
 $\bar{B} \rightarrow \psi X$ & $ 1.25 \pm 0.19  \pm 0.26  $ & 
$ 1.31 \pm 0.12  \pm 0.27  $ & 
$ 1.12 \pm 0.04  \pm 0.06  $ & 
$ 1.14 \pm 0.07  $ \\ 
 $\bar{B} \rightarrow \psi X$ (direct) & $ 0.95 \pm 0.27  $ & 
 & 
$ 0.81 \pm 0.08  $ & 
$ 0.82 \pm 0.08  $ \\ 
 $\bar{B} \rightarrow \psi$'$ X$ & $ 0.50 \pm 0.18  \pm 0.12  $ & 
$ 0.36 \pm 0.09  \pm 0.13  $ & 
$ 0.34 \pm 0.04  \pm 0.03  $ & 
$ 0.35 \pm 0.05  $ \\ 
 $\bar{B} \rightarrow \chi_{c1} X$ & $ 1.23 \pm 0.41  \pm 0.29  $ & 
 & 
$ 0.40 \pm 0.06  \pm 0.04  $ & 
$ 0.42 \pm 0.07  $ \\ 
 $\bar{B} \rightarrow \chi_{c1} X$ (direct) &  & 
 & 
$ 0.37 \pm 0.07  $ & 
$ 0.37 \pm 0.07  $ \\ 
 $\bar{B} \rightarrow \chi_{c2} X$ &  & 
 & 
$ 0.25 \pm 0.10  \pm 0.03  $ & 
$ 0.25 \pm 0.10  $ \\ 
 $\bar{B} \rightarrow \eta_{c} X$ &  & 
 & 
$ <0.90 $  (90\% C.L.) & 
 $ <0.90 $  (90\% C.L.) \\ 
  $\bar{B} \rightarrow p X$ & $ 8.2 \pm 0.5  \pm 1.2  $ & 
$ 8.0 \pm 0.5  \pm 0.3  $ & 
 & 
$ 8.0 \pm 0.5  $ \\ 
 $\bar{B} \rightarrow \bar{\Lambda} X$ & $ 4.2 \pm 0.5  \pm 0.6  $ & 
$ 3.8 \pm 0.4  \pm 0.6  $ & 
 & 
$ 4.0 \pm 0.5  $ \\ 
 $\bar{B} \rightarrow \Xi ^+ X$ & $ <0.51 $ (90\% C.L.)& 
 $ 0.27 \pm 0.05  \pm 0.04  $ & 
 & 
$ 0.27 \pm 0.06  $ \\ 
 $\bar{B} \rightarrow \Lambda _c^- X$ & $ 6.8 \pm 2.7  \pm 1.4  \pm 0.9  $ & 
$ 6.1 \pm 1.1  \pm 0.9  \pm 0.8  $ & 
 & 
$ 6.3 \pm 1.3  \pm 0.9  $ \\ 
 $\bar{B} \rightarrow \Sigma_c^0 X$ &  & 
 & 
$ 0.53 \pm 0.19  \pm 0.16  \pm 0.16  $ & 
$ 0.53 \pm 0.25  \pm 0.07  $ \\ 
 $\bar{B} \rightarrow \Sigma_c^0 \bar{N}$ &  & 
 & 
$ <0.17 $  (90\% C.L.) & 
 $ <0.17 $  (90\% C.L.) \\ 
  $\bar{B} \rightarrow \Sigma_c^{++} X$ &  & 
 & 
$ 0.50 \pm 0.18  \pm 0.15  \pm 0.15  $ & 
$ 0.50 \pm 0.23  \pm 0.07  $ \\ 
 $\bar{B} \rightarrow \Sigma_c^{++} \bar{\Delta}^{--}$ &  & 
 & 
$ <0.12 $  (90\% C.L.) & 
 $ <0.12 $  (90\% C.L.) \\ 
  $\bar{B} \rightarrow \Xi_c^+ X$ &  & 
 & 
$ 1.5 \pm 0.7  $ & 
$ 1.5 \pm 0.7  $ \\ 
 $\bar{B} \rightarrow \Xi_c^0 X$ &  & 
 & 
$ 2.4 \pm 1.3  $ & 
$ 2.4 \pm 1.3  $ \\ 
\end{tabular} 
\let\tabbodyfont\small 
\end{table} 
\begin{table}[htb] 
\caption{World average $B^-$ branching fractions [\%]} 
\label{kh3} 
\begin{tabular}{ll} 
Mode & Branching Fraction \\ 
\hline 
$B^- \rightarrow D^0 \pi ^-$ & $0.50 \pm 0.05 \pm 0.02 $ \\ 
$B^- \rightarrow D^0 \rho ^-$ & $1.37 \pm 0.18 \pm 0.05 $ \\ 
$B^- \rightarrow D^{0} \pi ^+ \pi ^- \pi ^-$ & $1.28 \pm 0.35 \pm 0.05 $ \\ 
$B^- \rightarrow D^{*0} \pi ^-$ & $0.52 \pm 0.08 \pm 0.02 $ \\ 
$B^- \rightarrow D^{*0} \rho ^-$ & $1.51 \pm 0.30 \pm 0.06 $ \\ 
$B^- \rightarrow D_J^{(*)0} \pi ^-$ & $0.13 \pm 0.05 \pm 0.01 $ \\ 
$B^- \rightarrow D^{*+} \pi ^- \pi ^- \pi ^0$ & $1.69 \pm 0.76 \pm 0.07 $ \\ 
$B^- \rightarrow D_J^{(*)0} \rho ^-$ & $0.33 \pm 0.21 \pm 0.01 $ \\ 
$B^- \rightarrow D^{*0} \pi ^- \pi ^- \pi ^+$ & $0.95 \pm 0.27 \pm 0.04 $ \\ 
$B^- \rightarrow D^{*0} a_1 ^-$ & $1.89 \pm 0.53 \pm 0.08 $ \\ 
$B^- \rightarrow D^+ \pi^- \pi ^- $ & $<0.14 $  (90\% C.L.)\\ 
$B^- \rightarrow D^{*+} \pi ^- \pi ^-$ & $0.20 \pm 0.07 \pm 0.01 $ \\ 
$B^- \rightarrow D^{**0}(2420) \pi^- $ & $0.16 \pm 0.05 \pm 0.01 $ \\ 
$B^- \rightarrow D^{**0}(2420) \rho^- $ & $<0.14 $  (90\% C.L.)\\ 
$B^- \rightarrow D^{**0}(2460) \pi^- $ & $<0.13 $  (90\% C.L.)\\ 
$B^- \rightarrow D^{**0}(2460) \rho^- $ & $<0.47 $  (90\% C.L.)\\ 
$B^- \rightarrow D^0 D_s^-$ & $1.36 \pm 0.28 \pm 0.33 $ \\ 
$B^- \rightarrow D^0 D_s^{*-}$ & $0.94 \pm 0.31 \pm 0.23 $ \\ 
$B^- \rightarrow D^{*0} D_s^-$ & $1.18 \pm 0.36 \pm 0.29 $ \\ 
$B^- \rightarrow D^{*0} D_s^{*-}$ & $2.70 \pm 0.81 \pm 0.66 $ \\ 
$B^- \rightarrow \psi K^-$ &$ 0.102 \pm 0.014 $ \\ 
$B^- \rightarrow \psi ' K^-$ &$ 0.070 \pm 0.024 $ \\ 
$B^- \rightarrow \psi K^{*-}$ &$ 0.174 \pm 0.047 $ \\ 
$B^- \rightarrow \psi ' K^{*-}$ & $<0.30 $  (90\% C.L.)\\ 
$B^- \rightarrow \psi K^- \pi ^+ \pi ^-$ &$ 0.140 \pm 0.077 $ \\ 
$B^- \rightarrow \psi ' K^- \pi ^+ \pi ^-$ &$ 0.207 \pm 0.127 $ \\ 
$B^- \rightarrow \chi_{c1} K^-$ &$ 0.104 \pm 0.040 $ \\ 
$B^- \rightarrow \chi_{c1} K^{*-}$ & $<0.21 $  (90\% C.L.)\\ 
$B^- \rightarrow \psi \pi ^-$ &$ 0.0057 \pm 0.0026 $ \\ 
$B^- \rightarrow \psi \rho ^-$ & $<0.077 $  (90\% C.L.)\\ 
$B^- \rightarrow \psi a_1 ^-$ & $<0.120 $  (90\% C.L.)\\ 
\end{tabular} 
\end{table} 
\begin{table}[htb] 
\caption{World average $\bar{B}^0$ branching fractions [\%]} 
\label{kh4} 
 \begin{tabular}{ll} 
Mode & Branching Fraction \\ 
\hline 
$\bar{B}^0 \rightarrow D^+ \pi ^-$ & $0.31 \pm 0.04 \pm 0.02 $ \\ 
$\bar{B}^0 \rightarrow D^+ \rho ^-$ & $0.84 \pm 0.16 \pm 0.07 $ \\ 
$\bar{B}^0 \rightarrow D^+ \pi ^- \pi ^- \pi ^+$ & $0.83 \pm 0.24 \pm 0.07 $ \\ 
$\bar{B}^0 \rightarrow D^{*+} \pi ^-$ & $0.28 \pm 0.04 \pm 0.01 $ \\ 
$\bar{B}^0 \rightarrow D^{*+} \rho ^-$ & $0.73 \pm 0.15 \pm 0.03 $ \\ 
$\bar{B}^0 \rightarrow D^{*+} \pi ^- \pi ^- \pi ^+$ & $0.80 \pm 0.14 \pm 0.03 $ \\ 
$\bar{B}^0 \rightarrow D^{*+} a_1^-$ & $1.27 \pm 0.30 \pm 0.05 $ \\ 
$\bar{B}^0 \rightarrow D^{0} \pi ^+ \pi^- $ & $<0.17 $  (90\% C.L.)\\ 
$\bar{B}^0 \rightarrow D^{**+}(2460) \pi^- $ & $<0.22 $  (90\% C.L.)\\ 
$\bar{B}^0 \rightarrow D^{**+}(2460) \rho^- $ & $<0.49 $  (90\% C.L.)\\ 
$\bar{B}^0 \rightarrow D^+ D_s^-$ & $0.74 \pm 0.22 \pm 0.18 $ \\ 
$\bar{B}^0 \rightarrow D^+ D_s^{*-}$ & $1.14 \pm 0.42 \pm 0.28 $ \\ 
$\bar{B}^0 \rightarrow D^{*+} D_s^-$ & $0.94 \pm 0.24 \pm 0.23 $ \\ 
$\bar{B}^0 \rightarrow D^{*+} D_s^{*-}$ & $2.00 \pm 0.54 \pm 0.49 $ \\ 
$\bar{B}^0 \rightarrow \psi K^0$ &$ 0.075 \pm 0.021 $ \\ 
$\bar{B}^0 \rightarrow \psi ' K^0$ & $<0.08 $  (90\% C.L.)\\ 
$\bar{B}^0 \rightarrow \psi \bar{K}^{*0}$ &$ 0.153 \pm 0.028 $ \\ 
$\bar{B}^0 \rightarrow \psi ' \bar{K}^{*0}$ &$ 0.151 \pm 0.091 $ \\ 
$\bar{B}^0 \rightarrow \psi K^{-} \pi ^+$ &$ 0.117 \pm 0.058 $ \\ 
$\bar{B}^0 \rightarrow \psi ' K^- \pi ^+$ & $<0.11 $  (90\% C.L.)\\ 
$\bar{B}^0 \rightarrow \chi_{c1} K^0$ & $<0.27 $  (90\% C.L.)\\ 
$\bar{B}^0 \rightarrow \chi_{c1} \bar{K}^{*0}$ & $<0.21 $  (90\% C.L.)\\ 
$\bar{B}^0 \rightarrow \psi \pi ^0$ & $<0.006 $  (90\% C.L.)\\ 
$\bar{B}^0 \rightarrow \psi \rho ^0$ & $<0.025 $  (90\% C.L.)\\ 
$\bar{B}^0 \rightarrow \psi \omega ^0$ & $<0.027 $  (90\% C.L.)\\ 
\end{tabular} 
\end{table} 
\begin{table}[htb]
\caption{Longitudinal polarization of $\psi$ mesons from $B \to \psi K^*$
decays.}
\label{Tpsipolex}
\begin{tabular}{cc}
Experiment & ${\left({\Gamma_L\over{\Gamma}}\right)}$\\ \hline
CLEO II & $ 0.80\pm 0.08 \pm 0.05$ \\
ARGUS \cite{argpol} & $0.97 \pm 0.16\pm 0.15$ \\
CDF \cite{cdfpolar} & $ 0.65 \pm 0.10\pm 0.04$\\
\hline
Average & $ 0.78 \pm 0.07$\\
\end{tabular}
\end{table}

\clearpage
\begin{table}
\caption{Isospin amplitudes and phase shifts for hadronic $D$ decay modes, 
calculated using the isospin decomposition and the updated branching 
fractions calculated in this review.}
\label{Isospind}
\begin{tabular}{llc}
 \qquad Decay Mode & Ratio of isospin amplitudes &  
$\delta=\delta_I-\delta_{I^\prime}$ \qquad \\
\hline
 \qquad $K\pi$ & $|A_{1/2}|/|A_{3/2}|=4.12\pm 0.40$ &
 $88^{\circ}\pm 8^{\circ}$ \qquad \\
 \qquad $K^{\star}\pi$ & $|A_{1/2}|/|A_{3/2}|=5.23\pm 0.59$ &
 $90^{\circ}\pm 16^{\circ}$ \qquad \\
 \qquad $K\rho$ & $|A_{1/2}|/|A_{3/2}|=3.22\pm 0.64$ &
 $10^{\circ}\pm 47^{\circ}$ \qquad \\
 \qquad $K^{\star}\rho$ & $|A_{1/2}|/|A_{3/2}|=4.93\pm 1.95$ &
 $33^{\circ}\pm 57^{\circ}$ \qquad \\
 \qquad $KK$ & $|A_{1}|/|A_{0}|=0.58\pm 0.12$ &
 $47^{\circ}\pm 13^{\circ}$ \qquad \\
 \qquad $\pi\pi$ & $|A_{2}|/|A_{0}|=0.63\pm 0.13$ &
 $81^{\circ}\pm 10^{\circ}$ \qquad \\
\end{tabular}
\end{table}
\begin{table}
\caption{Comparisons of measured branching fractions for
Cabibbo favoured $D$ decays to predictions from the BSW 
model, the values in parentheses take into account 
isospin phase shifts}
\label{Cabfavd}
\begin{tabular}{lll}
 \qquad\qquad Decay Mode \qquad\qquad\qquad & 
Branching fraction(\%)\cite{bfd}
 & \qquad BSW model (\%) \qquad\qquad \\
\hline
\qquad\qquad $D^0\rightarrow K^-\pi^+$ \qquad\qquad\qquad & 
$3.76\pm 0.15$ & \qquad\qquad $5.0\qquad(3.8)$ 
\qquad\qquad \\
\qquad\qquad $D^0\rightarrow\overline{K^0}\pi^0$ \qquad\qquad\qquad & 
$1.99\pm 0.26$ & \qquad\qquad $0.8\qquad(2.0)$ 
\qquad\qquad \\
\qquad\qquad $D^0\rightarrow\overline{K^0}\eta^0$ \qquad\qquad\qquad & 
$0.74\pm 0.16$ & \qquad\qquad $0.3$ 
\qquad\qquad \\
\qquad\qquad $D^0\rightarrow\overline{K^0}\rho^0$ \qquad\qquad\qquad & 
$1.10\pm 0.17$ & \qquad\qquad $0.3\qquad(0.9)$ 
\qquad\qquad \\
\qquad\qquad $D^0\rightarrow K^-\rho^+$ \qquad\qquad\qquad & 
$9.8\pm 1.2$ & \qquad\qquad $8.7\qquad(8.1)$ 
\qquad\qquad \\
\qquad\qquad $D^0\rightarrow\overline{K^0}\omega$ \qquad\qquad\qquad & 
$1.7\pm 0.5$ & \qquad\qquad $0.3$ 
\qquad\qquad \\
\qquad\qquad $D^0\rightarrow K^{\star-}\pi^+$ \qquad\qquad\qquad & 
$5.1\pm 0.6$ & \qquad\qquad $2.6\qquad(2.3)$ 
\qquad\qquad \\
\qquad\qquad $D^0\rightarrow\overline{K^{\star0}}\pi^0$ \qquad\qquad\qquad & 
$2.7\pm 0.5$ & \qquad\qquad $1.0\qquad(1.2)$ 
\qquad\qquad \\
\qquad\qquad $D^0\rightarrow K^{\star-}\rho^+$ \qquad\qquad\qquad & 
$5.9\pm 2.4$ & \qquad\qquad $17.1\quad(15.3)$ 
\qquad\qquad \\
\qquad\qquad $D^0\rightarrow\overline{K^{\star0}}\rho^0$ \qquad\qquad\qquad & 
$1.4\pm 0.3$ & \qquad\qquad $1.9\qquad(3.6)$ 
\qquad\qquad \\
\qquad\qquad $D^+\rightarrow\overline{K^0}\pi^+$ \qquad\qquad\qquad & 
$2.44\pm 0.43$ & \qquad\qquad $2.5\qquad(2.5)$ 
\qquad\qquad \\
\qquad\qquad $D^+\rightarrow\overline{K^0}\rho^+$ \qquad\qquad\qquad & 
$7.3\pm 2.5$ & \qquad\qquad $11.9\quad(11.9)$ 
\qquad\qquad \\
\qquad\qquad $D^+\rightarrow\overline{K^{\star0}}\pi^+$ \qquad\qquad\qquad & 
$2.1\pm 0.4$ & \qquad\qquad $0.1\qquad(0.1)$ 
\qquad\qquad \\
\qquad\qquad $D^+\rightarrow\overline{K^{\star0}}\rho^+$ \qquad\qquad\qquad & 
$2.2\pm 1.5$ & \qquad\qquad $12.3\quad(12.3)$ 
\qquad\qquad \\
\qquad\qquad $D^+\rightarrow\overline{K^0}a_1^+$ \qquad\qquad\qquad & 
$7.9\pm 2.0$ & \qquad\qquad $3.2$ 
\qquad\qquad \\
\end{tabular}
\end{table}
\begin{table}
\caption{Measurements of
branching fractions for Cabibbo favoured $D_s$ decays compared to 
predictions from the BSW model}
\label{Cabfavds}
\begin{tabular}{lll}
 \qquad\qquad Decay Mode \qquad\qquad\qquad & 
Branching fraction(\%)\cite{bfd}
 & \qquad BSW model (\%) \qquad\qquad \\
\hline
\qquad\qquad $D_s^+\rightarrow\phi\pi^+$ \qquad\qquad\qquad & 
$3.6\pm 0.9$ & \qquad\qquad $2.7$
\qquad\qquad \\
\qquad\qquad $D_s^+\rightarrow\overline{K^0}K^+$ \qquad\qquad\qquad & 
$3.6\pm 1.1$ & \qquad\qquad $1.5$
\qquad\qquad \\
\qquad\qquad $D_s^+\rightarrow\eta\pi^+$ \qquad\qquad\qquad & 
$1.9\pm 0.6$ & \qquad\qquad $2.8$
\qquad\qquad \\
\qquad\qquad $D_s^+\rightarrow\eta\rho^+$ \qquad\qquad\qquad & 
$10.3\pm 3.2$ & \qquad\qquad $5.2$
\qquad\qquad \\
\qquad\qquad $D_s^+\rightarrow\eta^\prime\pi^+$ \qquad\qquad\qquad & 
$5.0\pm 1.9$ & \qquad\qquad $1.6$
\qquad\qquad \\
\qquad\qquad $D_s^+ \rightarrow\overline{K^{\star0}}K^+$ \qquad\qquad\qquad & 
$3.4\pm 0.9$ & \qquad\qquad $1.8$ 
\qquad\qquad \\
\qquad\qquad $D_s^+ \rightarrow K^{\star+}\overline{K^0}$ \qquad\qquad\qquad & 
$4.3\pm 1.4$ & \qquad\qquad $0.7$ 
\qquad\qquad \\
\qquad\qquad $D_s^+\rightarrow\phi\rho^+$ \qquad\qquad\qquad & 
$6.7\pm 2.3$ & \qquad\qquad $16.8$ 
\qquad\qquad \\
\end{tabular}
\end{table}
\begin{table}
\caption{Measurements of branching fractions for
Cabibbo suppressed $D$ decays to predictions from the BSW 
model, the values in parentheses take into account isospin phase shifts}
\label{Cabsupd}
\begin{tabular}{lll}
 \qquad\qquad Decay Mode \qquad\qquad\qquad & 
Branching fraction(\%)\cite{bfd}
 & \qquad BSW model (\%) \qquad\qquad \\
\hline
\qquad\qquad $D^0\rightarrow \pi^-\pi^+$ \qquad\qquad\qquad & 
$0.15\pm 0.01$ & \qquad\qquad $0.26\quad(0.18)$ 
\qquad\qquad \\
\qquad\qquad $D^0\rightarrow\pi^0\pi^0$ \qquad\qquad\qquad & 
$0.08\pm 0.02$ & \qquad\qquad $0.03\quad(0.10)$ 
\qquad\qquad \\
\qquad\qquad $D^0\rightarrow K^-K^+$ \qquad\qquad\qquad & 
$0.43\pm 0.03$ & \qquad\qquad $0.38\quad(0.30)$ 
\qquad\qquad \\
\qquad\qquad $D^0\rightarrow\overline{K^0}K^0$ \qquad\qquad\qquad & 
$0.11\pm 0.05$ & \qquad\qquad $0.\qquad(0.08)$ 
\qquad\qquad \\
\qquad\qquad $D^0\rightarrow K^{\star+}K^-$ \qquad\qquad\qquad & 
$0.31\pm 0.08$ & \qquad\qquad $0.37$ 
\qquad\qquad \\
\qquad\qquad $D^0\rightarrow K^{\star-}K^+$ \qquad\qquad\qquad & 
$0.18\pm 0.10$ & \qquad\qquad $0.14$ 
\qquad\qquad \\
\qquad\qquad $D^0\rightarrow\phi\rho^0$ \qquad\qquad\qquad & 
$0.18\pm 0.05$ & \qquad\qquad $0.08$ 
\qquad\qquad \\
\qquad\qquad $D^+\rightarrow\pi^+\pi^0$ \qquad\qquad\qquad & 
$0.25\pm 0.07$ & \qquad\qquad $0.10\quad(0.10)$ 
\qquad\qquad \\
\qquad\qquad $D^+\rightarrow\overline{K^0}K^+$ \qquad\qquad\qquad & 
$0.68\pm 0.19$ & \qquad\qquad $0.97\quad(0.97)$ 
\qquad\qquad \\
\qquad\qquad $D^+\rightarrow\overline{K^{\star0}}K^+$ \qquad\qquad\qquad & 
$0.50\pm 0.10$ & \qquad\qquad $0.37$ 
\qquad\qquad \\
\qquad\qquad $D^+\rightarrow\phi\pi^+$ \qquad\qquad\qquad & 
$0.66\pm 0.08$ & \qquad\qquad $0.26$ 
\qquad\qquad \\
\qquad\qquad $D^+\rightarrow\overline{K^{\star0}}K^{\star+}$ 
\qquad\qquad\qquad & $2.6\pm 1.1$ & \qquad\qquad $1.91$ 
\qquad\qquad \\
\end{tabular}
\end{table}
\begin{table}
\let\tabbodyfont\scriptsize
\caption{Ingredients for Factorization Tests}
\label{Ingred}
\label{TFactst}
\begin{tabular}{cccc}
\multicolumn{2}{c}{$ f_{\pi} = 131.74 \pm 0.15  $ MeV} \\
\multicolumn{2}{c}{$ f_{\rho} = 215 \pm 4  $ MeV} \\
\multicolumn{2}{c}{$ f_{a_1} = 205 \pm 16  $ MeV} \\
\multicolumn{2}{c}{$ V_{ud}$ \protect{\cite{PDG}}$ = 0.9744  \pm 0.0010 $}     \\
Charm & Bottom \\ 
\hline
$ a_1 = 1.10 \pm 0.03$ &
$ \vert c_1 \vert = 1.12 \pm 0.1$  \\
$ f_+^{DK}(0) = 0.76 \pm 0.03  $  &
$ {{d {\cal B}}\over{dq^2}}(B \to D^* \ell ~\nu)\vert_{q^2=m_{\pi}^2} =
(0.237 \pm 0.026)\:\times 10^{-2} $ GeV$^{-2}$ \\
$ A_1^{DK^*}(0) =  0.56 \pm 0.04  $   &
$ {{d {\cal B}}\over{dq^2}}(B \to D^* \ell ~\nu)\vert_{q^2=m_{\rho}^2} =
(0.250 \pm 0.030)\:\times 10^{-2} $ GeV$^{-2}$ \\
$ A_2^{DK^*}(0) = 0.39 \pm 0.08  $   &
$ {{d {\cal B}}\over{dq^2}}(B \to D^* \ell ~\nu)\vert_{q^2=m_{a_1}^2}) =
(0.335 \pm 0.033)\:\times 10^{-2} $ GeV$^{-2} $ \\
$ V^{DK^*}(0) =  1.1  \pm 0.2   $   &
$ {{d {\cal B}}\over{dq^2}}(B \to D^* \ell ~\nu)\vert_{q^2=m_{D_s}^2} =
(0.483 \pm 0.033)\:\times 10^{-2} $ GeV$^{-2} $\\
$ {{d {\Gamma}}\over{dq^2}}(D \to K \ell^+ \nu)\vert_{q^2=m_{\pi}^2} =
 9.90 \pm 0.78 \times 10^{10} s^-1$ GeV$^{-2}$  &
$ {{d {\cal B}}\over{dq^2}}(B \to D^* \ell ~\nu)\vert_{q^2=m_{D_s^*}^2} =
(0.507 \pm 0.035)\:\times 10^{-2} $ GeV$^{-2} $ \\
$ {{d {\Gamma}}\over{dq^2}}(D \to K^* \ell^+ \nu)\vert_{q^2=m_{\pi}^2} =
 4.38 \pm 0.98 \times 10^{10} s^-1$ GeV$^{-2}$ & \\
\end{tabular}
\let\tabbodyfont\small
\end{table}
\begin{table}
\caption{Test of factorization by comparing hadronic and semileptonic
decay rates.}
\label{Restestd}
\label{Tfactc}
\begin{tabular}{llcc}
&  & $R_{Exp}$ (GeV$^2$) & $R_{Theo}$ (GeV$^2$)  \\ 
\hline
Charm & $D^0 \to K^-\pi^+    $ & $1.19 \pm 0.14$  &  $1.18 \pm 0.03$   \\
 & $D^0 \to K^{*-}\pi^+ $ & $3.09 \pm 0.82$  &  $1.18 \pm 0.03$   \\
\hline
Bottom & $\bar{B}^0 \to D^{*+}\pi^- $
& $1.14\pm 0.21$
  & $1.22 \pm 0.15$   \\
 & $\bar{B}^0 \to D^{*+}\rho^- $
& $2.80\pm 0.69$ 
& $3.26 \pm 0.42$   \\
 & $ \bar{B}^0 \to D^{*+} a_1^- $
& $3.6\pm 0.9$  
& $3.0 \pm 0.50$
\end{tabular}
\end{table}
\begin{table}[htb]
\caption{Measured ratios of decay rates for
color suppressed to external spectator diagrams.}
\label{Tcolsupd}
\begin{tabular}{ll}
  \qquad\qquad\qquad\qquad Ratio & \qquad\qquad Branching Ratio\cite{bfd}
 \qquad\qquad	\\
\hline
\qquad\qquad $\Gamma(D^{0}\rightarrow \pi^0\pi^0)/
\Gamma(D^{0}\rightarrow \pi^-\pi^+)$ \qquad\qquad & \qquad\qquad
$0.56 \pm 0.15 $ \qquad\qquad \\
\qquad\qquad $\Gamma(D^{0}\rightarrow \overline{K^0}\pi^0)/
\Gamma(D^{0}\rightarrow K^{-}\pi^+)$ \qquad\qquad & \qquad\qquad
$0.53 \pm 0.07 $ \qquad\qquad \\
\qquad\qquad $\Gamma(D^{0}\rightarrow \overline{K^0}\rho^0)/
\Gamma(D^{0}\rightarrow K^{-}\rho^+)$ \qquad\qquad & \qquad\qquad
$0.11 \pm 0.02 $ \qquad\qquad \\
\qquad\qquad $\Gamma(D^{0}\rightarrow \overline{K^{\star0}}\pi^0)/
\Gamma(D^{0}\rightarrow K^{\star-}\pi^+)$ \qquad\qquad & \qquad\qquad
$0.53 \pm 0.11 $ \qquad\qquad \\
\qquad\qquad $\Gamma(D_s^+\rightarrow \overline{K^{\star0}}K^+)/
\Gamma(D_s^+\rightarrow \phi\pi^+)$ \qquad\qquad & \qquad\qquad
$0.95 \pm 0.10 $ \qquad\qquad \\
\qquad\qquad $\Gamma(D_s^+\rightarrow \overline{K^0}K^+)/
\Gamma(D_s^+\rightarrow \phi\pi^+)$ \qquad\qquad & \qquad\qquad
$1.01 \pm 0.16 $ \qquad\qquad \\
\end{tabular}
\end{table}
\begin{table}[htb]
\let\tabbodyfont\scriptsize
\caption{Measured and predicted branching fractions of color 
suppressed $B$ decays.}
\label{Tbrcolcomp}
\begin{tabular}{lcccc}
Decay Mode & U. L. (\%) & BSW (\%) &
 $\cal{B}$ (BSW) & RI~model(\%)  \\ \hline
$\bar{B^0} \to D^{0} \pi^0$     &$<0.048$ & $0.012$
 & $0.20 a_2^{2} (f_{D}/220 \rm{MeV})^2$ & $0.0013 - 0.0018$   \\
$\bar{B^0} \to D^{0} \rho^0$    &$<0.055$ & $0.008$
 & $0.14 a_2^{2} (f_{D}/220 \rm{MeV})^2$ & $0.00044$   \\
$\bar{B^0} \to D^{0} \eta$      &$<0.068$ & $0.006$
& $0.11 a_2^{2} (f_{D}/220 \rm{MeV})^2 $             &              \\
$\bar{B^0} \to D^{0} \eta^{'}$  &$<0.086$ & $0.002$
& $ 0.03 a_2^{2}(f_{D}/220 \rm{MeV})^2$  &              \\
$\bar{B^0} \to D^{0} \omega $   &$<0.063$ & $0.008$
& $0.14 a_2^{2}(f_{D}/220 \rm{MeV})^2$   &              \\
$\bar{B^0} \to D^{*0} \pi^0$    &$<0.097$ & $0.012$
& $ 0.21 a_2^{2}(f_{D*}/220 \rm{MeV})^2$ & $0.0013-0.0018$   \\
$\bar{B^0} \to D^{*0} \rho^0$   &$<0.117$  & $0.013$
& $ 0.22 a_2^{2}(f_{D*}/220 \rm{MeV})^2$ & $0.0013 -0.0018$   \\
$\bar{B^0} \to D^{*0} \eta$     &$<0.069$ & $0.007$
& $0.12 a_2^{2}(f_{D*}/220 \rm{MeV})^2$   &   \\
$\bar{B^0} \to D^{*0} \eta^{'}$ &$<0.27$  & $ 0.002$
& $0.03 a_2^{2}(f_{D*}/220 \rm{MeV})^2$   &   \\
$\bar{B^0} \to D^{*0} \omega$   &$<0.21$  & $0.013$
& $ 0.22 a_2^{2}(f_{D*}/220 \rm{MeV})^2$  &
\end{tabular}
\let\tabbodyfont\small
\end{table}
\begin{table}[htb]
\caption{Ratios of normalization modes to determine the sign of
$a_2/a_1$. The magnitude of $a_2/a_1$ is the value in the 
BSW model which agrees with our result for $B\to \psi$ modes.}\label{Tbswexpc}
\begin{tabular}{ccccc}
Ratio &$a_2/a_1 =-0.23 $ & $a_2/a_1 =0.23 $ & Experiment & RI~ model \\ \hline
$R_1 $& 0.51  & 1.64 & $1.60 \pm 0.30$ &$1.20-1.28$  \\
$R_2 $& 0.72  & 1.33 & $1.61 \pm 0.39$ &$1.09-1.12$  \\
$R_3 $& 0.49  & 1.68 & $1.85 \pm 0.40$ &$1.19-1.27$  \\
$R_4 $& 0.68  & 1.37 & $2.10 \pm 0.61$ &$1.10-1.36$
\end{tabular}
\end{table}
%
%
\newsavebox{\figone}
\savebox{\figone}{\psfig{file=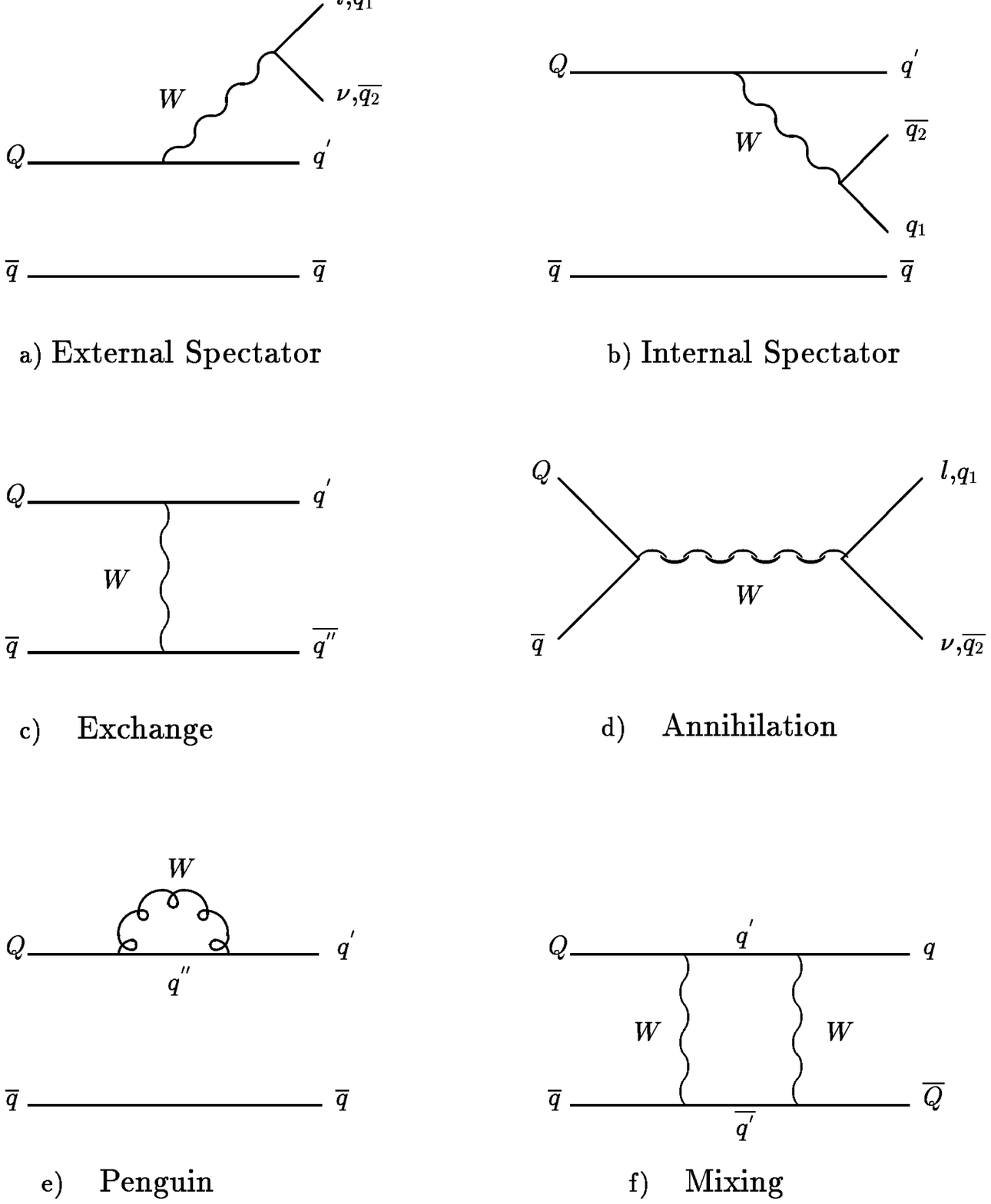,width=16.5cm}}
\newpage
\clearpage
\begin{figure}[htb]
 \begin{picture}(450,450)(0,0)
  \put(0,-200){\usebox{\figone}}
 \end{picture}
 \medskip
 \caption{Decay diagrams for mesons containing a $c$ or a $b$ quark.}
 \label{Fig1}
\end{figure}
\newsavebox{\figtwo}
\savebox{\figtwo}{\psfig{file=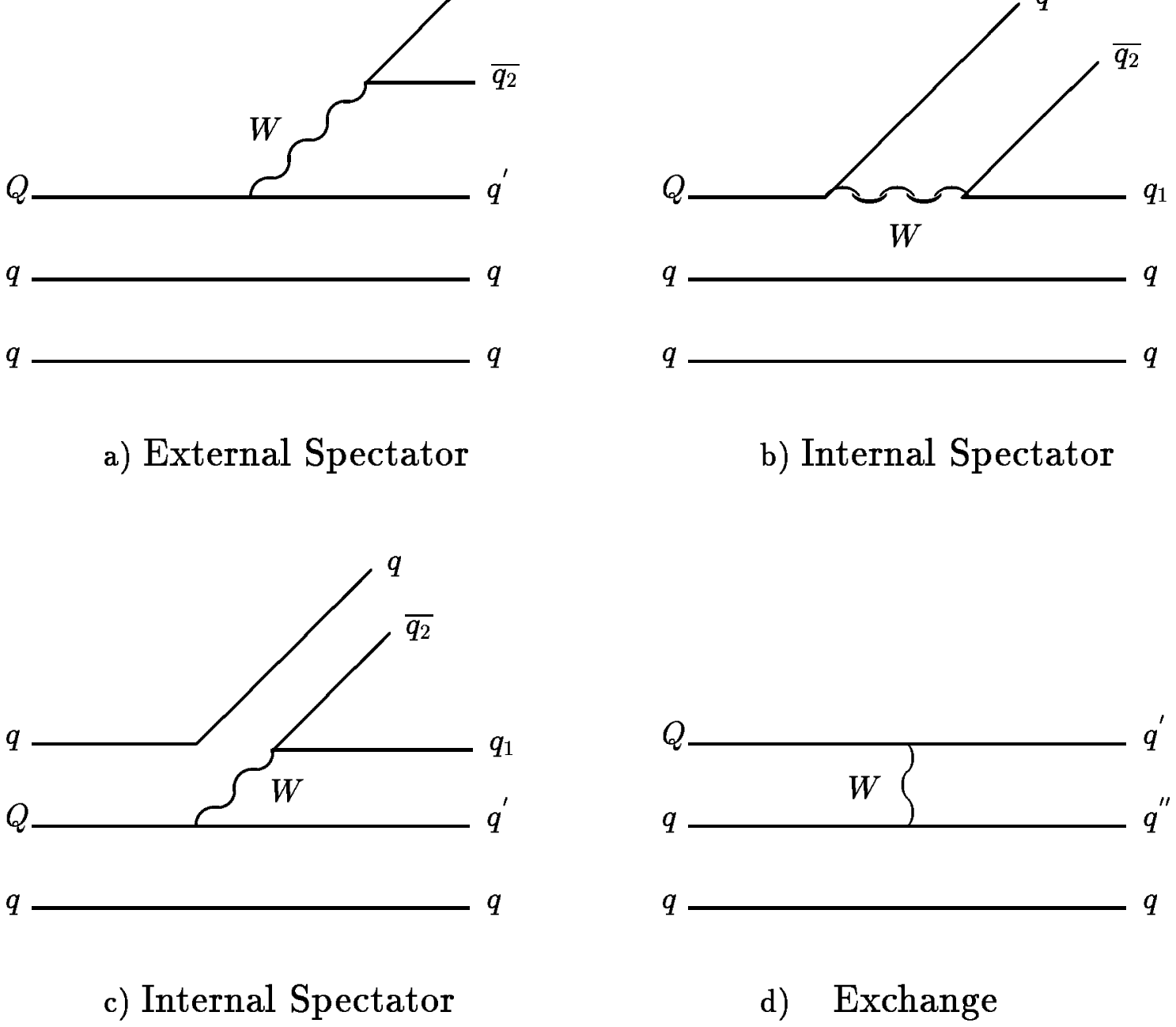,width=16.5cm}}
\newpage
\clearpage
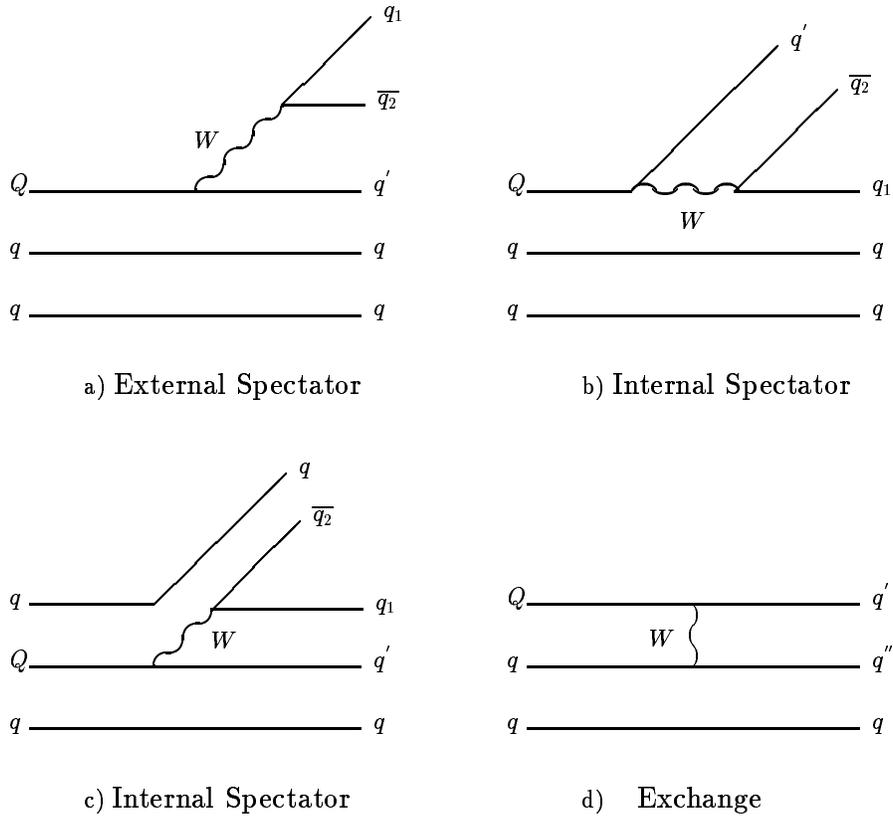
\begin{figure}[htb]
 \begin{picture}(450,450)(0,0)
  \put(0,-300){\usebox{\figtwo}}
 \end{picture}
 \caption{Hadronic decay mechanisms for baryons containing a $c$ or a $b$
quark.}
 \label{Fig2}
\end{figure}
\newpage
\begin{figure}[htb]
\unitlength 1.0in
\begin{center}
\begin{picture}(3.0,3.0)(0.0,0.0)
\put(-0.6,-1.){\psfig{bbllx=0pt,bblly=0pt,bburx=567pt,bbury=567pt,width=3.8in,height=3.8in,file=bexcl_mball.ps}}
\end{picture}
\bigskip
\bigskip
\caption{Beam constrained mass distributions from CLEO~II
for (a) $B^-$ events and (b) $\bar{B^0}$ events.}
\label{FBM}
\end{center}
\end{figure}

\begin{figure}[htb]
\unitlength 1.0in
\begin{center}
\begin{picture}(2.0,2.0)(0.0,0.0)
\put(-0.6,0.01){\psfig{width=3.8in,height=1.8in,file=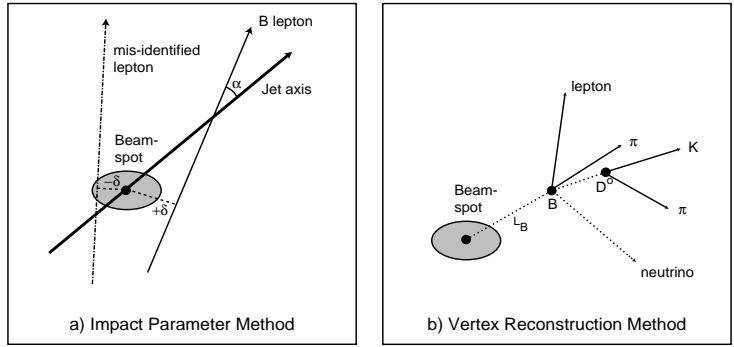}}
\end{picture}
\bigskip
\bigskip
\caption{B hadron 
lifetime measurements using the impact parameter method (a) and
decay length method (b).}
\label{impact}
\end{center}
\end{figure}
\newsavebox{\figfour}
\savebox{\figfour}{\psfig{file=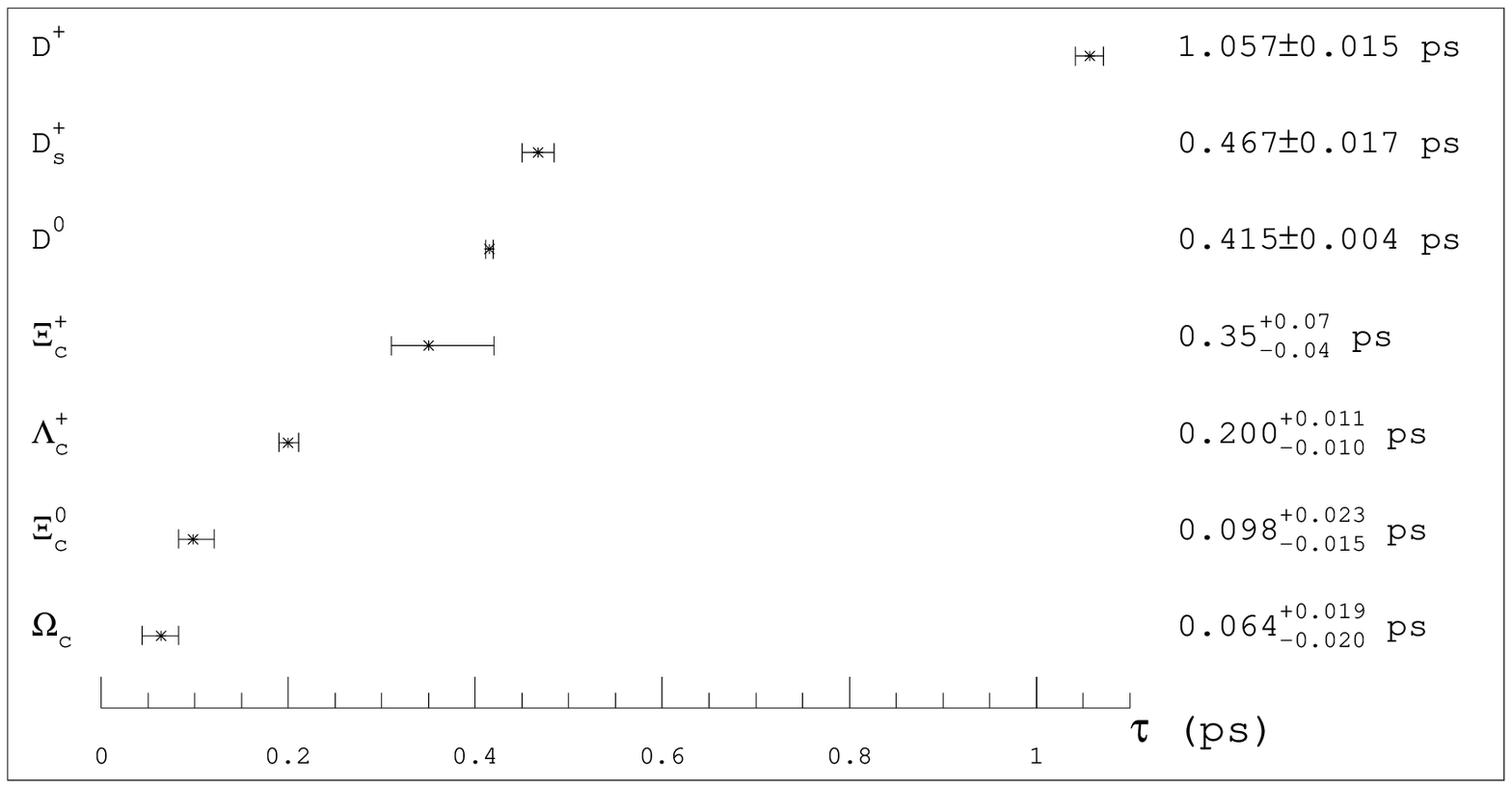,width=10.cm}}
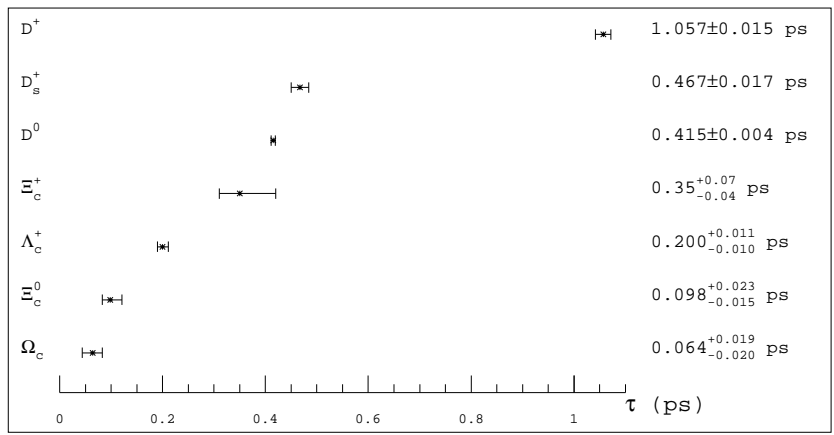
\begin{figure}[htb]
 \begin{picture}(450,450)(0,0)
  \put(40,-160){\usebox{\figfour}}
 \end{picture}
 \bigskip
 \caption{Summary of lifetime measurements  of charm hadrons.}
 \label{clife}
\end{figure}

\begin{figure}[htb]
\unitlength 1.0in
\begin{center}
\begin{picture}(3.0,3.0)(0.0,0.0)
\put(-1.,0.5){\psfig{file=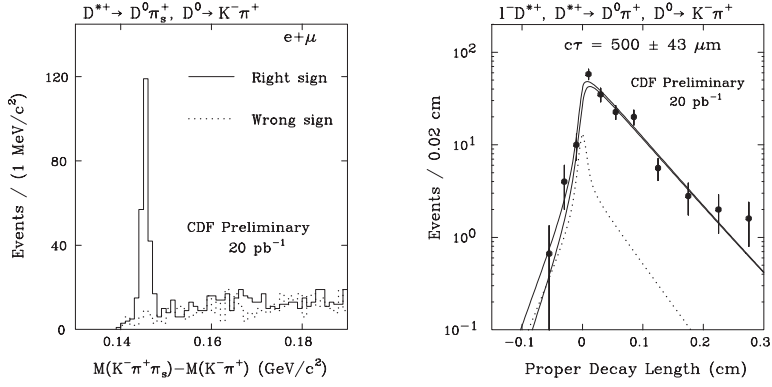,width=4in}}
\end{picture}
\vskip -10 mm
\caption{
$B^0$ lifetime measurement by CDF using $D^* - \ell$ correlations.
(a) $D^{*+}-D^0$ mass difference spectrum.
(b) Decay length distribution for right sign $D^*-\ell$
combinations.}
\label{cdf_dstarlnu}
\end{center}
\end{figure}

\begin{figure}[htb]
\unitlength 1.0in
\begin{center}
\begin{picture}(3.0,3.0)(0.0,0.0)
\put(-0.6,-0.81){\psfig{width=4.8in,height=8.8in,file=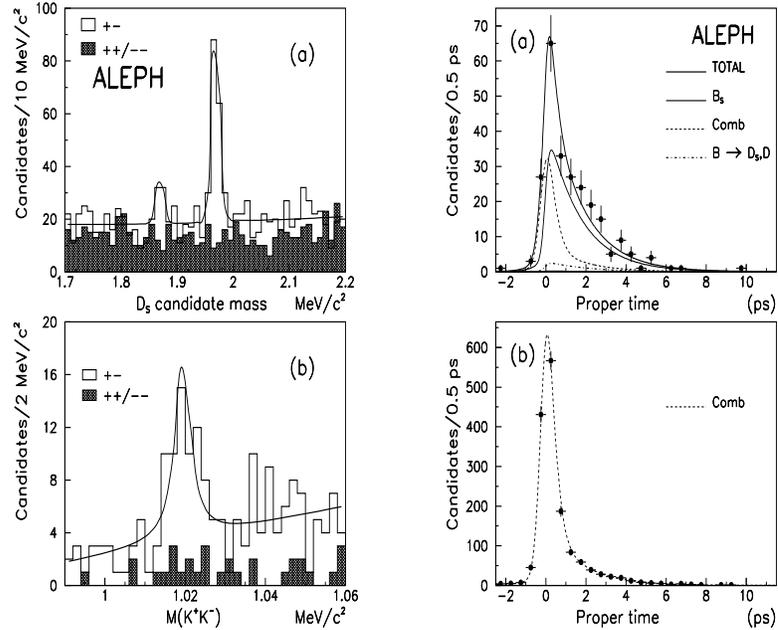}}
\end{picture}
\bigskip
\bigskip
\caption{$B_s$ lifetime measurement by ALEPH.
a) $K^-K^+\pi^+$ invariant 
mass distribution for right-sign $D_s^+\ell^-$ combinations.
Wrong sign (++ and --) are shown as a shaded histogram.
b) $K^-K^+$ invariant mass distribution for right-sign and
wrong-sign $D_s^+\ell^+$ combinations.
c) Proper time distribution of the right-sign $D_s^+\ell^-$ sample.
d) Proper time distribution of the combinatorial background.}
\label{aleph_bs_life}
\end{center}
\end{figure}
\newpage
\begin{figure}[htb]
\unitlength 1.0in
\begin{center}
\begin{picture}(3.0,2.8)(0.0,0.0)
\put(-1.5,-1.7){\psfig{bbllx=0pt,bblly=0pt,bburx=567pt,bbury=567pt,file=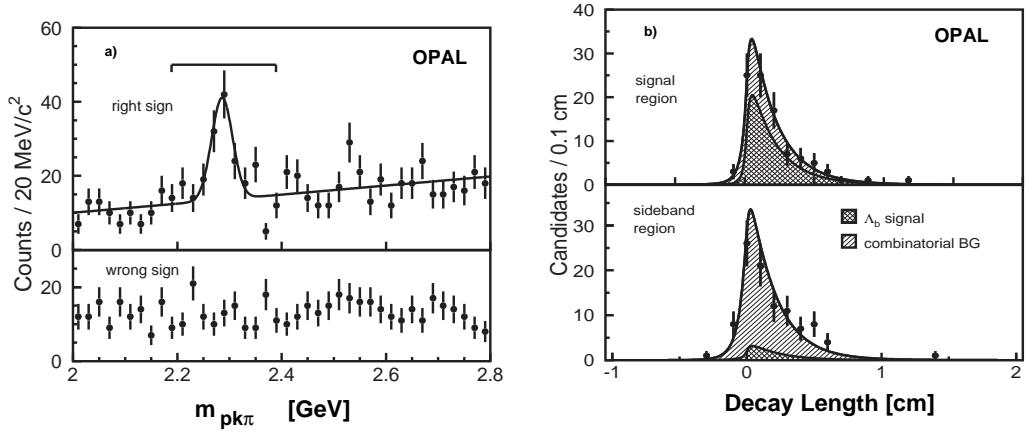,width=5.8in}}
\end{picture}
\caption{$\Lambda_b$ lifetime measurement by OPAL.
a) $pK\pi^-$ invariant mass distribution for right-sign and wrong-sign
$\Lambda_c\ell$ combinations.
b) Decay length distribution of the right-sign $\Lambda_c\ell^-$ sample
in the signal region and a sideband control region.}
\label{opal_lambdab_life}
\end{center}
\end{figure}
\newpage
\bigskip
\begin{figure}[htb]
\unitlength 1.0in
\begin{center}
\begin{picture}(2.5,3.2)(0.0,0.0)
\put(0.61,0.11){\psfig{bbllx=50pt,bblly=80pt,bburx=550pt,bbury=750pt,%
           height=2.4in,file=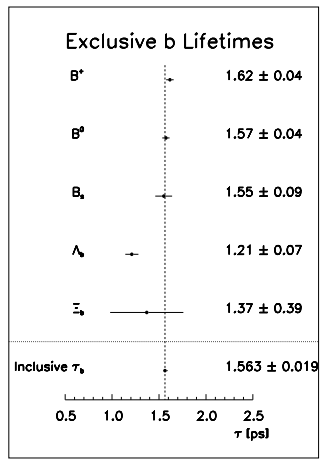}}
\end{picture}
\caption{Summary of exclusive $b$ hadron lifetime measurements.}
\label{blifetime}
\end{center}
\end{figure}

\newpage
\newsavebox{\figthree}
\savebox{\figthree}{\psfig{file=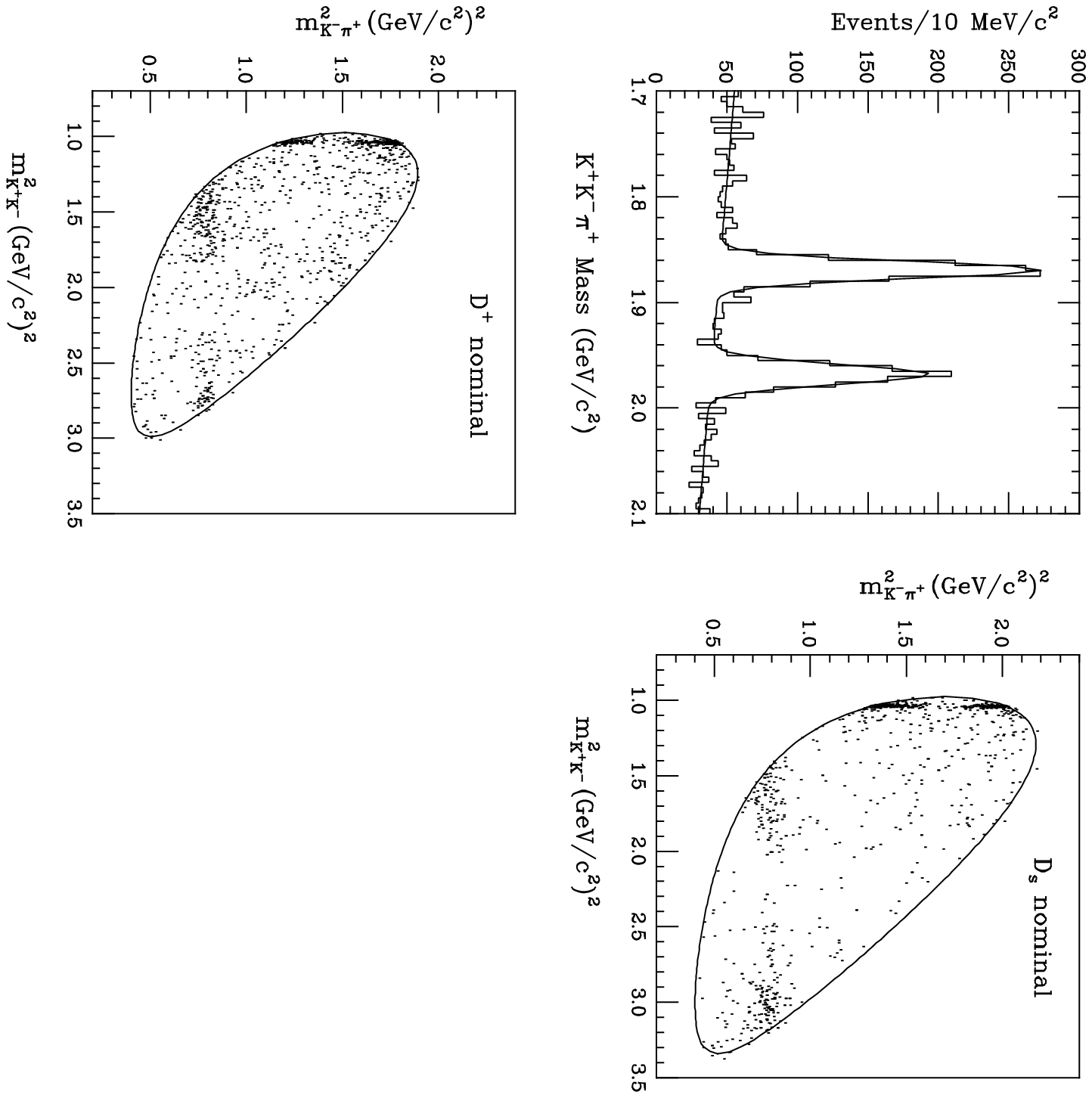,width=16.5cm}}
\begin{figure}[htb]
 \begin{picture}(450,450)(0,0)
  \begin{rotate}{270}
   \put(-80,-790){\usebox{\figthree}}
  \end{rotate}
 \end{picture}
 \medskip
 \caption{The $K^{\pm}K^{\mp}\pi^{\pm}$ invariant 
mass distribution and Dalitz plots in the $D^+$ and $D_s$ mass region
from E687.}
 \label{Kkpi_dalitz}
\end{figure}

\begin{figure}[htb]
\begin{center}
\vskip 10mm
\unitlength 1.0in
\begin{picture}(3.,2.8)(0,0)
\put(-0.7,0.3){\psfig{width=4.5in,height=3.0in,%
bbllx=0pt,bblly=0pt,bburx=567pt,bbury=567pt,file=bdds_fig9_dswithfits.ps}}
\end{picture}
\caption{$B\to D_s X$ momentum spectrum in CLEO~II data.
The solid histogram
 is the sum of the two components. The two dotted histograms
indicate the two body components from $\bar{B}\to D^{(*)} D_s^{(*)-}$ 
and $\bar{B}\to D^{(**)} D_s^{(*)-}$. 
The dash-dotted histogram shows the 
contribution of the three body process.}
\label{Fdsmomdata}
\end{center}
\end{figure}

\begin{figure}[htb]
\begin{center}
\unitlength 1.0in
\begin{picture}(3.,3.)(0,0)
\put(-2.0,-.8){\psfig{height=10.0in,file=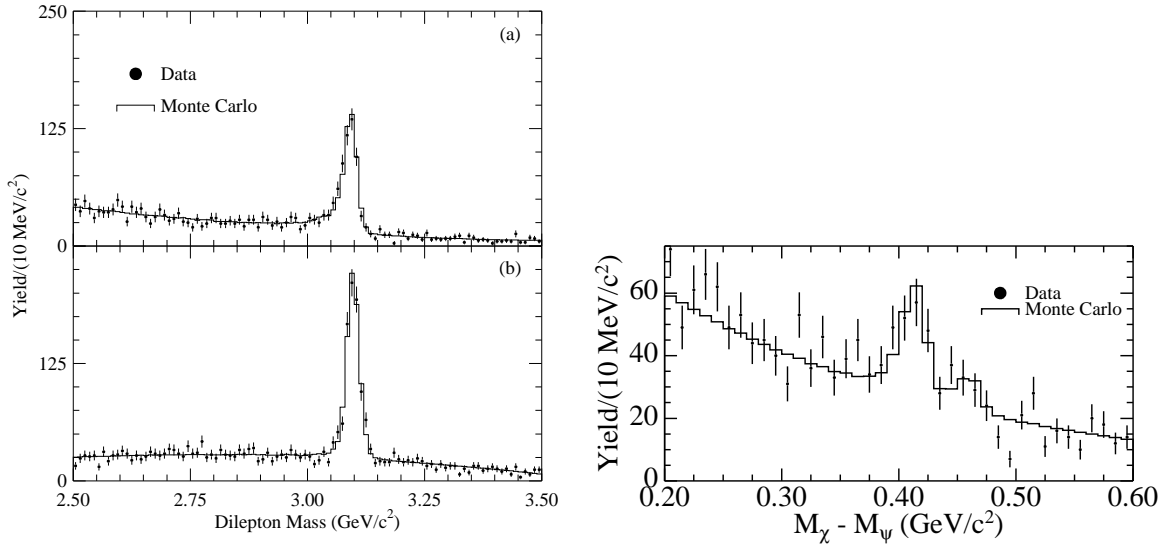}}
\end{picture}
\vskip 2mm
\caption{$B\to {\rm Charmonium} ~X$ invariant
mass spectra from CLEO II: (a) $\psi \to e^+ e^-$
channel and (b) $\psi \to \mu^+ \mu^-$ channel. (c) $\psi \gamma
- \psi$ mass difference showing the $\chi_{c1}$ and $\chi_{c2}$ signals.}
\label{Fpsi}
\end{center}
\end{figure}

\begin{figure}[htb]
\begin{center}
\unitlength 1.0in
\begin{picture}(3.,3.)(0,0)
\put(-1.9,-.8){\psfig{height=10.0in,file=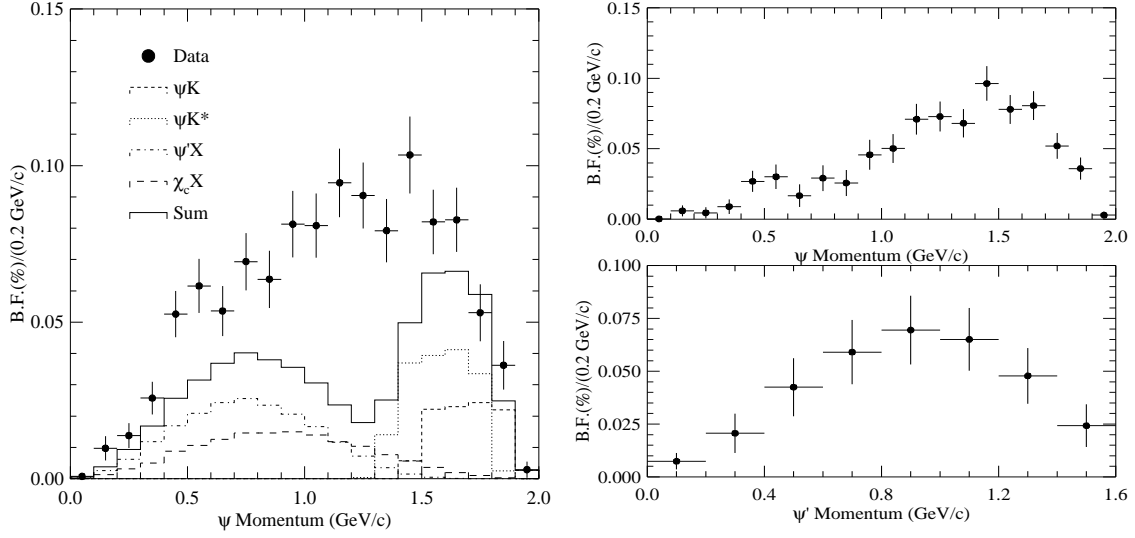}}
\end{picture}
\vskip 2 mm
\caption{$B\to {\rm Charmonium}~X$ momentum spectra in CLEO~II data.
(a) Inclusive $B \to \psi X$ production with  contributions from  individual
decay channels overlaid.
(b) Direct $B \to \psi X$ production.
(c) $B \to  \psi$'$X$.}
\label{Fpsimomdata}
\end{center}
\end{figure}

\begin{figure}[htb]
\begin{center}
\unitlength 1.0in
\begin{picture}(3.,2.5)(0,0)
\put(-1.1,-0.1)
{\psfig{bbllx=0pt,bblly=0pt,bburx=567pt,bbury=567pt,%
width=6.0in,height=4.0in,file=btobaryon.ps}}
\end{picture}
\vskip 10 mm
\caption{
Decay diagrams for $B$ meson decays to baryons: (a) External spectator
diagram (b) W Exchange diagram
(c) External spectator diagram which produces $D N \bar{N} X$ 
and $D Y \bar{Y} X$ final states (d) Internal spectator diagram which  
produces $DN\bar{N}X$ and $DY\bar{Y}X$ final states.}
\label{btobaryon}
\end{center}
\end{figure}

\begin{figure}[htb]
\begin{center}
\unitlength 1.0in
\begin{picture}(4.,4.0)(0,0)
\put(-1.1,-0.7)
{\psfig{bbllx=0pt,bblly=0pt,bburx=567pt,bbury=567pt,%
width=5.5in,height=4.5in,file=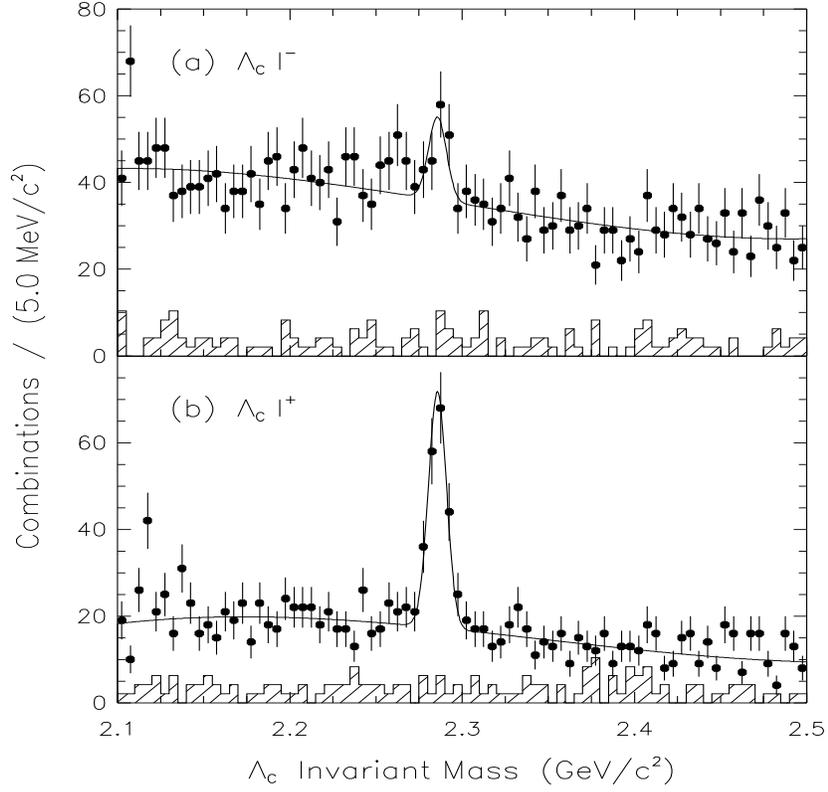}}
\end{picture}
\vskip 2 mm
\caption{$\Lambda_c -$ lepton correlation in $B$ decay (CLEO~II).
(a) The $p K^-\pi^+$ invariant 
mass spectrum for $\Lambda_c^+-\ell^-$ combinations.
(b) The $p K^-\pi^+$ invariant mass 
spectrum for $\Lambda_c^+-\ell^+$ combinations.}
\label{lambdalep}
\end{center}
\end{figure}

\begin{figure}[p]
\begin{center}
\unitlength 1.0in
\begin{picture}(3.,3.)(0,0)
\put(-.35,0.0){\psfig{width=2.5in,height=2.5in,%
file=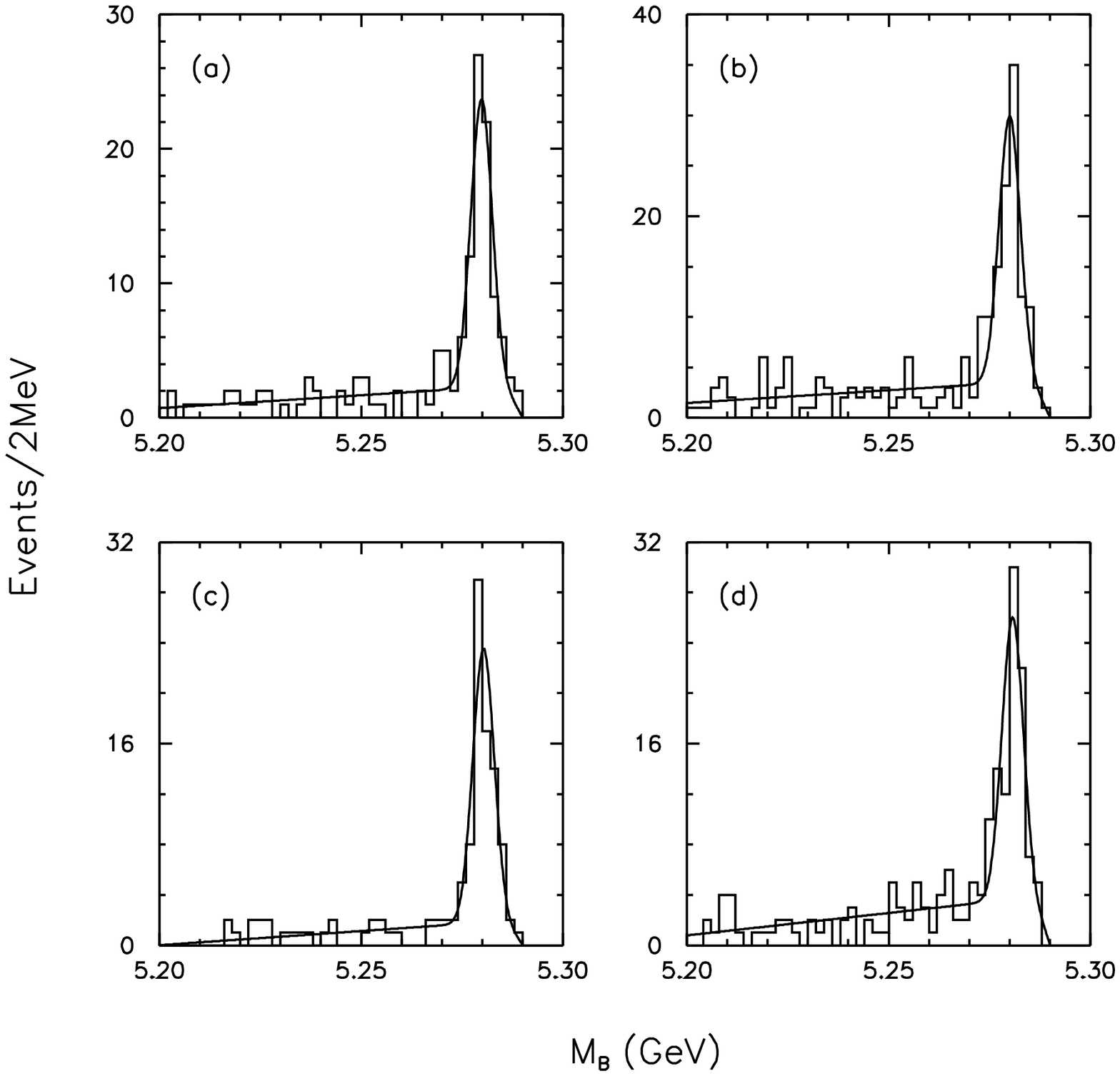}}
\end{picture}
\bigskip
\bigskip
\vskip 15 mm
\caption[]{Beam constrained mass
 distributions (CLEO~II) for: 
(a) $B^- \to D^{*0} \pi^-$ decays, 
 (b) $B^- \to D^{*0} \rho^-$ decays,
 (c) $\bar{B}^0 \to
D^{*+} \pi^-$ decays,  and
  (d) $\bar{B}^0
\to D^{*+} \rho^-$ decays.}
\label{dspi}
\end{center}

\vskip 2 mm
\begin{center}
\unitlength 1.0in
\begin{picture}(2.2,2.2)(0.0,0.0)
\put(-1.,-0.9){\psfig{width=4.in,height=4.in,file=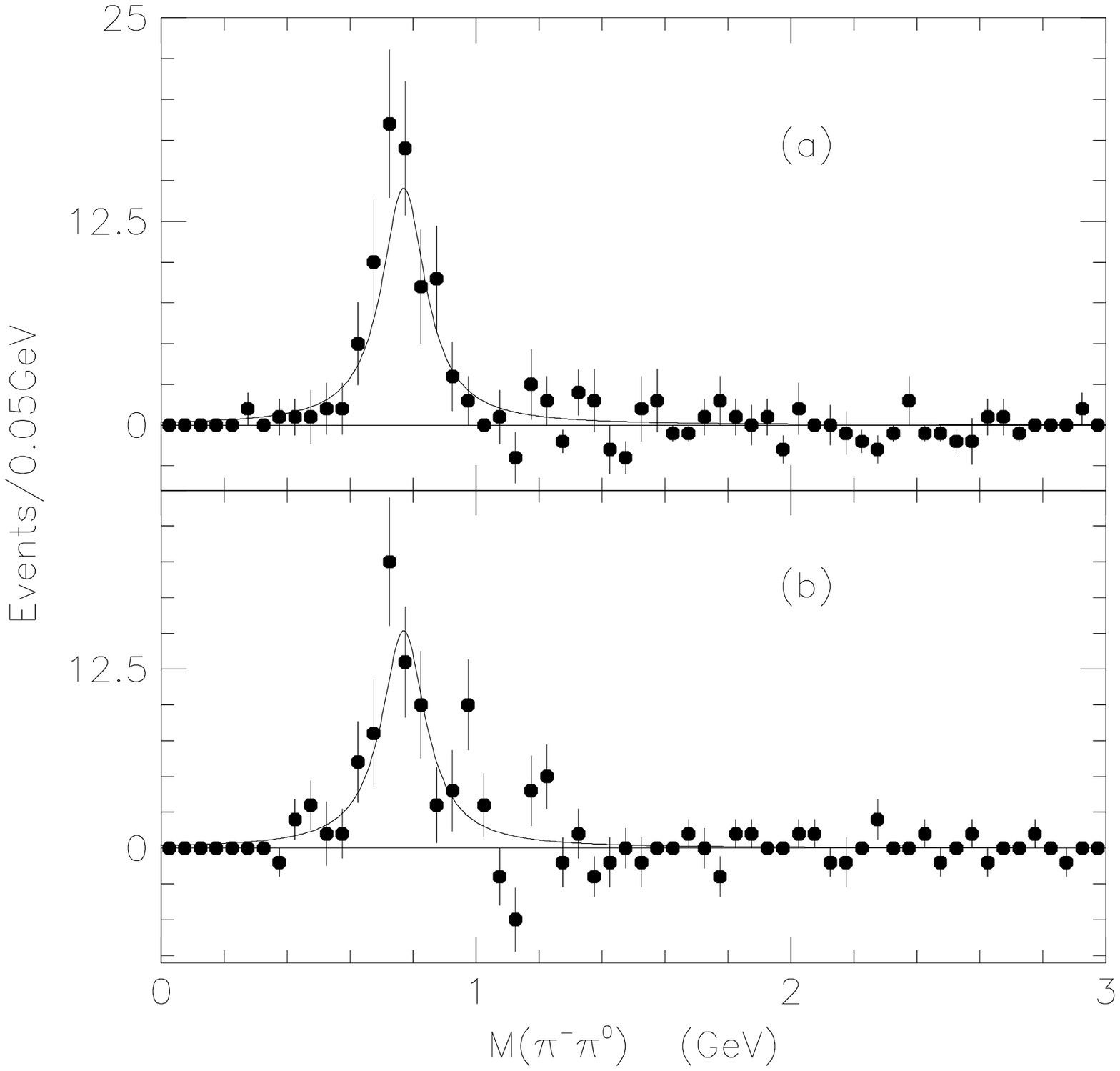}}
\end{picture}
\vskip 10 mm
\caption[]{Resonant substructure for $B\to D^* \rho^-$ (CLEO~II) for:
 (a) the
$\pi^0\pi^-$ invariant mass spectrum for 
$ \bar{B}^0 \to D^{*+} \pi^0\pi^-$.
 (b) the
$\pi^0\pi^-$ invariant mass spectrum for
$ {B}^- \to D^{*0} \pi^0\pi^-$.}
\label{subs}
\end{center}
\end{figure}
\newpage
\begin{figure}[htb]
\unitlength 1.0in
\begin{center}
\begin{picture}(3.0,3.0)(0.0,0.0)
\put(-1.1,-1.){\psfig{width=4.5in,height=4.5in,file=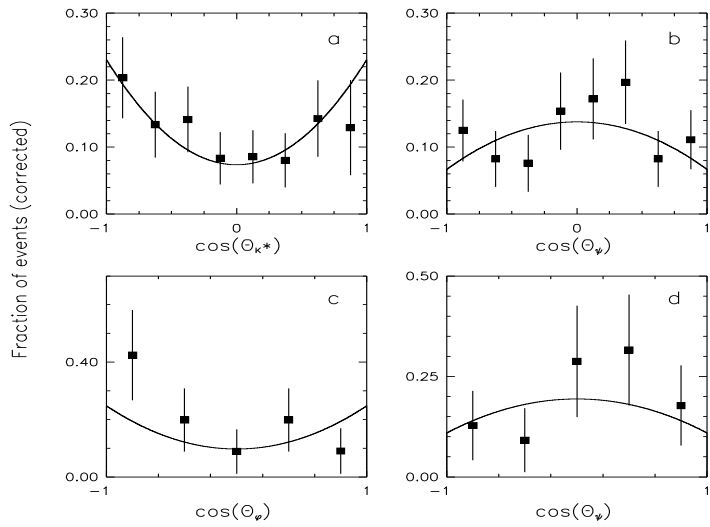}}
\end{picture}
\vskip 15 mm
\caption[]{
Distributions of the efficiency corrected
$\psi$ and $K^*$ helicity angles in reconstructed
$B \to \psi K^*$ decays from CDF.
The smooth curves are projections of the
unbinned maximum likelihood fit described in the text.}
\label{expol}
\end{center}
\end{figure}

\begin{thebibliography}{999}
\bibitem{comment_f}
 the $q\overline{q}$ assignement of the f mesons is still controversial, 
see for example the note in the Review of Particle Properties: 
Montanet,L. \etal,\PRref\PRD&50&1670(1994).
\bibitem{mcdd}{Two models are considered: non-resonant
 $B\to D \pi^- \pi^0$ and $ B\to D^{**}(2460) \pi^-$. Both  give very similar
$\pi^- \pi^0$ mass spectra and comparable limits on the non-rho contamination
in the signal region.}
\bibitem{mcdrho}{CLEO~II considers two models: (1) The $\pi^- \pi^0$ system
is produced non-resonantly following a phase space distribution or
(2) the $D^{*+} \pi^- \pi^0$ is produced from the decay
$D^{**} \pi^-$. Both give similar limits on the non-$\rho$ fraction.}
\bibitem{wex}{These modes are also accessible via the W-exchange graph which is
expected to be small in B decay.}
\bibitem{qcd}{The error is due to
the uncertainty in the scale at which to evaluate the Wilson
coefficient.}
\bibitem{width}{Since the form factor for $B\to D^* \ell \nu$ is slowly
varying, the width of the $\rho^-$ meson does not significantly modify the
result.}
\bibitem{psicomment}{The branching ratio for the modes $B\to \psi K$ and 
$B\to \psi K^{*}$ can be accommodated by the value $\xi \sim 0$ while 
$\xi \sim 1/3$ gives a branching ratio that is about a factor of 4 too low.}
\bibitem{volkie}{Private communication, Volker Rieckert; The result of a fit
including $R_4$ is not significantly different. We find in this case $a_2/a_1
\; = \; 0.282 \pm 0.07 \pm 0.06$.}
\end{thebibliography}
\end{document}